\documentclass[11pt]{article}
\usepackage{verbatim,amsmath,amssymb}
\usepackage{epsfig,color}
\usepackage[font=small,labelfont=bf]{caption}
\usepackage{geometry}
\usepackage{setspace}
\usepackage{natbib}
\usepackage{amsmath,amssymb,amsthm,mathrsfs,amsfonts,dsfont}   
\usepackage{hyperref}
\usepackage{float}
\usepackage{graphicx}
\usepackage{scalerel}
\usepackage{todonotes}
\usepackage{longtable}
\usepackage{empheq} 

\makeatletter
\newcommand\footnoteref[1]{\protected@xdef\@thefnmark{\ref{#1}}\@footnotemark}
\makeatother

\definecolor{mc}{rgb}{1,0,1} 
\definecolor{darkblue}{rgb}{0,0,1}
\definecolor{ured}{rgb}{0.9,0.3,0}
\definecolor{ugreen}{rgb}{0.0,0.5,0.0}
\definecolor{darkgreen}{rgb}{0,0.7,0}
\hypersetup{pdftex=true, colorlinks=true, breaklinks=true, linkcolor=darkblue, menucolor=darkblue, pagecolor=darkblue, citecolor=darkblue, urlcolor=darkblue}

\newcommand{\res}[1]{{{\textcolor{black}{#1}}}}

\newcommand{\otc}[1]{{\textcolor{black}{#1}}}

\newcommand{\resc}[1]{{{\textcolor{black}{#1}}}}

\newtheoremstyle{rem}
{6pt}
{6pt}
{\small}
{}
{\bf}
{:}
{.5em}
{}

\theoremstyle{rem}
\newtheorem{remarko}{Remark}[section]

\newcommand{\remark}[1]{{\remarko{{#1}}}}

\input{neco_updated.tex}

\renewcommand{\thefootnote}{\fnsymbol{footnote}}

\usepackage[font=small]{caption}

\begin{document}

\begin{center}
\Large{\bf{A finite strain model for fiber angle plasticity of textile fabrics based on isogeometric shell finite elements}}\\

\end{center}

\begin{center}
\large{Thang X. Duong$^{a}$ and Roger A. Sauer$^{b,c,d}$\footnote{corresponding author, email: roger.sauer@rub.de}}\\
\vspace{4mm}

\small{$^a$\textit{Institute of Engineering Mechanics and Structural Analysis, University of the Bundeswehr Munich,  85577 Neubiberg, Germany}}

\small{$^b$\textit{Institute for Structural Mechanics, Ruhr University Bochum, 44801 Bochum, Germany }}

\small{$^c$\textit{Department of Structural Mechanics, Gda{\'n}sk University of Technology,  80-233 Gda{\'n}sk, Poland}}

\small{$^d$\textit{Mechanical Engineering, Indian Institute of Technology Guwahati, Assam 781039, India }}

\end{center}


\renewcommand*{\thefootnote}{\arabic{footnote}}
\setcounter{footnote}{0}

\begin{center}
\small{\textbf{This is a personal version of the journal  article published in \\\textit{Journal of the Mechanics and Physics of Solids}, \href{https://doi.org/10.1016/j.jmps.2025.106158}{DOI: 10.1016/j.jmps.2025.106158}}}\\
\small{Submitted on 28 Dec 2024; Revised on 14 Mar  2025; Accepted on 14 Apr 2025.}
\end{center}

%
%
%
\rule{\linewidth}{.15mm}
{\bf Abstract:} 
This work presents a shear  elastoplasticity model for textile fabrics within the theoretical framework of anisotropic Kirchhoff-Love shells with bending of embedded fibers proposed by \cite{shelltextile}.  The  plasticity model aims at capturing the rotational inter-ply frictional sliding between fiber families in textile composites  undergoing large deformation.  Such effects are usually dominant in dry textile fabrics such as woven and non-crimp fabrics. The model explicitly uses relative angles between fiber families as strain measures for the kinematics. The plasticity model is formulated directly with surface invariants without resorting to thickness integration.   Motivated by  experimental observations from the picture frame test, a yield function is proposed with isotropic hardening and a simple evolution equation.  A classical return mapping algorithm is employed to solve the elastoplastic problem within the isogeometric finite shell element  formulation of \cite{shelltextileIGA}. The verification of the implementation is facilitated by the analytical solution  for the picture frame test. The proposed plasticity model is calibrated from the picture frame test and is then validated by the bias extension test, considering available experimental data for different samples from the literature. Good agreement between model prediction and experimental data is obtained. Finally, the applicability of the elastoplasticity model to 3D shell problems is demonstrated.
 

\vspace{-4mm}
\rule{\linewidth}{.15mm}
%
%
%
%
\vspace{-0.7cm}

\section*{List of important symbols}
\begin{longtable}[l]{ l l }
$ \bone $ & $=\bA_\alpha\otimes\bA^\alpha + \bN\otimes\bN = \ba_\alpha\otimes\ba^\alpha + \bn\otimes\bn$; 3D identity tensor \\
$a$ & model parameter of yield function $f_\mry$ \\
$\ba_\alpha$, $\ba^\alpha$  & current co- and contravariant tangent vectors at  $\bx\in\sS$; $\alpha = 1, 2$ \\ 
$\bahat_\alpha$, $\bahat^\alpha$  & co- and contravariant tangent vectors at  $\bxhat\in\hat{\sS}$; $\alpha = 1, 2$ \\ 
$\bA_\alpha$, $\bA^\alpha$ & initial co- and contravariant tangent vectors at  $\bX\in\sS_0$; $\alpha = 1, 2$ \\  
$ \ba_{\alpha,\beta}$ & parametric derivative of $\ba_\alpha$ w.r.t.~$\xi^\beta$ \\ 
$ \ba_{\alpha;\beta}$ & covariant derivative of $\ba_\alpha$ w.r.t.~$\xi^\beta$ \\  %
$ a_{\alpha\beta} $, $ a^{\alpha\beta}$  & current co- and contravariant surface metric at $\bx\in\sS$ \\ 
$ \hat{a}_{\alpha\beta} $, $ \hat{a}^{\alpha\beta}$  & co- and contravariant surface metric at $\bxhat\in\hat{\sS}$ \\ 
$A$ & model parameter of yield function $f_\mry$ \\
$ A_{\alpha\beta}$, $ A^{\alpha\beta}$ & initial co- and contravariant surface metric at $\bX\in\sS_0$ \\ %
$\alpha_\mrp$ & accumulated plastic angle strain \\
%
$b$ & model parameter of yield function $f_\mry$ \\
$ b_{\alpha\beta}$ & current covariant components of the out-of-plane curvature tensor  at $\bx\in\sS$ \\ %
$ \bar{b}_{\alpha\beta}$ & current covariant components of the in-plane fiber curvature tensor  at $\bx\in\sC\subset\sS$ \\ %
$ \bar{b}_{\alpha\beta}^i$ & current components $ \bar{b}_{\alpha\beta}$ indexed by fiber family $i$ \\ %
$B$ & model parameter of yield function $f_\mry$ \\
$ B_{\alpha\beta}$ & initial covariant components of the out-of-plane  curvature tensor  at $\bX\in\sS_0$ \\ %
$ \bar{B}_{\alpha\beta}$ & initial covariant components of the in-plane  curvature tensor \resc{at}  $\bX\in\sC_0\subset\sS_0$ \\ %
$ \beta_\bullet $ & material parameters for fiber bending and torsion \\
$c$ & model parameter of yield function $f_\mry$ \\
$\bc,\,\bc_i$ & current in-plane fiber director vector of fiber $\sC$ (or $\sC_i$) at fiber point $\bx\in\sC\subset\sS$\\%
$ \bc_{,\alpha}$ & parametric derivative of $\bc$ w.r.t.~$\xi^\beta$ \\ 
$ \bar{\bc}_{,\alpha}$ & projection of $\bc_{,\alpha}$ onto the current tangent plane\\ 
$\bc_\mathrm{0}$ & initial in-plane fiber director vector of fiber $\sC_0$ at fiber point $\bX\in\sC_0\subset\sS_0$ \\%
$ \bc_{0,\alpha}$ & parametric derivative of $\bc_0$ w.r.t.~$\xi^\beta$ \\ 
$c_\alpha$, $c^\alpha$ & co- and contravariant components of vector $\bc$ at fiber point $\bx\in\sC\subset\sS$ \\%
$\bar{c}_1,\bar{c}_2, \bar{c}_3$ & integration constants \\
$c_{0}^{\alpha}$& contravariant components of vector $\bc_0$ at fiber point $\bX\in\sC_0\subset\sS_0$ \\%
$c_{\alpha;\beta}$, $c^{\alpha}_{;\beta}$  & covariant derivatives of $c_\alpha$ and $c^\alpha$\\%
$c^{\alpha\beta\gamma\delta}$ & components of the material tangent associated with $\tauab$\\
$c^{\alpha\beta\gamma\delta}_\mra$ & components of the material tangent associated with fiber angle stress $\tauab_\mra$\\
$C$ & model parameter of yield function $f_\mry$ \\
$ \bC $ & right Cauchy-Green tensor of the shell mid-surface\\
$ \mC_{e,\alpha} $ & auxiliary shape function array for the in-plane bending contribution \\
$ \sC$ & the curve representing a fiber embedded in the current shell surface $\sS$\\
$\sC_0$ & initial configuration of fiber curve $\sC$ embedded in shell surface $\sS_0$\\
$ \sC_{i} $ & the curve of fiber family $i$; $i=1,...,n_\mrf$ \\
$ \sC_{I} $ & the curve of fiber family $I$ within a pair; $I=1,2$\\
$ \sC^\gamma_\alpha $ & components of shape function $ \mC_{e,\alpha} $ associated with $\bell\otimes\bc$\\
$ \mathscr{C}$ &  interval of the cycling loading curve\\
$ \mathscr{C}_\mre$ &  elastic phase of  loading interval $ \mathscr{C}$\\
$ \mathscr{C}_\mrp$ &   plastic phase  of  loading interval$ \mathscr{C}$\\
%
$ \dif \bullet $ & infinitesimal element of quantity $\bullet$ \\
$ \delta \bullet $ & variation of $\bullet$ \\
$ \delta_\alpha^\beta  $ & surface Kronecker delta  \\ %
$ \Delta \bullet $ & linearization of $\bullet$ \\
%
%
$ \epsilon_\bullet $ & material parameters for fiber stretching and shearing \\
$\varepsilon_{12}$ & shear stress component of the classical infinitesimal strain tensor\\
$ \bE $ & Green-Lagrange strain tensor of the shell mid-surface \\
$ E_{\alpha\beta}$ & covariant components of tensor $\bE$  at surface point $\bx\in\sS$ \\ %
$E^\mre_{\alpha\beta}$ & elastic part of $E_{\alpha\beta}$\\
%
$ \bff $ & prescribed surface loads \\
$ \mf_\bullet $ & finite element force vectors\\
$ f_\mry$, $ f_\mry^{n+1}$ &yield function, evaluated at time $t=n+1$ \\
$ f_\mry^\mathrm{trial}$ & trial yield function in the predictor-corrector algorithm \\
$ f_\mathrm{iso}$ &isotropic hardening function \\
$ \bF $ & deformation gradient of the shell mid-surface\\
$ \bF_{\!\mre} $ & elastic part of deformation gradient $\bF$\\
$ \bF_{\!\mrp} $ & plastic part of deformation gradient $\bF$\\
%
%
$ g_{12}^{\alpha\beta} $ & contravariant components of a structure tensor induced by fibers $I=1$ and $2$  \\
$ g_{12}^{\alpha\beta\gamma\delta} $ & contravariant components of the tangent tensor induced by $ g_{12}^{\alpha\beta} $   \\
$ G_\mathrm{in} $ & inertial virtual work \\
$ G_\mathrm{ext} $ & external virtual work \\
$ G_\mathrm{int} $ & internal virtual work \\
$ \Gamma^\gamma_{\alpha\beta} $ & current surface Christoffel symbols of the second kind on $\sS$\\
$ \Gamma^\mrc_{\alpha\beta} $ & components of vectors $\ba_{\alpha,\beta}$ in direction $\bc$ \\
$ \Gamma^{\uell}_{\alpha\beta} $ & components of vectors $\ba_{\alpha,\beta}$ in  direction $\bell$\\
$h$ & $=\sign (\tau)$; sign function of shear stress $\tau$\\
$h_\mre$ & $=\sign (\tau_\mre)$; sign function of shear stress $\tau_\mre$\\
$\bullet_i$,  $\bullet_j$ &  global fiber family index, taking value $i,j=1,...,n_\mrf$\\
$\bullet_I$, $\bullet_J$  &  local fiber family index within a pair of two fiber families, taking value $1$ or $2$\\
$ I_1 $ & first invariant of tensor $ \bC $ \\
$ \bI $ & $= \bA_\alpha\otimes\bA^\alpha = \bA^\alpha\otimes\bA_\alpha $; surface identity tensor on $\sS_0$\\
$J$  & $= \det_{\!\mrs}\bF$; area stretch at $\bx\in\sS$  \\
$J_\mre$  & $= \det_{\!\mrs}\bF_{\!\mre}$; elastic part of area stretch  $J$  \\
$J_\mrp$  & $= \det_{\!\mrs}\bF_{\!\mrp}$; plastic part of area stretch  $J$  \\
$ \kappa_\mrn $  & normal curvature of current fiber configuration $\sC$ at $\bx\in\sC\subset\sS$ \\ %
$ \kappa_\mrn^0 $  & normal curvature of reference fiber configuration $\sC_0$ at $\bX\in\sC_0\subset\sS_0$\\ %
$ \kappa_\mrg $& geodesic curvature of current fiber configuration $\sC$ at $\bx\in\sC\subset\sS$ \\ %
$ \kappa_\mrg^0 $ & geodesic curvature of reference fiber configuration $\sC_0$ at $\bX\in\sC_0\subset\sS_0$ \\[0.55mm] %
$ K_{\alpha\beta}$ & covariant components of the relative out-of-plane curvature tensor $\sS$ w.r.t.~$\sS_0$  \\ %
$ K^\mre_{\alpha\beta} $& elastic part of $ K_{\alpha\beta} $ \\
$ \bar{K}_{\alpha\beta}$ & covariant components of the relative in-plane curvature tensor$\sS$ w.r.t.~$\sS_0$ \\ 
$K_\mrn$ & nominal change in normal curvature $\kappa_\mrn$ at point $\bx$ of fiber $\sC\subset\sS$\\%
$K_\mrg$ & nominal change in geodesic curvature $\kappa_\mrg$ at point $\bx$ of fiber $\sC\subset\sS$\\%
$ \bell,\,\bell_i,\, \bell_I $ & current  unit tangent vector of fiber $\sC$ (or $\sC_i$, $\sC_I$) at fiber point $\bx\in\sC\subset\sS$\\ 
$\bellhat_I$, $ \bellhat_J $ & unit tangent vector of fiber indexed $I$ or $J$ at fiber point $\bxhat$ in $\hat{\sS}$\\ 
$\bar{\bellhat}_I$ &current  unit tangent vector of fiber indexed $I$ in $\bar{\hat\sS}$ in loading interval $\mathscr{C}$\\ 
$\ell_\alpha$, $\ell^\alpha$ & current co- and contravariant components of $\bell$ in $\sS$; $\alpha = 1, 2$ \\%
$\ell_I^\alpha$ &  contravariant components of $\bell$ indexed by fiber index $I$ in $\sS$; $I = 1, 2$ \\%
$\ellab$ & $=\ell^\alpha\,\ell^\beta$; current contravariant components of structural tensor $\bell\otimes\bell$ in $\sS$ \\%
$\ellab_I$ & current contravariant components $\ellab$ indexed by fiber family $I$;   $I = 1, 2$\\%
$\lambda$ & stretch of fiber $\sC$ at fiber point $\bx\in\sC\subset\sS$ \\
$\bL$ & initial unit tangent vector of fiber $\sC_0$ at fiber point $\bX\in\sC_0\subset\sS_0$\\
$L_0$ &  reference length unit \\
$\bL_I$, $\bL_J$ & fiber direction $\bL$ indexed by fiber family $I$ or $J$, where $I,J = 1, 2$ \\
$\bL^I$   & dual fiber direction to  $\bL_I$ of fiber family $I$\\
$L_\alpha$, $L^\alpha$ & co- and contravariant components of $\bL$; $\alpha = 1, 2$ \\%
$L^\gamma_{,\alpha}$ & parametric derivative of $L^\gamma$ w.r.t.~$\xi^\beta$ \\ 
$L^{\alpha\beta}$, $L_i^{\alpha\beta}$ & $ = L^\alpha\,L^\beta$; contravariant components of  tensor $\bL\otimes\bL$; possibly with fiber index  $i$\\%
$ \sL^\gamma_\alpha $ & components of shape function $ \mC_{e,\alpha} $ associated with $\bone - 2\,\bell\otimes\bell$\\
$\Lambda$ & $= \lambda^2$; square of \resc{the} stretch of fiber $\sC\subset\sS$ \\%
%
%
$\mu$ &   shear modulus of shell $\sS$ \\
$\mu_0$ &  a stress measure, i.e.~a reference stress unit \\
$\mu_\mrf$ &  shear modulus of fabrics \\
$m_\tau$, $m_\nu$, $\bar{m}$ & components of  moments causing out-of-plane, drilling, and in-plane bending \\ %
%
$\bM$ & stress couple vector associated with out-of-plane bending on a cutting \resc{boundary}\\ %
$\bMbar$ & stress couple vector associated with in-plane bending on a cutting \resc{boundary}\\ %
$\bMbar_i$ & stress couple vector $\bMbar$ indexed by fiber family $i$\\ %
$ M^{\alpha\beta} $ & contravariant components of out-of-plane bending  stress couple tensor  \\ %
$ M_0^{\alpha\beta} $ &  components $M^{\alpha\beta} $  scaled by $J$ \\ %
$ \bar{M}^{\alpha\beta} $ & contravariant components of  in-plane bending stress couple tensor   \\ %
$ \bar{M}_{0}^{\alpha\beta} $ &  components $ \bar{M}^{\alpha\beta} $  scaled by $J$ \\
 $ \bar{M}_{0i}^{\alpha\beta} $ & components $ \bar{M}_{0}^{\alpha\beta} $ indexed by fiber family $i$\\
$ \bn $ & current unit surface normal vector of $\sS$ at $\bx\in\sS$ \\ 
$ \hat{\bn} $ & unit surface normal vector of $\hat{\sS}$ at $\bxhat\in\hat{\sS}$ \\ 
$n_\mrf$ & total number of fiber families embedded in $\sS$  \\
$ \bN $ & initial unit surface normal vector of $\sS_0$ at $\bX\in\sS_0$ \\ %
$\mN_e$ & shape function array of finite element $e$ \\
$\mN_{e,\alpha}$ & parametric derivative of $\mN_e$ w.r.t $\xi^\alpha$\\
$ \sN^\gamma_\alpha $ & components of shape function $ \mC_{e,\alpha} $ associated with $\bell\otimes\bn$\\
%
%
$ \nu_\alpha $, $\nu^\alpha$ & co- and contravariant components of unit normal on a cutting boundary\\
$ \xi^\alpha $ & curvilinear coordinates; $\alpha = 1, 2$ \\ %
$p$ & external surface pressure (following surface deformation)\\
$ \sP $ & parametric domain spanned by $\xi^1$ and $\xi^2$ \\
$\phi$, $\phi^n$  & relative fiber angle change during deformation, evaluated at time $t=n$ \\
$\phi_\mre$, $\phi_\mre^n$ & elastic part of $\phi$, evaluated at time $t=n$ \\
$\phi_\mrp$, $\phi_\mrp^n$ & plastic part of $\phi$, evaluated at time $t=n$ \\
$\bar\phi$, $\bar\phi_\mre$, $\bar\phi_\mrp$  & relative angles defined analogously to $\phi$, $\phi_\mre$, $\phi_\mrp$ but interval-wise in $\mathscr{C}$\\
$\phi_\mry$ & yield angle \\
$\hat\Psi$ & \resc{Helmholtz} free energy per unit intermediate configuration area \\
%
$ q$, $ q^n$ & isotropic hardening variable,  evaluated at time $t=n$ \\
$ \tilde{q}$ & an explicit expression of $q$ in terms of interval-wise relative plastic angles $\bar\phi_\mrp$\\
$ q^0$ & value of variable $q$ at the start of loading interval  $\mathscr{C}$ \\
$Q$ & isotropic hardening variable offset by a constant \\
$Q^0$ & value of variable $Q$ at the start of loading interval  $\mathscr{C}$ \\
$ \mathbb{R}^3 $ & three dimensional vector space \\
$ \rho $ & areal mass density of surface $\sS$ \\
$s$ & arc-length parameter coordinate of fiber $\sC$ \\%
$\sS$ & current configuration of the shell surface \\ %
$\hat{\sS}$ & intermediate (fictitious) configuration of the shell surface \\ %
$\sS_0$ & reference configuration of the shell surface \\ %
$\bar{\sS}_0$ & the shell configuration at the start of loading interval  $\mathscr{C}$\\ %
$\bar{\hat\sS}$ & intermediate configuration in between $\sS$ and $\bar{\sS}_0$ within  loading interval  $\mathscr{C}$\\ %
$\partial\sS$ & boundary of $\sS$ \\
$\bullet_{\alpha\beta}^\mathrm{sym}$ & $ = \frac{1}{2}( \bullet_{\alpha\beta} + \bullet_{\beta\alpha})$; symmetrization of  $\bullet_{\alpha\beta}$\\
$ \sigma_{12} $ & $=\tau/J$; Cauchy shear stress work-conjugate to  fiber angle change $\phi_\mre$ \\
$\sigab$ & contravariant components of Cauchy stress tensor  of the shell \\
%
$ t $ & time variable \\ %
$ \bt $ & external load vector acting on shell boundary \\ %
$T_\mrg$ & nominal change in geodesic torsion $\tau_\mrg$ at point $\bx$ of fiber $\sC\subset\sS$\\%
$ \tau $ & Kirchhoff shear stress work-conjugate to  fiber angle change $\phi_\mre$ \\
$\tau^0_{12}$ &  stress $\tau$ at the start of loading interval $\mathscr{C}$\\ 
$\tau_{\mre}$ &  stress $\tau$ at the end of elastic phase  $\mathscr{C}_\mre$ of loading interval $\mathscr{C}$\\ 
$\bar{\tau}$ &  Kirchhoff shear stress contribution from loading interval $\mathscr{C}$\\ 
$ \tau^{n} $,  $ \tau^{n+1} $  & Kirchhoff shear stress evaluated at time $t=n$ and $t=n+1$\\
$ \tau^\mathrm{trial} $ & trial shear stress $\tau$ in the predictor corrector algorithms  \\
$ \bT $ & traction vector acting on a cutting boundary \\
$\tau_\mrg$& geodesic torsion of current fiber configuration $\sC$ at $\bx\in\sC\subset\sS$  \\%
$\tau_\mrg^0$ & geodesic torsion of reference fiber configuration $\sC_0$ at $\bX\in\sC_0\subset\sS_0$  \\%
$ \tau^{\alpha\beta} $ & contravariant components of the Kirchhoff stress tensor  of the shell \\
$ \tau^{\alpha\beta}_\mra $ &  components of Kirchhoff stress tensor  induced by fiber angle change  $\phi_\mre$\\
$\theta$ &   relative angle between current fibers $I$ and $J$ in $\sS$ \\
$\theta_{IJ}$ & $=\bell_I\cdot\bell_J$; angle cosine between current fibers $I$ and $J$ in $\sS$ \\
$\Theta$ &  relative angle between initial fibers $I$ and $J$ in $\sS_0$ \\
$\Theta_{IJ}$ & $=\bL_I\cdot\bL_J$; angle cosine between initial fibers $I$ and $J$ in $\sS_0$ \\
$\hat\theta $ & relative angle between current fibers $I$ and $J$ in $\hat{\sS}$\\
$\hat\theta_{IJ}$ & $=\bellhat_I\cdot\bellhat_J$; angle cosine between current fibers $I$ and $J$ in $\hat{\sS}$ \\
$\bar{\hat\theta}_{12}$ & angle cosine $\hat\theta_{12}$  in  $\bar{\hat\sS}$ within loading interval $\mathscr{C}$\\ 
$\theta^0_{12}$ &  angle cosine $\theta_{12}$ at the start of loading interval $\mathscr{C}$\\ 
$\theta^\mathrm{max}_{12}$ &  angle cosine $\theta_{12}$ at the end of loading interval $\mathscr{C}$\\ 
%
%
$ \bv $ & material velocity, i.e.~the material time derivative of $\bx$ \\ 
$ \dot{\bv} $ & material acceleration, i.e.~the material time derivative of $\bv$ \\ 
$ \sV $ & space of admissible variations $\delta\bx$ \\
$ W $ & strain energy density function per reference area \\
$ W_\mathrm{fib\mbox{-}angle}$ & part of $W$ that is induced by fiber angle change \\
$ W_\mathrm{fib\mbox{-}stretch}$ & part of $W$ that is induced by fiber stretching \\
$ W_\mathrm{fib\mbox{-}bending}$ & part of $W$ that is induced by fiber bending \\
$ W_\mathrm{fib\mbox{-}torsion}$ & part of $W$ that is induced by fiber torsion \\
$ \bx $ & current position of a surface point on the current shell surface $\sS$ \\ 
$ \bx_{,\alpha}$ & parametric derivative of $\bx$ w.r.t.~$\xi^\beta$ \\ 
$ \bxhat $ & position of a surface point on the intermediate  shell surface $\hat{\sS}$ \\ 
$ \mx_e $ & array of positions of nodes in finite element $e$ \\ 
$ \bX $ & initial position of $\bx$ on the reference shell surface $\sS_0$ \\
\end{longtable}

\setcounter{table}{0}

\section{Introduction}


This work is concerned with dry textile fabric sheets that are formed by two (or more) families of fiber bundles -- called warp and weft yarns, 
which are loosely linked together by weaving, \res{resulting in woven fabrics. 
These} fabric structures are widely used in the automotive, marine and aeronautic industry due to their high specific stiffness-to-weight ratio. Our present contribution aims at formulating a general shear elastoplasticity model for such dry textile fabrics  in  the framework of nonlinear Kirchhoff-Love shells and rotation-free \res{isogeometric  discretization. Such  fabric models are required for accurately simulating the local deformations during mechanical loading. A particular motivation is the description and optimization of draping processes that are crucial for the production of fiber reinforced composites.}

\res{ In this work,} we  focus on the angle plasticity of  woven fabrics, which is the plasticity induced by rotational sliding between warp and weft yarns when the fabric undergoes deformation. Our model assumes that the plastic shear behavior of the fabric is governed by two main mechanisms: At small deformation,  friction between warp and weft yarns is the dominant mechanism, while at large deformation, the fabric behavior is dominated by  \textit{locking} between tows (i.e.~warp and weft yarns), \res{also denoted \textit{yarn-yarn locking}}  \citep{Cao2008a}. 
Locking denotes the state when the rotation of tows becomes restricted due to their in-plane compression.





In order to formulate constitutive models   for shells (including elastoplasticity), two main approaches are usually distinguished in the literature:



The first is characterized by \textit{numerical thickness integration} that numerically reduces (or pro-jects) the underlying 3D continuum to a 2D one.
  Examples of this type are found, \res{among others, by \cite{kiendl15} and \cite{NingLiu2022}.
This} approach uses existing 3D constitutive models  and obtains their surface counterparts by numerical thickness integration. 
  The integrand is usually the 3D stress function,  but the method extends to other quantities, such as strain, energy  and algorithms. The \textit{numerical thickness integration} approach has been widely used for shells, since the computational procedure is straightforward, as long as the underlying 3D material model exists.  An example is the thickness integration of the classical $J_2$-\resc{plasticity} model with isotropic hardening, see e.g.~\cite{simo2006computational}. The method has also been applied to anisotropic large strain \resc{elastoplasticity} \citep{Schieck99}  as well as viscoplasticity  \citep{Sansour2001}. \res{Examples in the field of isogeometric Kirchhoff-Love shells are  \citet{Marreddy2018}, \citet{Huynh2020} and \citet{Alaydin2021}.}
  

The second is a \textit{direct surface} approach that avoids numerical integration.  Now the shell equations are directly formulated (or theoretically derived) in surface form, which allows to make the computational formulation more efficient.  
The approach is also referred to \resc{as}  the \textit{stress-resultant} approach in the literature, see e.g.~\cite{simo1992}.  In order to obtain this,  various modeling techniques can be used.  Depending on the treatment of the thickness, one can further subdivide the \textit{direct surface} approach   into \textit{analytical thickness integration}  and  \textit{surface invariant-based} approaches.


In  the \textit{analytical thickness integration} approach, 
the stress resultant is obtained by  (exact or approximate) analytical integration over the thickness of an existing 3D constitutive model,  
or the stress resultant is proposed directly in surface form such that  the 3D model is recovered (exactly or approximately). Therefore, this approach can usually be distinguished by the thickness variable that is considered as an extrinsic parameter (i.e.~a geometrical quantity that is not part of the material properties). \res{An example for  $J_2$-\resc{plasticity} is given by \cite{simo1992}, which is generalized from the two-yield-surface model of \cite{Shapiro61}. Other examples are found in  \cite{Skallerud2001}, \cite{Zeng2001} and \cite{Jaka2012}.
As noted in \citet{simo1992}, the yield surface is expressed in terms of the stress resultants, which in turn are obtained from an  analytical thickness integration. 
Since such an integration is quite complicated 
when applied to elastoplasticity even for simple $J_2$-\resc{plasticity}, a  highly complicating expression for the stress resultants is obtained.}

The \textit{surface invariant-based} approach can also be categorized into the \textit{direct surface} approach. 
However, in contrast to the \textit{analytical thickness integration}, the stress resultant is usually not constructed from an existing 3D material model.  Instead, it is derived from a surface energy density function that is constructed directly in terms of surface invariants without resorting to a thickness variable. 
Examples of this type are  the shell models of \cite{canham70} and \cite{helfrich73}. 
Facilitated by the representation theorem \citep{Rivlin1949}, the  invariant-based formalism -- see e.g.~\cite{Spencer1984,Rubin2007,Andrey2018} --  usually plays an important role in reducing the modeling complexity. From another perspective, \textit{surface-invariant-based} models are usually distinguished by considering the thickness variable as an intrinsic parameter of the material model (i.e.~the thickness variable becomes part of the material properties). 

In applications of the approaches discussed above, the \textit{thickness integration}  approach (either numerical or analytical) usually \resc{finds} its advantages when the shell thickness can be well-defined and well-measured, like for a metal sheet at engineering scale, so that  thickness integration can be performed for a 3D material model.  The \textit{surface invariant-based} approach, on the other hand,  finds its advantage in shell structures where the shell thickness is not well-defined to be measured accurately. An example are  lipid bilayers that behave like a liquid in-plane  and like a solid out-of-plane \citep{zimmerberg06}. An other example is  graphene, where the shell thickness is at the atomic scale with bending resistance induced by electromagnetic interaction forces \citep{Lu_2009}. In particular for dry textile shells as considered in this work, the thickness is usually not well-defined or measured accurately either, since the thickness scale is equivalent to the scale of a yarn, which consists of   fibers  only loosely compacted \citep{Allaoui12}. Further, the dominant mechanism for  angle plasticity  is  frictional contact between yarns. For each neighboring pair of fiber families, there is a yarn-yarn contact interface along the thickness direction. Accurately accounting for all contact conditions in the thickness integration can become complicating. 


Apart from  using \textit{direct surface} material models that  avoid numerical thickness integration, the efficiency in shell computations can be further improved by using Kirchhoff-Love shell formulations based on rotation-free isogeometric discretization \citep{hughes05} that requires only three displacement degrees of \res{freedom} per control point. Such a shell formulation was first proposed \res{by \cite{kiendl09} for linear shells, and later extended to
 various types of nonlinear shells by \cite{kiendl15}, \cite{tepole15} and \cite{solidshell} among many others.}

In shell formulations for dry textile fabrics, it is required to account for  in-plane bending of embedded fibers  as it becomes significant  in certain \res{loading scenarios 
\citep{Barbagallo17}. 
In this case, a gradient shell theory (e.g.~\cite{Steigmann2018}) can be used. 
Corresponding computational formulations can be found in
e.g.~\cite{Schulte2020} and \cite{witt_finite_2021}. 
Alternatively} to gradient theory,  in-plane bending of embedded fibers can also be captured by  Kirchhoff-Love shell theory extended to the geodesic curvature of  embedded fibers. As demonstrated by \cite{shelltextile}, such a generalized theory  can be formulated directly from  Kirchhoff-Love assumptions similarly to out-of-plane bending, without resorting to a general gradient shell theory.  Its rotation-free isogeometric  discretization and an efficient implementation was detailed in \citet{shelltextileIGA}. It is worth to mention here, that despite the additional kinematics, no extra degrees of freedom are required beyond the three displacement degrees of freedom  in  standard rotation-free shell formulations. 


For finite element computations of textile fabrics where dissipation is not of interest,  hyperelasticity is sufficient and \res{more efficient. 
However, elastoplasticity is} a more realistic behavior of textile fabrics.  In this context, the first  \textit{surface invariant-based}  elastoplasticity model was proposed by \cite{Denis2018} for 2D plane problems. Their model considers two fiber families with an initial relative angle of $90$ degrees and is  based on a fractional derivative  model with 20 parameters for both kinematic and isotropic hardening. It uses nested yield surfaces in order to account for the asymptotic paths of unloading curves observed in their cyclic load experiments \res{for carbon fabrics.}


\res{In this} contribution, we propose an angle elastoplasticity model directly in terms of surface invariants for textiles modeled by the (anisotropic) Kirchhoff-Love shell formulation of  \cite{shelltextile}.  \res{That is, this work extends the hyperelasticity framework of \cite{shelltextile} and its computational counterpart \citep{shelltextileIGA} to angle elastoplasticity. Therefore, the} proposed model has no limitation on the number of fiber families and initial angle between them. In order to obtain a simple but still applicable model for  loading-unloading computations,  we focus on reproducing the isotropic hardening observed in the shear behavior of fabrics. To the best of our knowledge, this is the first time such an elastoplasticity model is formulated and applied to an isogeometric shell formulation.

Our model is a  \textit{surface invariant-based}  formulation. It uses  surface invariants induced  from  fiber angles and proposes  a yield function with isotropic hardening in terms of these invariants. The expression of the yield function uses 7 physical constants  in order to capture the three deformation phases usually observed in the shear response of woven fabrics.  Since we formulate the direct surface model based on invariants,   only one (scalar) internal variable appears in the model.  It can be shown that the adopted surface invariants are induced by strain tensors of the first displacement gradient, and  correspond to a multiplicative  split of the surface deformation gradient  $\bF$  into elastic and plastic parts, $\bF_{\!\mre}$ and $\bF_{\!\mrp}$ \citep{sauer2019decomF}. Unlike classical 3D isotropic plasticity where $\det\bF_{\!\mrp} = 1$, the plastic deformation  in our formulation assumes no length change in fiber direction, which is a characteristic of many deforming textiles.

For the purpose of verifying finite element implementations, we further provide an analytical solution of the \resc{proposed} model.


In summary, our contribution contains the following novelties and merits:
  
$\bullet$   It proposes a new shear elastoplasticity  model for textile fabrics

$\bullet$   It has no limitation on the number of fiber families and the initial angles between them.

$\bullet$ Its  strain energy is directly formulated in surface form without any thickness integration.

$\bullet$  It is based explicitly on the relative fiber angles and their elastoplastic split.

$\bullet$  It proposes a yield function with isotropic hardening for dry textile fabrics.

$\bullet$  It contains the analytical solution for elastoplasticity in the  picture frame test.

$\bullet$ It contains model verification, calibration, and validation of the proposed model.

The remaining sections are organized as follows: Sec.~\ref{s:geometry} \res{summarizes the hyperelastic fabric model of \cite{shelltextile} and extends its kinematics and constitutive equations to admit angle elstoplasticity.  Sec.~\ref{s:modelelastoplasticity} then presents the proposed angle elastoplasticity model in detail.  Its analytical solution for the picture frame test is provided in Sec.~\ref{s:ana_picF}. With this, the finite element implementation is verified  and the influence of model parameters is  illustrated in Sec.~\ref{s:model_verification}.}
Sec.~\ref{s:model_verifi_cali_valid} then discusses the calibration and validation of the proposed model. A three dimensional example is shown in Sec.~\ref{s:three_dim_eg}.
 \res{Sec.}~\ref{s:conclude} concludes the paper.

\section{\res{Generalized \resc{Kirchhoff}--Love shell with embedded fibers}} \label{s:geometry} 

\res{This section first summarizes the generalized hyperelastic \resc{Kirchhoff}--Love shell theory with embedded fibers of \cite{shelltextile}. Sec.~\ref{s:angle_measure_sec} and \ref{s:Variations} then extend it to angle elastoplasticity by proposing a multiplicative split of the deformation gradient that preserves fiber length during plasticity. Adapting this enhanced kinematics to the constitutive equations is demonstrated in Sec.~\ref{s:constitute}.}


\subsection{Surface description}{\label{s:geodescription}}
\res{This subsection summarizes the essential geometrical objects used to describe thin shells with embedded fibers under large deformations within the anisotropic Kirchhoff-Love shell theory of \cite{shelltextile}.}

\begin{figure}[!htp]
\begin{center} \unitlength1cm
\begin{picture}(0,3.9)
\put(-4.5,0.0){\includegraphics[width=0.53\textwidth]{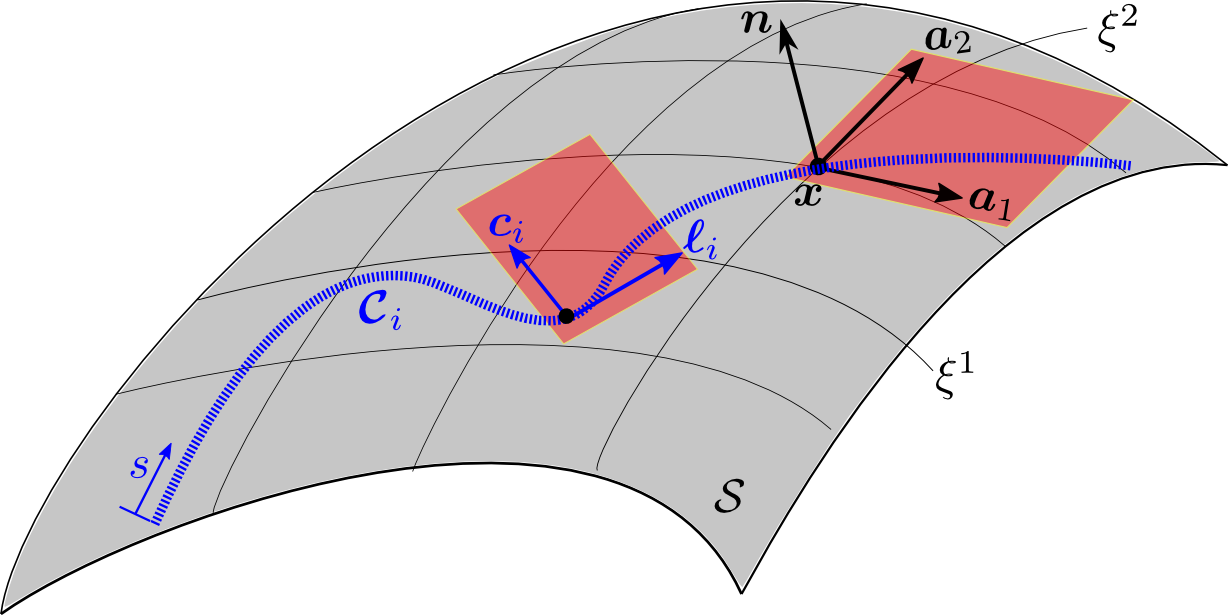}}
\end{picture}
\caption{Shell surface $\sS$ with an embedded fiber bundle along curve $\sC$ \citep{shelltextile}. Multiple fiber families are distinguished by index $i=1,2,3,..., n_\mrf$. \res{Vectors $\bell_i$, $\bc_i$, and $\bn$ maintain unit length during deformations by definition.} The red planes illustrate tangent planes. 
}
\label{f:kin0}
\end{center}
\vspace{-0.7cm}
\end{figure} 

A mathematical surface $\sS$ representing the mid-surface of the shell  in $\mathbb{R}^3$  at any time $t$  is defined by the one-to-one mapping from a point $ (\xi^1\,,\xi^2)$ in  2D parameter space $\sP$ to the point $\bx\in\sS$ as (see Fig.~\ref{f:kin0})
\eqb{l}
\bx=  \bx(\xi^\alpha,t)\,.
\label{e:shell_surface_x}
\eqe
The two tangent vectors at $\bx\in\sS$
\eqb{l} 
 \ba_\alpha:=\ds\pa{\bx}{\xi^\alpha} = \bx_{,\alpha}~,
 \label{e:tangentsl}
\eqe
 with $\alpha=1,2$ define the unit normal vector
\eqb{l}
\bn := \ds\frac{\ba_1\times\ba_2}{\norm{\ba_1\times\ba_2}}~.
\label{e:normalmsl}
\eqe
\res{In Eq.~\eqref{e:tangentsl}, the comma  denotes the parametric derivative. From Eqs.~\eqref{e:tangentsl} and \eqref{e:normalmsl},} follow the components of the surface metric
\eqb{l} 
a_{\alpha\beta}:=\ba_\alpha\cdot\ba_\beta~,
\label{e:a_ab}
\eqe
and the components of the out-of-plane surface curvature
\eqb{l} 
b_{\alpha\beta} := \bn\cdot\ba_{\alpha,\beta} = -\bn_{,\beta}\cdot\ba_{\alpha}~.
\label{e:b_ab}
\eqe
Here, the latter identity follows from  $\bn\cdot\ba_\alpha = 0$. The \resc{contravariant} metric components 
\eqb{l} 
a^{\alpha\beta} :=  [a_{\alpha\beta}]^{-1}~,
\eqe
relate the dual tangent vectors $\ba^\alpha$  to the tangent vectors $\ba_\alpha$  by\footnote{In this paper, the summation convention is applied to repeated Greek indices taking values from 1 to 2.}
\eqb{l} 
\ba^{\alpha}=a^{\alpha\beta}\,\ba_\beta~,
\eqe
which satisfies $\ba^\alpha\cdot\ba_\beta = \delta^\alpha_\beta$, \res{with $\delta^\alpha_\beta$ being the Kronecker  delta.}


Note, that vectors $\ba_{\alpha,\beta}$ appearing in~\eqref{e:b_ab} have both in-plane and out-of-plane  components, i.e.
\eqb{l} 
\ba_{\alpha,\beta} = \Gamma^\gamma_{\alpha\beta}\,\ba_\gamma + b_{\alpha\beta}\,\bn~,
\label{e:Christopfel}
\eqe
where the in-plane components $\Gamma^\gamma_{\alpha\beta}:=\ba_{\alpha,\beta}\cdot\ba^\gamma$  denote the so-called Christoffel symbols.

Further, a fiber family \res{$\sC_i$} embedded in surface $\sS$ is characterized by its normalized fiber tangent vector, \res{see Fig.~\ref{f:kin0}, defined by 
\eqb{lll}
 \bell_i:=  \ds\pa{\bx}{s}~,
 \label{e:defined_bell}
\eqe
where $s$ denotes the arc length coordinate along $\sC_i$ such that $\dif s = \norm{\dif \bx}$. Fiber direction $\bell_i$ and surface normal \eqref{e:normalmsl} define the fiber director vector 
\eqb{lll}
\bc_i := \bn \times{\res{\bell_i}}~.
 \label{e:defined_bc}
\eqe
In the following derivation for a single fiber family where no ambiguities arise, fiber index $i$ is skipped to simplify the notation. Vectors  $\bell$ and $\bc$} can be expressed in  bases $\{\ba^1,\ba^2,\bn\}$ and $\{\ba_1,\ba_2,\bn\}$ as
 \begin{empheq}[ ]{align}
{\bell}  &= \ell_\alpha\,\ba^\alpha = \ell^\alpha\,\ba_\alpha~,\label{e:bell_cv_def_1}\\[2mm]
\bc &= c_\alpha\,\ba^\alpha =  c^\alpha\,\ba_\alpha ~.
\label{e:bell_cv_def}
\end{empheq}
\res{Here,  $\ell_\alpha$ and $\ell^\alpha$ are the covariant and contravariant components of vector $\bell$. Likewise,   $c_\alpha$ and  $c^\alpha$  for components of $\bc$.  }For the sake of concise expressions, we further define \res{the auxiliary quantities }
\eqb{lll} 
\Gamma_{\!\alpha\beta}^{\mrc} \dis \bc\cdot\ba_{\alpha,\beta} = c_\gamma\,\Gamma^\gamma_{\alpha\beta} ~, \quad\quad 
  \Gamma_{\!\alpha\beta}^{\uell} := \bell\cdot\ba_{\alpha,\beta}=\ell_\gamma\,\Gamma^\gamma_{\alpha\beta}~.
\label{e:Gama_c_and_ell}
\eqe
Similarly to Eq.~\eqref{e:b_ab}, the components of the (symmetric) in-plane fiber curvature tensor can be defined from the parametric derivatives of the fiber director, $\bc_{,\alpha}$,  as
\eqb{l}
\bar{b}_{\alpha\beta} := -\ds \frac{1}{2} ( \bc_{,\alpha}\cdot\ba_\beta + \bc_{,\beta}\cdot\ba_\alpha)~.
\label{e:binplane2} 
\eqe
According to this, only the in-plane components of vectors $\bc_{,\alpha}$ contribute to the in-plane fiber curvature tensor. We thus can rewrite Eq.~\eqref{e:binplane2} as
\eqb{l}
\bar{b}_{\alpha\beta} =  \ds  -\frac{1}{2} ( \bar\bc_{,\alpha}\cdot\ba_\beta + \bar\bc_{,\beta}\cdot\ba_\alpha) =  -\frac{1}{2} ( c_{\alpha;\beta} + c_{\beta;\alpha}) ~,
\label{e:binplane2b} 
\eqe
\res{where} 
\eqb{rlll}
\bar\bc_{,\alpha} := (\ba_\beta\otimes\ba^\beta)\,  \bc_{,\alpha} 
\label{e:coderivbcbar}
\eqe
denote the projection of vectors  $\bc_{,\alpha}$ onto the current tangent plane. \res{Further in Eq.~\eqref{e:binplane2b},  the semicolon denotes the covariant derivative, which is defined for the covariant and contravariant components of $\bc_{,\alpha}$ as
\eqb{llll}
 c_{\beta;\alpha} \dis \ba_{\beta}\cdot \bc_{,\alpha} =  c_{\beta,\alpha} - c_\gamma\,\Gamma^\gamma_{\beta\alpha}~.\\[3mm]
 c^{\beta}_ {;\alpha} \dis \ba^{\beta}\cdot \bc_{,\alpha} = c^{\beta}_{,\alpha} + c^\gamma\,\Gamma_{\gamma\beta}^{\alpha}~,
 \label{e:define_semi}
\eqe
respectively.  As a result from definitions \eqref{e:coderivbcbar} and \eqref{e:define_semi},} vectors $\bar\bc_{,\alpha}$  only have in-plane components, i.e.
 \eqb{rlll}
\bar\bc_{,\alpha} =  c^\beta_{;\alpha}\,\ba_\beta = c_{\beta;\alpha}\,\ba^\beta~.
 \eqe
 
According to \cite{shelltextile}, the curvature tensors \eqref{e:b_ab} and \eqref{e:binplane2} can induce useful invariants to measure curvatures of \resc{embedded} fiber such as
\eqb{llll}
\kappa_\mrn \dis b_{\alpha\beta}\,\ellab~,\quad\quad$with$\quad\ellab:=\ell^\alpha\,\ell^\beta~,\\[3mm]
\tau_\mrg \dis b_{\alpha\beta}\,\ell^\alpha\,c^\beta~,\\[3mm]
\kappa_\mrg \dis \bar{b}_{\alpha\beta}\,\ellab~.
\label{e:curvature_measures}
\eqe
They denote the normal curvature, geodesic torsion, and geodesic curvature of the embedded fiber, respectively. Note, that only the magnitude of the curvature measures \eqref{e:curvature_measures} is strictly invariant, while their sign depends on the direction of vectors $\bn$, $\bell$, and/or $\bc$.

\subsection{Surface deformation} \label{s:surface_deformation}
In order to measure shell deformations, \res{we distinguish between the reference surface configuration $\sS_0$  and the current surface configuration $\sS$. The former can be understood as the instance of configuration $\sS$ at time $t=0$. Therefore, all the geometrical objects defined in Sec.~\ref{s:geodescription} are valid on $\sS_0$, but we rename their symbols by uppercase (or 0-indexed) symbols. Namely, shell surface $\sS_0$ is defined from Eq.~\eqref{e:shell_surface_x}  as $\bX = \bx(\xi^\alpha,t=0)$.  Tangent vectors $\bA_\alpha$,  normal vector $\bN$,  metric components $A_{\alpha\beta}$, and out-of-plane  curvature  components $B_{\alpha\beta}$ are defined analogously from Eqs.~\eqref{e:tangentsl},  \eqref{e:normalmsl}, \eqref{e:a_ab},  and \eqref{e:b_ab},   respectively. Likewise, fiber family $\sC$, its normalized fiber direction $\bell$  and  fiber director $\bc$  are redenoted by  $\sC_0$,   $\bL$ and  $\bc_0$ on  $\sS_0$, respectively. Similarly  to  Eqs.~\eqref{e:bell_cv_def_1} and \eqref{e:bell_cv_def}, vectors $\bL$ and  $\bc_0$ can be expressed in bases $\{\bA^1,\bA^2,\bN\}$ and $\{\bA_1,\bA_2,\bN\}$  as   $\bL=L^\alpha\bA_\alpha=L_\alpha\bA^\alpha$, and $\bc_0 = c^0_\alpha\,\bA^\alpha$. With these,  in-plane  curvature tensor components $\bar{B}_{\alpha\beta}$ are defined analogously to \eqref{e:binplane2} as $\bar{B}_{\alpha\beta} := - \frac{1}{2} ( \bc_{0,\alpha}\cdot\bA_\beta + \bc_{0,\beta}\cdot\bA_\alpha)$~.}

The in-plane surface deformation is characterized by the surface deformation gradient 
\eqb{lll}
\bF \dis \ba_\alpha\otimes\bA^\alpha~.
\label{e:defgrad}
\eqe
It is used to define the right Cauchy-Green surface tensor
\eqb{l}
 \bC:=\bF^\mrT\,\bF = a_{\alpha\beta}\,\bA^\alpha\otimes\bA^\beta~, 
 \label{e:tensorbC}
 \eqe
 as well as the Green-Lagrange surface strain tensor $\bE:=(\bC - \bI)/2$, that has the components
\eqb{l}
E_{\alpha\beta} :=\ds \frac{1}{2}\,(a_{\alpha\beta} - A_{\alpha\beta})~.
\label{e:bEtensor}
\eqe
\res{Here, $\bI = \bA_\alpha\otimes\bA^\alpha$ denote the surface identity tensor on $\sS_0$}.  Further, $\bF$ pushes forward  the fiber direction according to
\eqb{lll}
\lambda\,\bell = \bF\,\bL~,
\label{e:DFdef}
\eqe
where $\lambda=\norm{\bF\,\bL}$ denotes the fiber stretch. \res{Note that fiber direction $\bL\in\sS_0$ is a normalized vector. Alternatively to definition \eqref{e:defined_bell}, current fiber direction $\bell$ can be computed from deformation map~\eqref{e:DFdef} by $\bell = \bF\,\bL/\lambda$. That is, by definition $\norm{\bell} = 1$  for all deformations.}

The out-of-plane bending deformation is characterized by the relative  surface curvature
\eqb{l}
 K_{\alpha\beta} :=  b_{\alpha\beta} - B_{\alpha\beta}~,
\label{e:Kten}
\eqe
and similarly the in-plane fiber bending deformation is characterized by the relative  in-plane curvature
\eqb{l}
 \bar{K}_{\alpha\beta} :=  \bar{b}_{\alpha\beta} - \bar{B}_{\alpha\beta}~,
\label{e:Ktenb}
\eqe
where components  $\bar{b}_{\alpha\beta}$ can be computed from definition \eqref{e:binplane2}. \res{Further, thanks to the deformation map \eqref{e:DFdef} and relation \eqref{e:defined_bc}, components $c_{\beta;\alpha}$ can be rewritten from Eq.~\eqref{e:define_semi}  as} \citep{shelltextile}
\eqb{lll}
c_{\beta;\alpha} \is   - \ell_{\beta}\,   (\lambda^{-1}\,c_\gamma\,  {L}^{\gamma}_{,\alpha} + \ell^\gamma\,\Gamma^\mrc_{\gamma\alpha})~,\\[2mm]
c^{\beta}_{;\alpha}~ \is   -  \ell^{\beta}\, (\lambda^{-1}\, c_\gamma\,  {L}^{\gamma}_{,\alpha} + \ell^\gamma\,\Gamma^\mrc_{\gamma\alpha})~.
\label{e:coderivbcDef}
\eqe
Note that for multiple fiber families,  fiber-related quantities -- such as Eq.~\eqref{e:DFdef}, \eqref{e:Ktenb} and  \eqref{e:coderivbcDef} -- are defined separately for each  fiber family. 

With \res{deformation and strain measures} \eqref{e:tensorbC}, \eqref{e:Kten} and \eqref{e:Ktenb}, various \res{kinematical} invariants useful for the construction of material models can be induced. For example \citep{shelltextile}
\eqb{llllll}
I_1 \dis A^{\alpha\beta}\,\auab~,\\[2mm]
\Lambda_i \dis \auab\,L_i^{\alpha\beta} \res{ = \lambda_i^2} ~, \\[2mm]
 T_\mrg  \dis b_{\alpha\beta}\,c_0^\alpha\,L^\beta - \tau_\mrg^0~,\\[2mm]
 K_\mrn  \dis\kappa_\mrn\,\lambda^2 - \kappa_\mrn^0~,\\[2mm]
 K_\mrg \dis\kappa_\mrg\,\lambda^2 - \kappa_\mrg^0~,
 \label{e:strains_invariant}
 \eqe 
are the surface shear \res{measure,} \res{square of the} fiber stretch, the \resc{nominal} change in geodesic fiber torsion, normal fiber curvature, and geodesic fiber curvature, respectively. \res{In Eq.~\eqref{e:strains_invariant}, we have defined  $L^{\alpha\beta}:= L^\alpha\,L^\beta$, $\tau_\mrg^0 := B_{\alpha\beta}\,L^\alpha\,c_0^\beta$,  $\kappa_\mrn^0:= B_{\alpha\beta}\,L^{\alpha\beta}$, and $\kappa_\mrg^0:= \bar{B}_{\alpha\beta}\,L^{\alpha\beta}$.}

\subsection{\res{Angle measures} for shear elastoplasticity} \label{s:angle_measure_sec}
%

This section presents \res{angle measures}  suitable for angle elastoplasticity of textile materials under large deformations. It  also shows, that various measures for relative fiber angles are invariants that can be properly induced by strain tensors of the displacement gradient.
\begin{figure}[H]
\begin{center} \unitlength1cm
\begin{picture}(0,9.5)
\put(-7.5,0.0){\includegraphics[width=0.95\textwidth]{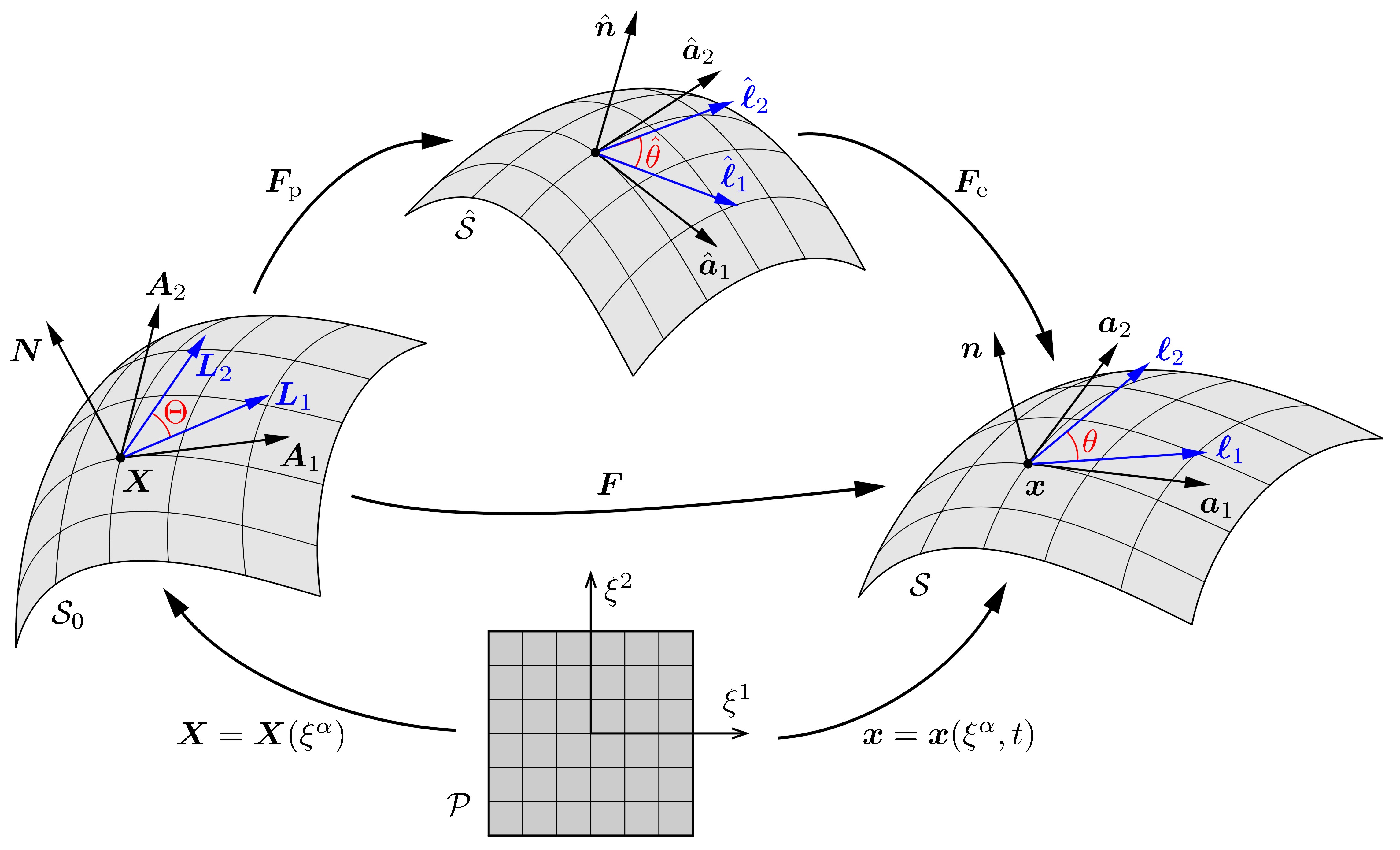}}
\put(-1.2,7.15){\small{$\bxhat$}}
\end{picture}

\caption{Split of the surface deformation into elastic and plastic parts \citep{sauer2019decomF}. This introduces the intermediate configuration $\hat\sS$ between reference and current configurations $\sS_0$ and $\sS$. The geometrical objects of Sec.~\ref{s:geodescription} are defined likewise in the three configurations. In particular, fiber angle measures  $\Theta_{12} := \bL_1\cdot\bL_2=\cos\Theta$,  $\theta_{12} := \bell_1 \cdot \bell_2 = \cos\theta$ and $\hat\theta_{12} := \bellhat_1\cdot\bellhat_2=\cos\hat\theta$ are defined here for a pair of two fiber families $\sC_1$ and $\sC_2$. 
 }
\label{f:kin}
\end{center}
\end{figure} 
To this end, an intermediate (fictitious) configuration $\hat\sS$ is introduced following \cite{sauer2019decomF}, see  Fig.~\ref{f:kin}. The geometrical objects of Sec.~\ref{s:geodescription} are  defined analogously  on $\hat\sS$ using a hat, giving  tangent vectors $\hat\ba_\alpha$,  normal vector $\hat\bn$, and metric components $\hat{a}_{\alpha\beta}$. 
With this, the deformation gradient \eqref{e:defgrad} is split into  plastic and elastic parts as  
\eqb{lll}
\bF =  \bF_{\!\mre}\,\bF_{\!\mrp} ~,
\label{e:DFdef_split}
\eqe
where
 \begin{empheq}[ ]{align}
\bF_{\!\mrp} &:=\bahat_\alpha\otimes\bA^\alpha~,\label{e:DFdef_splitp}\\[3mm]
\bF_{\!\mre} &:= \ba_\alpha\otimes\bahat^\alpha~\label{e:DFdef_splite}.
\end{empheq}
The area  stretch $J:=\det_\mrs\bF$ can then be expressed as
\eqb{lll}
J = J_\mre\,J_\mrp~,\quad $where$\quad J_\mre:=\det_\mrs\bF_{\!\mre} \quad$and$\quad J_\mrp:=\det_\mrs\bF_{\!\mrp}~.
\label{e:JeJp}
\eqe

In order to characterize angle plasticity,  we consider  a pair of fiber families. \res{For multiple pairs of fiber families in $\sS$, the kinematics in this section is simply replicated for each pair. We use uppercase Latin indices, such as $I,J$ -- taking value $1$ or $2$ -- to count the two fiber families locally within a pair. Summation is also implied on those repeated Latin indices unless otherwise stated. They should not be confused with the lowercase  Latin indices used to count all the fiber families appearing in $\sS$, and  with Greek indices used for the two curvilinear coordinates.}

Under deformations, the  directions of  fiber  family $\res{\sC_I}$  
in configurations $\sS_0$,  $\hat\sS$, and $\sS$ are $\bL_\res{I}$, $\bellhat_\res{I}$, and $\bell_\res{I}$, respectively. The relative angles cosines between  fibers \res{$I$} and \res{$J$} are defined by
 \begin{empheq}[ ]{align}
\Theta_\res{IJ} &:=  \bL_\res{I}\cdot\bL_\res{J}  ~,\label{e:defineGamma_0}\\[2mm]
\thetahat_\res{IJ}  &:= \bellhat_\res{I}\cdot\bellhat_\res{J} ~,\label{e:defineGamma_hat}\\[2mm]
\theta_\res{IJ}  &:=  \bell_\res{I}\cdot \bell_\res{J}
\label{e:defineGamma_current}
\end{empheq} 

in configurations $\sS_0$,  $\hat\sS$, and $\sS$, respectively.  Note that in these equations  $\Theta_\res{II}=\thetahat_\res{II}=\theta_\res{II}=1$ (no sum on \res{$I$}), since vectors $\norm{\bL_\res{I}}=\norm{\bellhat_\res{I}}=\norm{\bell_\res{I}} = 1$. Therefore, the mixed components 
become $\Theta_{12}=\cos\Theta$,  $\thetahat_{12}=\cos\thetahat$, and $\theta_{12}=\cos\theta$, where $\Theta$, $\thetahat$, and $\theta$ denote  the corresponding relative angles between fiber family $1$ and $2$ in  configurations $\sS_0$,  $\hat\sS$, and $\sS$, respectively.
%
 Further,   Eqs.~\eqref{e:defineGamma_0}-\eqref{e:defineGamma_current} induce surface metrics that can be arranged in matrix form, e.g.
\eqb{lll}
 [\Theta_\res{IJ}] =  \left[\begin{matrix}
1 & \Theta_{12}  \\ 
\Theta_{12} & 1 \\ 
\end{matrix} \right]~. 
\label{e:defined_angle_metric}
\eqe                       
For the angle plasticity model proposed in this work, we employ the following measures of the  relative fiber angle change during deformation
 \begin{empheq}[ ]{align}
\phi  &:=    \theta_{12}  - \Theta_{12}  ~,\label{e:definePhi1}\\[2mm]
\phi_{\mre} &:=  \theta_{12}  - \thetahat_{12} ~,\label{e:definePhi2}\\[2mm]
\phi_{\mrp} &:= \thetahat_{12} - \Theta_{12}~,
\label{e:definePhi3}
\end{empheq}
that  satisfy the additive split 
\eqb{lll}
\phi = \phi_{\mre} + \phi_{\mrp}~.
\label{e:definePhi_total}
\eqe
%
  %
 %
%
\res{In most textile composites, since fiber tensile stiffness usually dominates over  shear stiffness, we assume there is no change in fiber length during angle plasticity. We thus} pick deformation gradient $\bF_{\!\mrp}$  such that fiber lengths are preserved  in  configuration  $\hat\sS$. Namely,
\eqb{lll}
\bF_{\!\mrp} =  \bellhat_\res{I}\otimes\bL^\res{I}~,\label{e:DFdef_splitp_defined}
\eqe
where  $\bL^\res{I}$ is the fiber direction vector  dual to $\bL_\res{I}$  in  $\sS_0$, defined by
\eqb{lll}
\bL^\res{I} = \Theta^\res{IJ}\,\bL_\res{J}~, \quad $with$\quad [ \Theta^\res{IJ}]:= [\Theta_\res{IJ}]^{-1}~.
\label{e:defined_angle_dual}
\eqe
%
%
By comparing Eq.~\eqref{e:DFdef_splitp_defined} and \eqref{e:DFdef_splitp} we obtain an expression for $\bahat_\alpha$ in terms of the fiber directions, i.e.
\eqb{lll}
\bahat_\alpha = (\bL^\res{I}\cdot\bA_\alpha)\, \bellhat_\res{I} =   \bF_{\!\mrp} \,\bA_\alpha~.
\label{e:defined_bahat}
\eqe
%

\remark{The  \res{angle change} $\phi$, $\phi_{\mre}$ and $\phi_{\mrp}$ defined from Eq.~\eqref{e:definePhi1}-\eqref{e:definePhi3} are invariants that can be properly induced by strain tensors related to the displacement gradient. This is shown in Apendix~\ref{s:prove_invariant}. It should be noted, that similarly to the curvature measures \eqref{e:curvature_measures}, only the magnitude of these measures  is  invariant in a strict sense, since their sign still depends on the direction of the fiber directions, which is exchangeable.  However, it can easily be shown that (both the sign and magnitude of) the measures $\phi$, $\phi_{\mre}$ and $\phi_{\mrp}$ are frame invariant under superimposed rigid body motions of the shell surface. }

\subsection{\res{Variations and linearizations of kinematical quantities}} \label{s:Variations}
\res{This section recalls essential variations (and linearizations)  of some kinematical quantities mentioned above.  Quantities that are not listed here can be found elsewhere,~e.g in~\citet{shelltheo} and \citet{shelltextile}.}

We consider a variation of the surface deformation, denoted $\delta\bx$, which causes variations in \res{tangent vectors} $\ba_\alpha$, denoted $\delta\ba_{\alpha} = \delta\bx_{,\alpha}$. The variation of the kinematical quantities in Eqs.~\eqref{e:bEtensor}, \eqref{e:Kten} and \eqref{e:Ktenb}  requires $\delta\auab$, $\delta\buab$ and $\delta\bar{b}_{\alpha\beta}$. They are given by \citep{shelltheo,shelltextile} 
\eqb{rrll}
\delta{a}_{\alpha\beta} \is \delta\ba_\alpha\cdot\ba_\beta + \ba_\alpha\cdot\delta\ba_\beta~,\\[3mm]
\delta{b}_{\alpha\beta} \is \bn\cdot\delta\bd_{\alpha\beta}~, \quad$with$ \quad  \delta\bd_{\alpha\beta} := \delta \vauacb - \Gamma_{\alpha\beta}^\gamma\,\delta\ba_\gamma~,\\[3mm]
\bar{M}_0^{\alpha\beta}\,\delta\bar{b}_{\alpha\beta} \is  - \bar{M}_0^{\alpha\beta}\,(\delta\ba_\alpha \cdot \bar\bc_{,\beta}  + \ba_\alpha\cdot\delta\bar\bc_{,\beta})~,
\label{e:variations}
\eqe
where the last equation holds for any symmetric tensor $\bar{M}_0^{\alpha\beta}$.

\res{The variation of surface normal vector $\bn$, fiber direction $\bell$ and fiber director $\bc$ can be shown to be expressible in terms of $\delta\ba_\alpha$ (cf.~Eqs.~(204), (206), and (211) in \citet{shelltextile}) as 
 \begin{empheq}[ ]{align}
 \delta\bn &= -\ba^\alpha\,(\bn\cdot\delta\ba_\alpha) ~.
 \label{e:var_bn}\\[2mm]
\delta\bell &= ( \bn\otimes\bn + \bc\otimes\bc)\,\ella\,\delta\ba_\alpha~.
\label{e:deltaell}\\[2mm]
\delta\bc &= -(\bn\otimes\bc)\,\delta\bn - (\bell\otimes\bc)\,\delta\bell~.
\label{e:deltac1}
\end{empheq}
Consequently, the variation $\delta\phi$ following from Eq.~\eqref{e:definePhi1}, with \eqref{e:defineGamma_current} and \eqref{e:deltaell}, can be expressed fully in terms of the variation of the surface metric as}
\eqb{lll}
\delta\phi = \delta\theta_{12}  =  g_{12}^{\alpha\beta}\,\delta\auab~,
\label{e:var_delta_gamma_12}
\eqe
where we have defined 
\eqb{lll}
g^{\alpha\beta}_{12} := \ds   \frac{1}{2}\Big(\ell_{1}^{\alpha}\,\ell_{2}^{\beta} + \ell_{1}^{\beta}\,\ell_{2}^{\alpha}\Big) - \frac{\theta_{12} }{2}\,\Big( \ellab_1 + \ellab_2\Big)~,\quad\quad $with$\quad \ellab_\res{I} := \ell_\res{I}^\alpha\,\ell_\res{I}^\beta~$(no sum on \res{$I$})$.
\label{e:l12ab}
\eqe
\res{From this equation and \eqref{e:deltaell},  follows the linearization} 
\eqb{lll}
\Delta g^{\alpha\beta}_{12} = g^{\alpha\beta\gamma\delta}_{12}\,\Delta a_{\gamma\delta}~,
\eqe
\res{where}
\eqb{lll}
g^{\alpha\beta\gamma\delta}_{12}:= \ds \pa{g^{\alpha\beta}_{12}}{a_{\gamma\delta}} = -\Big(\ell_{1}^{\alpha}\,\ell_{2}^{\beta}\Big)^{\mathrm{sym}}\,\frac{1}{2}\Big(\ell^{\gamma\delta}_1 + \ell^{\gamma\delta}_2\Big) -  \frac{1}{2}\,\Big(\ellab_1 + \ellab_2\Big)\,g_{12}^{\gamma\delta} +  \frac{\theta_{12} }{2}\,\Big(\ellab_1\,\ell^{\gamma\delta}_1 + \ellab_2\,\ell^{\gamma\delta}_2\Big)~.
\label{e:l12abcd}
\eqe

\subsection{\res{Weak form}} \label{s:weak_form}
%

\res{This section presents the weak form and the material tangents for the generalized Kirchhoff-Love  shell with embedded fibers of \citet{shelltextile}. For all variation $\delta\bx$ within} the set of kinematically admissible variations $\sV$, the weak form follows from the principle of virtual work as \citep{shelltextile}
\eqb{l}
G_\mathrm{in} + G_\mathrm{int} - G_\mathrm{ext} = 0  \quad\forall\,\delta\bx\in\sV~,
\label{e:wfu}
\eqe
where 
 \begin{empheq}[ ]{align}
G_\mathrm{in} &= \ds\int_{\sS_0} \delta\bx\cdot\rho_0\,\dot\bv\,\dif A~,\label{e:Giie_2} \\[2mm]
G_{\mathrm{int}} &= \ds\frac{1}{2}\int_{\sS_0}\, \tau^{\alpha\beta} \,\delta{a}_{\alpha\beta}\,\dif A + \int_{\sS_0}\Mab_0\,\delta{b}_{\alpha\beta}\,\dif A + \sum_{i=1}^{n_\mrf} \int_{\sS_0}\Mbarab_{0i}\,\delta{\bar{b}}^i_{\alpha\beta}\,\dif A~,\label{e:Giie_2}\\[2mm]
G_\mathrm{ext} &=\ds\int_{\sS}\delta\bx\cdot\bff\,\dif a 
+ \ds\int_{{\partial\sS}} \delta\bx\cdot\bT\,\dif s + \ds \int_{\partial\sS} \delta\bn\cdot\bM\,\dif s + \ds \sum_{i=1}^{n_\mrf} \int_{\partial\sS} \delta\bc_i\cdot\bMbar_{\!i}\,\dif s~.
\label{e:Giie_3}
\end{empheq}
Here, \res{ $\dot\bv$ and $\rho_0$  denote the material acceleration and the mass density with unit [mass / reference area], while} 
$n_\mrf$,  $\tau^{\alpha\beta}$, $M_0^{\alpha\beta} $, and $\bar{M}_{0i}^{\alpha\beta}$ are   the number of fiber families, the components of the  effective stress tensor, the nominal stress couple tensor associated with out-of-plane bending, and the nominal stress couple tensor associated with in-plane fiber bending, respectively. \res{Eq.~\eqref{e:Giie_3} contains  the external body force  $\bff$, external boundary traction $\bT$,   external out-of-plane bending moment $\bM$, and external in-plane fiber bending moment $\bMbar_{\!i}$.  Apart from these \resc{Neumann} type boundary conditions,  displacements and rotations can be prescribed on the shell boundary. Further in Eqs.~\eqref{e:Giie_2}-\eqref{e:Giie_3}, the variations of tensor components  $\delta\auab$, $\delta\buab$, and $\delta{\bar{b}}_{\alpha\beta}$,  and vectors  $\delta\bn$ and $\delta\bc$ are found in Eq.~\eqref{e:variations},  \eqref{e:var_bn}, and \eqref{e:deltac1}, respectively.}

From the linearization of internal virtual work \eqref{e:Giie_2} follows\footnote{Apart from the Greek indices, summation is also implied on repeated fiber indices $i,j=1...{n_\mrf}$.}
\eqb{lllllll}
\Delta G_{\mathrm{int}} 
\is  \ds \int_{\sS_0}\, \left[ \ds c^{\alpha\beta\gamma\delta}\,\frac{1}{2}\,\delta\auab\,\frac{1}{2}\,\Delta\augd \right.
\plus \ds d^{\alpha\beta\gamma\delta}\,\frac{1}{2}\,\delta\auab\,\Delta\bugd 
\plus  \ds\left.\bar{d}_i^{\alpha\beta\gamma\delta}\,\frac{1}{2}\,\delta a_{\alpha\beta}\,\,\Delta\bar{b}^i_{\gamma\delta}\right.  \\[4mm] 
\plus ~~~~~~~\ds e^{\alpha\beta\gamma\delta}~\,\delta\buab\,\frac{1}{2}\,\Delta\augd 
\plus f^{\alpha\beta\gamma\delta} \,\delta\buab\,\Delta\bugd
\plus    \ds \bar{g}_i^{\alpha\beta\gamma\delta}\,\delta{b}_{\alpha\beta}\, \Delta\bar{b}^i_{\gamma\delta}   \\[4mm]
\plus~~~~~~~\ds\bar{e}_i^{\alpha\beta\gamma\delta}~\,\delta\bar{b}^i_{\alpha\beta}\,\frac{1}{2}\,\Delta\augd 
\plus   \ds  \bar{h}_i^{\alpha\beta\gamma\delta}\,\delta\bar{b}^i_{\alpha\beta}\, \Delta{b}^i_{\gamma\delta}
\plus \ds  \bar{f}_{ij}^{\alpha\beta\gamma\delta}\,\delta\bar{b}^i_{\alpha\beta}\, \Delta\bar{b}^j_{\gamma\delta} \\[4mm] 
%
%
%
\plus  ~~~~~~~~~ \ds \tauab\,\frac{1}{2}\,\Delta\delta\auab  \plus \Mab_0\,\Delta\delta\buab
\plus  \ds\bar{M}_{0i}^{\alpha\beta}\,\Delta\delta\bar{b}^i_{\alpha\beta} \bigg]\,\dif A~, 
\label{e:dxdxWS} 
\eqe
with the nine material tangents \citep{shelltextile}
\eqb{llllll}
c^{\alpha\beta\gamma\delta} \dis 2\ds\frac{\partial \tau^{\alpha\beta}}{\partial a_{\gamma\delta}}~,~~ \quad \quad
d^{\alpha\beta\gamma\delta} :=\ds\frac{\partial \tau^{\alpha\beta}}{\partial b_{\gamma\delta}}~,~~ \quad \quad
\bar{d}_i^{\alpha\beta\gamma\delta}  :=  \ds\frac{\partial \tau^{\alpha\beta}}{\partial \bar{b}^i_{\gamma\delta}}~,~~\quad\quad\\[4mm]
e^{\alpha\beta\gamma\delta} \dis 2\ds\frac{\partial M_0^{\alpha\beta}}{\partial a_{\gamma\delta}}~, \quad \quad
f^{\alpha\beta\gamma\delta} := \ds\frac{\partial M_0^{\alpha\beta}}{\partial b_{\gamma\delta}}~, \quad\quad
\bar{g}_i^{\alpha\beta\gamma\delta} := \ds\frac{\partial M_0^{\alpha\beta}}{\partial \bar{b}^i_{\gamma\delta}}~,\quad\quad\\[4mm]
\bar{e}_i^{\alpha\beta\gamma\delta}  \dis 2\ds\frac{\partial \bar{M}_{0i}^{\alpha\beta}}{\partial a_{\gamma\delta}}~, \quad\quad%
\bar{h}_i^{\alpha\beta\gamma\delta} := \ds\frac{\partial \bar{M}_{0i}^{\alpha\beta}}{\partial {b}_{\gamma\delta}}~\quad\quad%
\bar{f}_{ij}^{\alpha\beta\gamma\delta} :=  \ds\frac{\partial \bar{M}_{0i}^{\alpha\beta}}{\partial \bar{b}^j_{\gamma\delta}}~.
\label{e:cdef}
\eqe

\res{
\remark{As seen from internal virtual work~\eqref{e:Giie_2},  the bending terms contain the variations of curvature tensors $\delta\buab$ and $\delta\bar{b}_{\alpha\beta}$, which are defined from the second derivative of the displacement field, see Eq.~\eqref{e:variations}. Therefore,  a rotation-free shell formulation (with only three displacement degrees of freedom per node) requires at least $C^1$-continuity of the geometry in order to transfer moments through the shell structure. The continuity condition can be satisfied by  the isogeometric discretization \citep{hughes05}. Such a finite element procedure can be found in \cite{shelltextileIGA}, which is summarized here in  Appendix~\ref{s:fe}.}
}

\res{\subsection{Constitutive equations}}{\label{s:constitute}}

Given the work-conjugation pairs appearing in Eq.~\eqref{e:Giie_2},  and by using the classical procedure of \cite{coleman64} for hyperelasticity,  the stresses and  stress couples in Kirchhoff-Love shells can be obtained from the derivative of a stored energy function per reference area, denoted by $W$,  with respect to the corresponding work-conjugate kinematic variables \citep{shelltextile}. I.e.
\eqb{llrlrlr}
\tau^{\alpha\beta} = \ds2\,\pa{W}{a _{\alpha\beta}}~,\quad\quad
M_0^{\alpha\beta}  =  \ds\pa{W}{b _{\alpha\beta}}~,\quad\quad
\bar{M}_{0i}^{\alpha\beta} = \ds\pa{W}{\bar{b}^i _{\alpha\beta}}~.
\label{e:consticom}
\eqe
In the presence of  material anisotropy and angle elastoplasticity  as considered in this work,  
$W$ takes the form
\eqb{l}
W = W\big(a _{\alpha\beta}, b_{\alpha\beta},  \bar{b}^i _{\alpha\beta}; L_i^{\alpha\beta}, \phi_{\mrp}\big)~,
\label{e:WEK}
\eqe 
while the constitutive relations \eqref{e:consticom} remains unchanged, see \cite{sauer2019decomF,shelltextile} and   Remark~\ref{s:RogerW}. Here, $\phi_{\mrp}$ is chosen as the internal variable to describe the plastic angle change between two fiber families $i$ and $j$.  Alternatively to Eq.~\eqref{e:WEK}, $W$ can also be expressed in terms of invariants thanks to the representation theorem \citep{Rivlin1949}.  This is considered in the following. 

A simple form of a generalized  hyperelastic shell model for two-fiber-family dry fabrics is given by \citep{shelltextile} 
\eqb{llllll}
W =     
W_\mathrm{fib\mbox{-}stretch} + W_\mathrm{fib\mbox{-}bending} +  W_\mathrm{fib\mbox{-}torsion} + W_\mathrm{fib\mbox{-}angle} ~,
   \label{e:eg_Wsimple}
\eqe
which consists of the strain energies for  fiber stretching, out-of-plane and in-plane fiber bending, fiber torsion, and elastic angle change between the two fiber families, respectively. For angle elastoplasticity, the last term is a function of the elastic angle $\phi_{\mre}$ from Eq.~\eqref{e:definePhi2}, instead of the total angle $\phi$ as in hyperelasticity, see \cite{shelltextile}. A simple quadratic form for $W_{\mathrm{fib\mbox{-}angle}}$  is
\eqb{lll}
  W_{\mathrm{fib\mbox{-}angle}} \is  \ds\frac{1}{2}\,\mu_\mrf\,\phi_{\mre}^2~,
\label{e:eg_Wsimple_angle}
\eqe
where material parameter  $\mu_\mrf$ can be  identified as the shear modulus of  fabrics, see Remark~\ref{s:shearmodu}.

The other terms in Eq.~\eqref{e:eg_Wsimple} are taken as in the hyperelastic textile model of  \cite{shelltextile},
\eqb{llllll}
%
W_{\mathrm{fib\mbox{-}stretch}} \is \ds  {  \frac{1}{2}\, \epsilon_{\mrL}}\, \sum\limits_{i=1}^2  \big( {\lambda}_i - 1 \big)^{2} ~,\\[5mm]
%
 %
 W_{\mathrm{fib\mbox{-}bending}} \is \ds  \frac{1}{2} \,  \sum\limits_{i=1}^2 \Big[ \beta_\mrn\, (K^i_\mrn)^2  +   \beta_\mrg\, (K^i_\mrg)^2  \Big]~,\\[5mm]
 W_{\mathrm{fib\mbox{-}torsion}} \is \ds  \frac{1}{2}\, \beta_{\tau}\,   \sum\limits_{i=1}^2 {(T^i_\mrg)^2}, 
   \label{e:eg_W1}
\eqe
where  
$T_\mrg$, $K_\mrn$,  and $K_\mrg$ are defined in Eq.~\eqref{e:strains_invariant}.
 Quantities \res{$\epsilon_{\mrL}$, $\beta_{\mrn}$, $\beta_{\mrg}$, and $\beta_{\tau}$ are material parameters representing the  stiffness for fabric stretching, out-of-plane fabric bending, in-plane fabric bending, and fabric torsion, respectively}.  

From Eq.~\eqref{e:eg_W1} and \eqref{e:consticom},  the effective stress and moment components follow as
\eqb{lll}
\tau^{\alpha\beta} \is 
\ds{ \epsilon_{\mrL}}\, \sum\limits_{i=1}^2(\lambda_i-1)\,\frac{1}{\lambda_i}\,L_i^{\alpha\beta}  + \tau^{\alpha\beta}_\mra~,\\[5mm]
M_0^{\alpha\beta} \is  \ds \beta_\mrn\sum\limits_{i=1}^2\,K^i_\mrn\, L_i^{\alpha\beta} +   \beta_\tau\,\frac{1}{2} \sum\limits_{i=1}^2\,\, T^i_\mrg\, (c_{0i}^\alpha\,L_i^\beta + c_{0i}^\beta\,L_i^\alpha)~,\\[5mm]
\bar{M}_0^{\alpha\beta} \dis  \ds \sum\limits_{i=1}^2 \bar{M}_{0i}^{\alpha\beta} =    \ds \beta_\mrg\sum\limits_{i=1}^2\,K^i_\mrg\, L_i^{\alpha\beta}~,
\label{e:eg_tauab_W1}
\eqe
where 
$\tau^{\alpha\beta}_\mra := \ds 2\, {\partial W_{\mathrm{fib\mbox{-}angle}}}/{\partial\auab}$.
 The material tangents of \eqref{e:eg_tauab_W1} follow from Eq.~\eqref{e:cdef} as
\eqb{lll}
c^{\alpha\beta\gamma\delta} \is \ds 
+ \ds \sum\limits_{i=1}^{n_\mrf}\epsilon^i_{\mrL} \,\lambda_i^{-3}\,\,L_i^{\alpha\beta}\,L_i^{\gamma\delta} +c_{\mra}^{\alpha\beta\gamma\delta}\\[6mm]
f^{\alpha\beta\gamma\delta} \is  \ds \sum\limits_{i=1}^{n_\mrf} \beta^i_\mrn\, L_i^{\alpha\beta}\, L_i^{\gamma\delta} +   \sum\limits_{i=1}^{n_\mrf}  \beta^i_\tau\,  \big  (c_{0i}^\alpha\,L_i^\beta \big)^{\mathrm{sym}}\, \big(c_{0i}^\gamma\,L_i^\delta\big)^{\mathrm{sym}}~,\\[6mm]
\bar{f}^{\alpha\beta\gamma\delta}  \dis  \ds  \sum\limits_{i=1}^{n_\mrf} \bar{f}_{ii}^{\alpha\beta\gamma\delta} =  \ds \sum\limits_{i=1}^{n_\mrf} \beta^i_\mrg \, L_i^{\alpha\beta}\,L_i^{\gamma\delta} ~,\\[6mm]
d^{\alpha\beta\gamma\delta} \is e^{\alpha\beta\gamma\delta} = \bar{d}_i^{\alpha\beta\gamma\delta} = \bar{e}_i^{\alpha\beta\gamma\delta} = \bar{g}_i^{\alpha\beta\gamma\delta} = \bar{h}_i^{\alpha\beta\gamma\delta} = 0~,
\eqe
where $c_{\mra}^{\alpha\beta\gamma\delta}=\ds 2\, {\partial \tau^{\alpha\beta}_\mra}/{\partial\auab}$ denotes the material tangents associated with stress $\tau^{\alpha\beta}_\mra$. They will be derived in Sec.~\ref{s:modelelastoplasticity}.

\remark{\label{s:RogerW} By applying the classical procedure of \cite{coleman64} to thin shell elastoplasticity,  \cite{sauer2019decomF} show that the stresses and moments are given by (cf.~\cite{sauer2019decomF}, Eq.~(157))
\eqb{llrlrlr}
\sigab = \ds \,\frac{1}{J_\mre}\pa{\hat{\Psi}}{\otc{E^\mre_{\alpha\beta}}}~,\quad\quad
M^{\alpha\beta}  =  \ds\frac{1}{J_\mre}\pa{\hat  {\Psi}}{\otc{K^\mre_{\alpha\beta}}}~,
\label{e:consticomRs}
\eqe
\otc{in our notation,} where $\otc{E^\mre_{\alpha\beta}}:=(\auab - \hat{a}_{\alpha\beta})/2$ and $\otc{K^\mre_{\alpha\beta}}:= \buab - \hat{b}_{\alpha\beta}$ are the elastic strains, and where $\hat{\Psi}$  is is the Helmholtz free energy per unit intermediate configuration area. The energy per reference area then follows as $W = J_\mrp \, \hat \Psi$. Since 
\eqb{lll}
\ds \pa{W}{\otc{E^\mre_{\alpha\beta}}} = 2\pa{W}{\auab} , \quad \pa{J_\mrp}{\otc{E^\mre_{\alpha\beta}}} = 0,~\quad \pa{W} {\otc{K^\mre_{\alpha\beta}}} = \pa{W}{\buab}~, 
\eqe
(cf.~Eqs.~(78.1), (80.1), (90.1) in \citet{sauer2019decomF}), this leads to the nominal stresses $\tauab := J \, \sigab$ and $\Mab_0 := J\,\Mab$ as given in Eq.~\eqref{e:consticom} above. 
Working with $\hat\Psi$ offers advantages when adapting existing hyperelasticity models to inelastic materials, as the functional form $\hat \Psi = \hat\Psi (\otc{E^\mre_{\alpha\beta},K^\mre_{\alpha\beta}}) $ (as well as stresses and material tangents) can be picked the same as that of the purely elastic case $W = W(\otc{E_{\alpha\beta},K_{\alpha\beta}})$. See \cite{Namvu2019} for an application to thermoelastic shells.}

\section{Angle elastoplasticity of textiles}{\label{s:modelelastoplasticity}}
This section derives the stress response due to elastoplastic angle change. It follows from a given  stored energy function $ W_{\mathrm{fib\mbox{-}angle}}(\phi_{\mre})$, which for example is given by Eq.~\eqref{e:eg_Wsimple_angle}. According to Eq.~(\ref{e:consticom}.1) this stress is
\eqb{lll}
\tau^{\alpha\beta}_\mra = \ds 2\, \pa{W_{\mathrm{fib\mbox{-}angle}}}{\auab}~.
\label{e:tauab_eps_phi}
\eqe 
As seen from Eqs.~\eqref{e:definePhi_total}, \eqref{e:definePhi2} and \eqref{e:var_delta_gamma_12}, angle measure $\phi_\mre$ can be expressed in terms of the metric $\auab$.  Therefore Eq.~\eqref{e:tauab_eps_phi} can be rewritten as
\eqb{lll}
\tau^{\alpha\beta}_\mra =  \ds 2\, \tau\,\pa{\phi_\mre}{\auab}~,
\label{e:tauab_eps_phi_2}
\eqe 
where   we have defined the  (scalar) shear stress between fiber families 
\eqb{lll}
\tau := \ds \pa{W}{\phi_\mre}~,
\label{e:tau_def}
\eqe 
and  where  ${\partial \phi_\mre}/{\partial\auab}$  is determined below, see Remark~\ref{rm:dphi_da_ab}. In case of model~\eqref{e:eg_Wsimple_angle}, $\tau$ takes the simple expression
\eqb{lll}
\tau = \mu_\mrf\,\phi_\mre~.
\label{e:tau_eps_phi}
\eqe
The computation of $\phi_\mre$ requires the knowledge of $\phi$ and $\phi_\mrp$
according to Eq.~\eqref{e:definePhi_total}. While $\phi$ is observable from the current configuration (using Eq.~\eqref{e:definePhi1}), the internal variable $\phi_\mrp$ requires resorting to the intermediate configuration $\hat\sS$. This is discussed in the following subsection.

\remark{\label{s:shearmodu} 
\res{As seen from Eq.~\eqref{e:tauab_eps_phi_2}, scalar stress quantity $\tau$ defined by \eqref{e:tau_def} can be interpreted as a stress invariant induced by angle stress tensor \eqref{e:tauab_eps_phi}, while tensor $\partial{\phi_\mre}/\partial{\auab}$ is a structural tensor that is given explicitly in our plasticity model by Eq.~\eqref{e:l12ab} following from Remark~\ref{rm:dphi_da_ab}.  In case of multiple  pairs of fiber families, stress $\tau$ is defined separatedly for each pair and $\tau^{\alpha\beta}_\mra$ results from the summation over all pairs on the right hand side of Eq.~\eqref{e:tauab_eps_phi_2}. Stress $\tau$  is like the Kirchhoff stress in classical continuum mechanics, i.e.~having the unit [force/reference length].  It can be transformed to the physical Cauchy stress (in the current configuration) by $\sigma_{12} = \tau/J$,  with $J$ defined in Eq.~\eqref{e:JeJp}.} For an initial fiber angle of 90 degrees and  small angle changes, the quantity $\phi$ is equal to two times the \res{infinitesimal shear strain} component $\varepsilon_{12}$, i.e.~$\phi = \theta_{12}  = 2\, \varepsilon_{12}$. In linear elasticity, the shear stress then is $\sigma_{12} = 2\,\mu\, \varepsilon_{12}$, where $\mu$ is the shear modulus. In other words, in this case stress $\tau$ defined in Eq.~\eqref{e:tau_eps_phi} is the classical shear stress, while $\mu_\mrf$ is the classical shear modulus of fabrics. }

\subsection{\res{Underlying mechanism of angle elastoplasticity}} {\label{r:model_motive}}
\res{This section discusses the major underlying mechanisms of angle elastoplasticity from the picture frame test for woven fabrics, which motivates our choice of the yield function in Sec.~\ref{s:yield_surf}.}

\begin{figure}[H]
\begin{center} \unitlength1cm
\begin{picture}(0,11.9)

\put(-7.9,6.8){\includegraphics[width=1\textwidth]{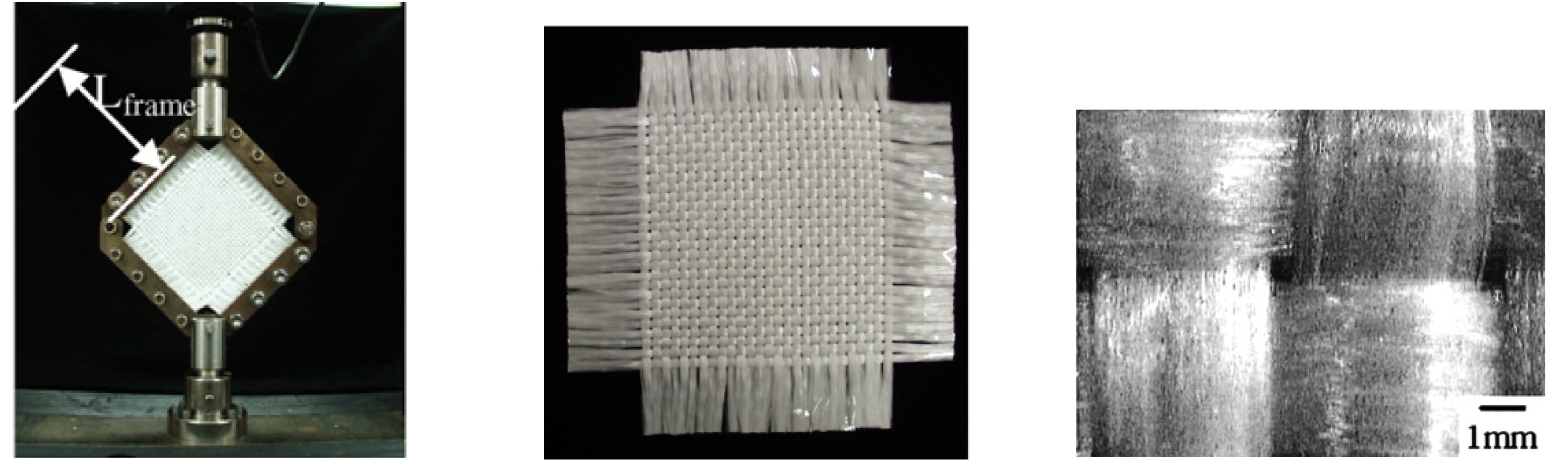}}
\put(-7.9,0.3){\includegraphics[width=1\textwidth]{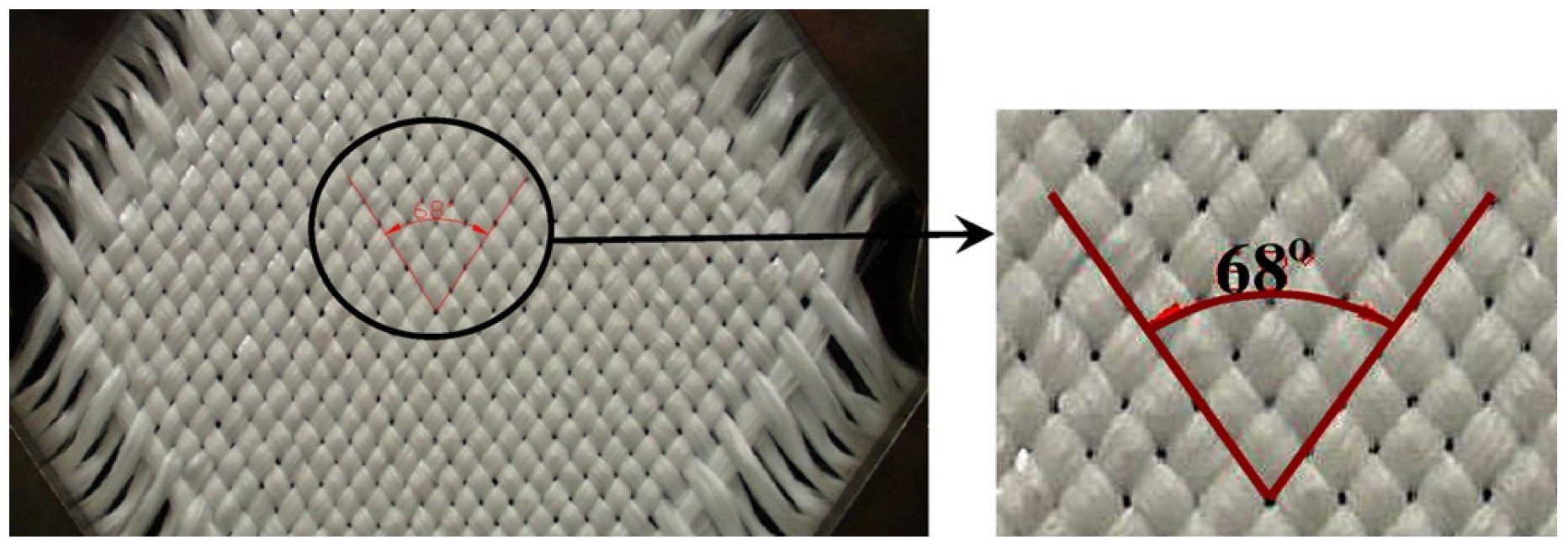}}

\put(-7.9,6.5){\small{(a)}}
\put(-2.5,6.5){\small{(b)}}
\put(2.9,6.5){\small{(c)}}
\put(-7.9,0.0){\small{(d)}}

\end{picture}
\caption[caption]{\res{Picture frame test  conducted at Hong Kong University of Science and Technology (HKUST):  (a)~frame setup, (b)~a woven fabric sample at initial configuration, (c) enlargement of a representative fabric \resc{cell}, (d)  fabric sample  at  deformed configuration. Here, the pictures  are taken from \cite{Cao2008a} and \cite{Zhu2007}, with permission from  Elsevier.}}
\label{f:Pic_HKUST}
\end{center}
\begin{center} \unitlength1cm
\begin{picture}(0,7.4)

\put(-4.5,0){\includegraphics[width=0.56\textwidth]{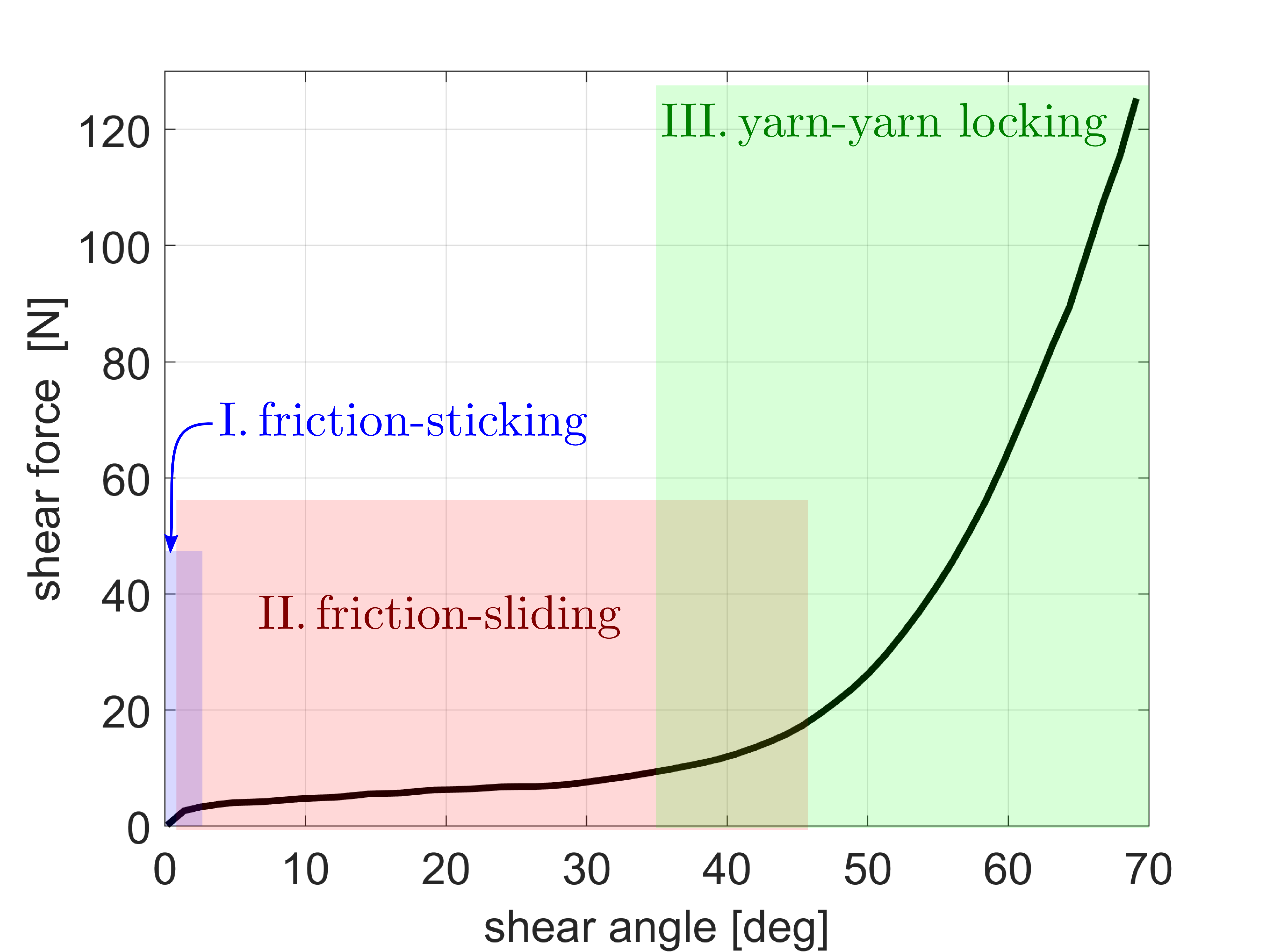}}

\end{picture}
\caption[caption]{\res{Picture frame test: A typical shear force curve versus shear angle, approximately subdivided \resc{into three} phases, where phases I-II result from rotational friction, and phase III is governed by yarn-yarn locking.  The  shear angle here is defined as $\gamma:=90^\circ- \theta$.}}
\label{f:PicF_phases}
\end{center}
\end{figure}

\res{Therefore we examine the experiments conducted at Hong Kong University of Science and Technology (HKUST) as seen in Fig.~\ref{f:Pic_HKUST}, whose experimental results were reported in \cite{Cao2008a} and \cite{Zhu2007}. In the experiment, a woven glass fabric --  with two fiber families with $\Theta=90^\circ$ (Fig.~\ref{f:Pic_HKUST}b) --  is mounted in a square frame (Fig.~\ref{f:Pic_HKUST}a), which is then deformed into a rhombus (Fig.~\ref{f:Pic_HKUST}d). }

\res{Angle shear resistance in woven fabrics is observed to be varied significantly by the frame design and preparation of specimens \citep{Cao2008a}. This leads to a scattering of data seen in the reported curves by different research groups, see also Fig.~\ref{f:fitting_plastic_shear} below. Nevertheless, a typical response of shear force versus shear angle change during deformation looks like in Fig.~\ref{f:PicF_phases}. Overall, the curve is a monotonically increasing function, and there are two major events that significantly alter the trajectory of the curve. The first one appears at small angles ($\gamma=1-2^\circ$, i.e.~almost at the start of deformation),  followed then by a low shear resistance period. Meanwhile, the second happens much later  ($\gamma=35$-$45^\circ$), which is followed by a drastic increase in shear resistance.}
\res{We assume there are two main physical mechanisms governing the above-mentioned shear response of woven fabrics: The first is elastoplastic angle change due to rotational friction between warp and weft yarns, which includes initial sticking and rotational sliding friction. The second is elastoplastic angle change due to  yarn-yarn locking \citep{Cao2008a}, where warp and weft yarns exert in-plane pressure on each other, constraining further relative rotations between them.}

\res{During deformation, one mechanism dominates over the other,  resulting in the three phases   marked in Fig.~\ref{f:PicF_phases}. In particular,  when inspecting the experimental data of Fig.~\ref{f:fitting_plastic_shear}, we can assume that  sticking and rotational sliding friction  dominate phase I (at $\gamma<2^\circ$, approximately) and phase II  ($\gamma=2$-$35^\circ$), respectively. Meanwhile yarn-yarn locking begins at around $\gamma=35$-$45^\circ$ and  becomes dominant  at larger shear angle $\gamma$ (phase III).  }

\res{It should be noted that, the characterization into phases here is only meant qualitatively (yet useful for our modeling purpose). For example, the extent of phase II depends on the initial clearance between yarns that facilitates the shearing of the fabric, see Fig.~\ref{f:Pic_HKUST}c. It can depends also on fiber density, fiber pretension, among other factors.}

%


\subsection{Proposed yield function and flow rule} \label{s:yield_surf}

\res{Motivated by the experimental observations from the picture frame test discussed in \res{Sec.}~\ref{r:model_motive}, this section proposes a corresponding  yield function and flow rule. Recall from Eq.~\eqref{e:tauab_eps_phi_2} (and Remark~\ref{s:shearmodu}), that   only one invariant of the angle stress tensor is required as the work-conjugated counterpart to the angle change. This helps to significantly reduce the modeling complexity of angle \resc{elastoplasticity} as is shown in the following.}

We assume the evolution of the internal variable $\phi_\mrp$ is governed by a yield function with an associated flow rule and evolution equations of independent hardening/softening variables. 
 
The yield function for angle plasticity \res{can be} written in the form
\eqb{lll}
\res{f_\mry(\tau,   q) = | \tau |  -  f_{\mathrm{iso}}(q)~,}
 \label{e:fy_general}
\eqe
where  \res{$f_{\mathrm{iso}}(q) \ge 0 $ models isotropic hardening, and $q$ denotes the hardening variable}. The yield function takes the following values 
\eqb{lll}
f_\mry  \left\{\begin{array}{rl} <0 & $ elastic $,\\
                    =0 & $ plastic $.\\
                    \end{array} \right.
 \label{e:fy_general_case}
\eqe
%


\res{To reproduce the shear behavior discussed in Sec.~\ref{r:model_motive}, we  propose   the scalar-valued function}
\eqb{lll}
 f_{\mathrm{iso}}(q) = \tau_\mry  + A \,\mathrm{asinh}(a\,q) + B \,\mathrm{tanh}(b\,q) +  C \,q^c~,
 \label{e:fiso}
\eqe
\res{for isotropic hardening. Here,}  $\tau_\mry$ is the initial yield stress, while $A,B,C$, and  $a$, $b$ and $c$ are model parameters. \res{In connection with the deformation phases shown in Fig.~\ref{f:PicF_phases}, parameter $\tau_\mry$ sets the initial sticking limit  of phase I, while the four parameters $A$, $a$, $B$ and $b$ reflect the  low plastic resistance  of phase II, and the two parameters $C$ and $c$ capture the increased hardening of phase III by a power law.}

\begin{figure}[H] 
\begin{center} \unitlength1cm
\begin{picture}(0,6.6)

\put(-8.2,0){\includegraphics[width=0.52\textwidth]{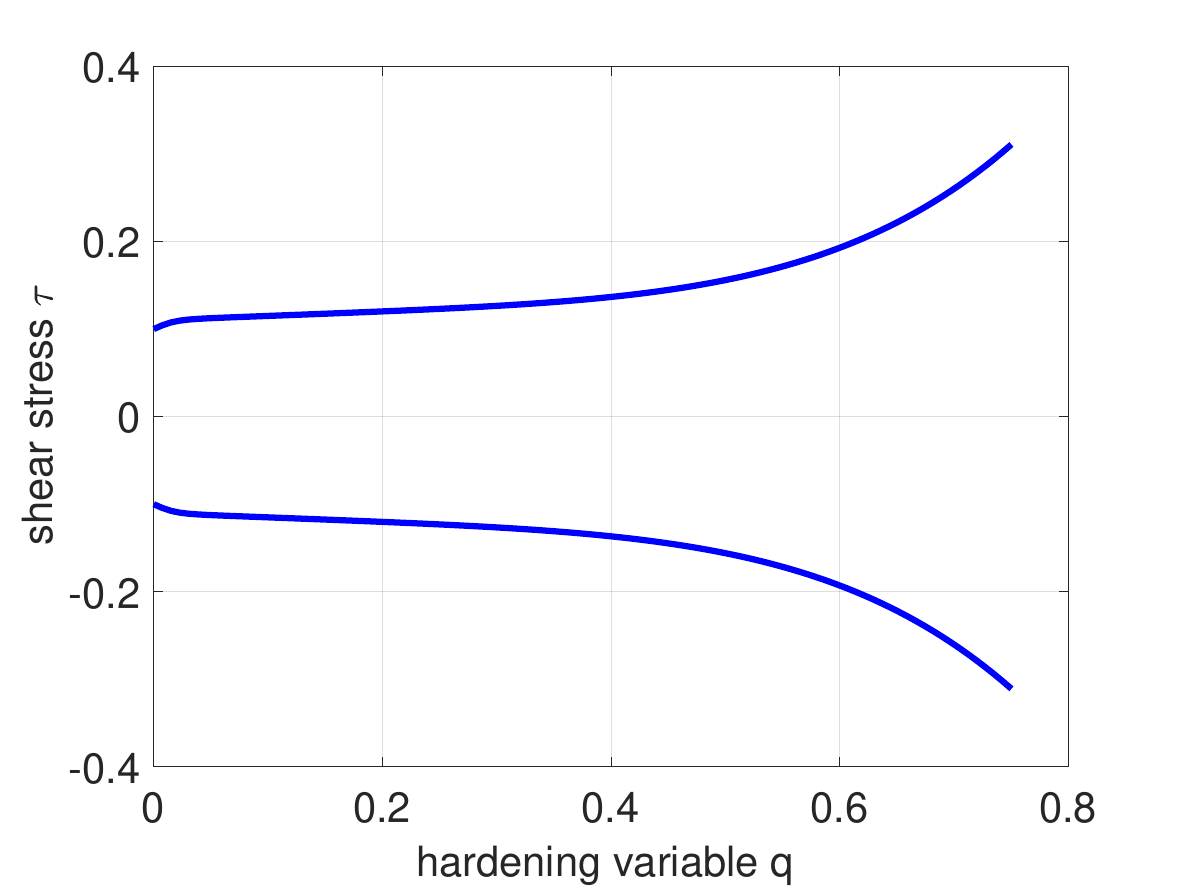}}
\put(0.1,0){\includegraphics[width=0.52\textwidth]{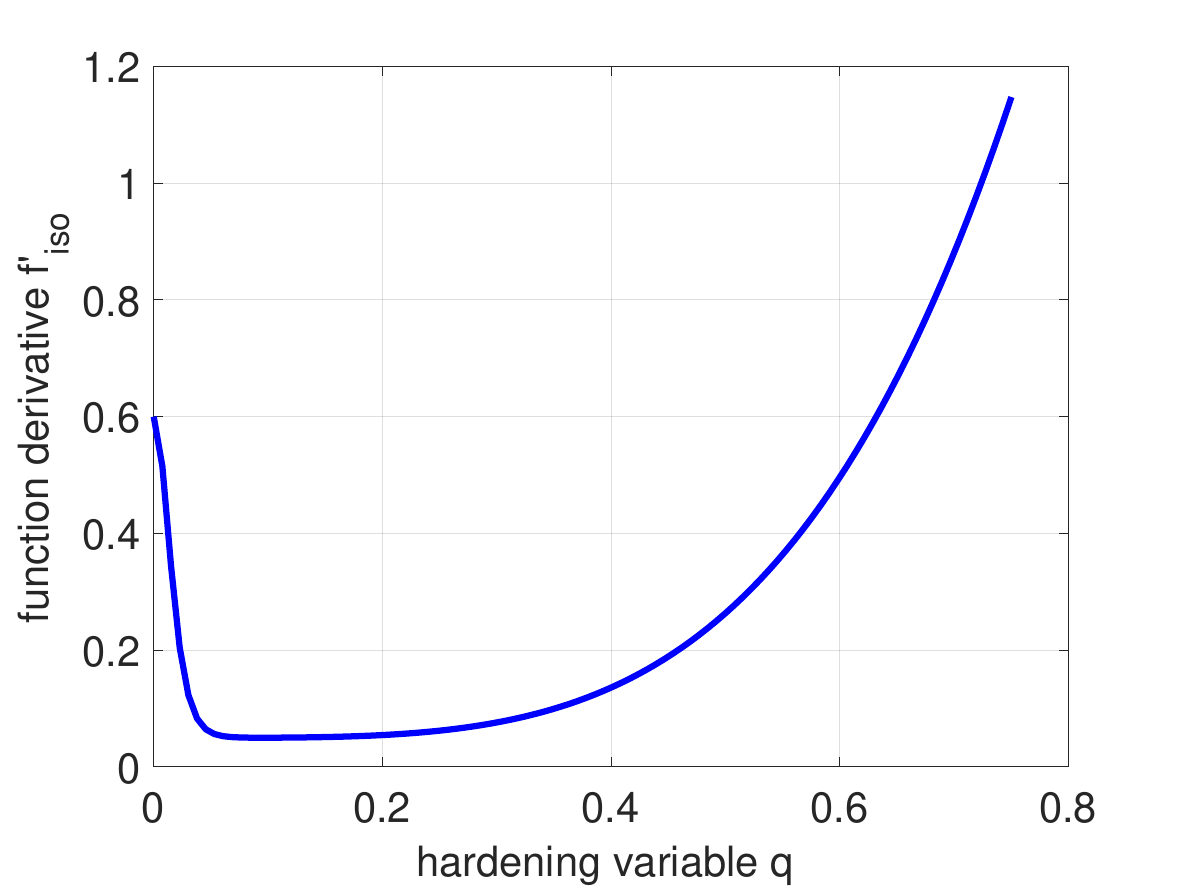}}

\put(-8.0,0.1){{\small{(a)}}}
\put(0.2, 0.1){{\small{(b) }}}

\end{picture}
\caption{\res{Characteristics of the proposed yield function: (a)~yield surface  $f_\mry = 0$ from Eq.~\eqref{e:fy_general} and (b)~function derivative  $f'_{\mathrm{iso}}(q)$ from Eq.~\eqref{e:fiso} versus hardening variable $q$. Here, we have used $\tau_\mry=0.1\,\mu_0$, $A=0.05\, \mu_0$, $a=  1$, $B= 0.01\, \mu_0$, $b=   55$, $C= 0.7\, \mu_0$ and $c= 5$, where $\mu_0$ denotes a stress measure.} }
\label{f:yields_surf}
\end{center}
\end{figure}

With Eq.~\eqref{e:fiso},  Eq.~\eqref{e:fy_general} becomes
\eqb{lll}
f_\mry(\tau,  q) = h\,\tau -  \tau_\mry  - A \,\mathrm{asinh}(a\,q) - B \,\mathrm{tanh}(b\,q) -  C \,q^c~, \quad $with$ \quad h:=\sign(\tau)~,
 \label{e:fy_textile}
\eqe
\res{which is plotted in Fig.~\ref{f:yields_surf}a.}

We further assume the associated flow rule 
\eqb{lll}
\dot{\phi}_\mrp = \ds {\otc{\dot{\alpha}_\mrp}}\,\pa{f_\mry}{\tau} ~, 
\label{e:flowWFa}
\eqe
where ${\otc{\alpha_\mrp}}$ denotes the accumulated plastic angle change.  Inserting \eqref{e:fiso} into Eq.~\eqref{e:flowWFa} gives
\eqb{lll}
\dot{\phi}_\mrp = h\, {\otc{\dot{\alpha}_\mrp}}~.
\label{e:flowWF}
\eqe
For the hardening variable $q$, we assume the simple evolution law
\eqb{lll}
 \dot{q} \is {\otc{\dot{\alpha}_\mrp}} ~.
\label{e:evoldotWF}
\eqe

\remark{ \res{The parameter set [$\tau_\mry,\,A,\,a,\,B,\,b,\,C,\,c$] appearing in Eq.~\eqref{e:fy_textile} is chosen such that it leads to a monotonically increasing hardening function $f_{\mathrm{iso}}(q)$ in accordance with the experimental behavior of Fig.~\ref{f:PicF_phases}, i.e. $f'_{\mathrm{iso}}(q) > 0$ as seen Fig.~\ref{f:yields_surf}b.}}

\remark{\res{Hardening function \eqref{e:fiso} is not only motivated from reproducing the experimental observation discussed in Sec.~\ref{r:model_motive}. It is also chosen such that it provides continuous differentiability as seen in Fig.~\ref{f:yields_surf}b.  The latter is to obtain a well-behaved and smooth tangent matrix, which facilitates robustness of the Newton-\resc{Raphson} algorithm.}}

\subsection{Time discretization and predictor-corrector algorithm}{\label{s:predictor_corrector}}
A time integration scheme is required for solving Eq.~\eqref{e:flowWF} and \eqref{e:evoldotWF}. Here, we employ the backward Euler  scheme for both  variables $\phi_\mrp$ and $q$, i.e.
\eqb{lll}
\phi_\mrp^{n+1} \is \phi_\mrp^{n} + h^{n+1}\,\Delta{\otc{\alpha_\mrp^{n+1}}}~,\\[2mm]
 q^{n+1} \is q^n + \Delta{\otc{\alpha_\mrp^{n+1}}}~.
 \label{e:timedisWF}
\eqe

With this, the evaluation of Eq.~\eqref{e:fy_textile} is based on a classical predictor-corrector algorithm:\footnote{{Henceforth, unless indicated otherwise, variables without time step superscript are evaluated at the current time step $n+1$.}}

1.~The \textit{trial step}  is to determine whether the material point is in an elastic or plastic state by  assuming the former, i.e.~setting the increment of the plastic strain to $\Delta{\otc{\alpha_\mrp}} = 0$. This gives the current elastic  trial  angle
\eqb{lll}
\phi^{\mathrm{trial}}_\mre =\phi^{n+1} - \phi^{n}_\mrp
\label{e:phi_step_trial}
\eqe
due to Eq.~(\ref{e:timedisWF}.1) and Eq.~\eqref{e:definePhi_total}. Thus the current trial stress \eqref{e:tau_eps_phi} becomes
\eqb{lll}
\tau^{\mathrm{trial}} :=  \mu_\mrf\,\phi^{\mathrm{trial}}_\mre  ~. 
\label{e:tau_step_trial}
\eqe
The verification of the elastic assumption is then based on the value of the yield function~\eqref{e:fy_textile} computed in this step,
\eqb{lll}
f_\mry^{\mathrm{trial}} \dis h^{\mathrm{trial}}\,\tau^{\mathrm{trial}} -  \tau_\mry  - A \,\mathrm{asinh}(a\,q^n) - B \,\mathrm{tanh}(b\,q^n) -  C \,(q^n)^c~,\\[3mm]
h^{\mathrm{trial}}\dis \sign(\tau^{\mathrm{trial}} )~.
\label{e:alg_fytrial}
\eqe

2.~If $f_\mry^{\mathrm{trial}} \leq 0$, then  the material is indeed in an \textit{elastic state}. Thus
\eqb{llll}
\tau \is \tau^{\mathrm{trial}}~,\\[3mm]
\phi_\mre \is \phi^{\mathrm{trial}}_\mre~.
\label{e:elastic_stress_strain_trial}
\eqe
Taking the variation of the last equation in \eqref{e:elastic_stress_strain_trial} and considering Eq.~\eqref{e:phi_step_trial} and \eqref{e:var_delta_gamma_12}, gives
\eqb{llll}
\delta\phi_\mre \is  \delta\theta_{12} ^{n+1} =    g_{12}^{\alpha\beta}\,\delta\auab~,
\label{e:delta_phie_trial}
\eqe
such that Eq.~\eqref{e:tauab_eps_phi} leads to the stress components 
\eqb{lll}
\tauab_\mra = 2\, \tau^{\mathrm{trial}}\,g_{12}^{\alpha\beta}~.
\label{e:alg_stress_elastic}
\eqe
The corresponding material tangents  follows from Eq.~\eqref{e:cdef} as
\eqb{lll}
c_\mra^{\alpha\beta\gamma\delta}  = 4\, \otc{\mu_\mrf}\,g_{12}^{\alpha\beta}\,g_{12}^{\gamma\delta} ~ + ~ 4\,\tau^{\mathrm{trial}}\,g_{12}^{\alpha\beta\gamma\delta}~,
\label{e:alg_tangent_elastic}
\eqe
where $g_{12}^{\alpha\beta}$ and $g_{12}^{\alpha\beta\gamma\delta}$ are defined by Eq.~\eqref{e:l12ab} and \eqref{e:l12abcd}, respectively. \\

3.~If $f_\mry^{\mathrm{trial}} > 0$, then the state is plastic. An evaluation of plastic strain $\Delta{\otc{\alpha_\mrp}}>0$ is then required for the correction of the trial state. The computation of plastic strain $\Delta{\otc{\alpha_\mrp}}$ is found such that Eq.~\eqref{e:fy_textile} satisfies
\eqb{lll}
f_\mry^{n+1} \is h^{n+1}\, \tau^{n+1} - \tau_\mry  - A \,\mathrm{asinh}\big(a\,q^{n+1}) - B \,\mathrm{tanh}\big(b\,q^{n+1}\big) -  C \,\big(q^{n+1}\big)^c = 0~,
\label{e:fszero}
\eqe
where $\tau^{n+1}$ is now computed from
\eqb{lll}
\tau^{n+1} = \mu_\mrf\, \big(\phi^{n+1} - \phi^{n+1}_\mrp\big) ~.
\label{e:tauWFs1}
\eqe
By considering Eq.~(\ref{e:timedisWF}.1),  \eqref{e:tau_step_trial} and \eqref{e:phi_step_trial}, Eq.~\eqref{e:tauWFs1} becomes
\eqb{lll}
\tau^{n+1} = \tau^{\mathrm{trial}} - h^{n+1}\,\mu_\mrf\,\Delta{\otc{\alpha_\mrp}}~.
\label{e:tauWFs}
\eqe
Given that parameters $\mu_\mrf>0$ and $\Delta{\otc{\alpha_\mrp}}>0$, Eq.~\eqref{e:tauWFs} leads to 
\eqb{lll}
h^{n+1} = h^{\mathrm{trial}}~.
\label{e:h_np1}
\eqe
Inserting Eq.~\eqref{e:h_np1}, \eqref{e:tauWFs} and (\ref{e:timedisWF}.2)  into \eqref{e:fszero}  gives
\eqb{lll}
g(\Delta{\otc{\alpha_\mrp}}) \dis h^{\mathrm{trial}}\,\tau^{\mathrm{trial}} - \mu_\mrf\,\Delta{\otc{\alpha_\mrp}}  -  \tau_\mry  \\[2mm]
\mi  A\,\mathrm{asinh}\big[a\,\big(q^{n} + \Delta{\otc{\alpha_\mrp}}\big)\big] - B\,\mathrm{tanh}\big[b\,\big(q^{n} + \Delta{\otc{\alpha_\mrp}}\big)\big] - C \,\big(q^{n} + \Delta{\otc{\alpha_\mrp}}\big)^c = 0~,
\label{e:fsinh}
\eqe
which is solved for $\Delta{\otc{\alpha_\mrp}}$ using Newton's method. For this, the derivative
\eqb{lll}
g'= - \ds \mu_\mrf -\frac{A\,a}{\sqrt{1 + a^2\,(q^n + \Delta{\otc{\alpha_\mrp}})^2}}  - B\,b\,\mathrm{sech}^2\big[b\,\big(q^n + \Delta{\otc{\alpha_\mrp}}\big)\big] - C\,c\,\big(q^{n} + \Delta{\otc{\alpha_\mrp}}\big)^{c-1} ~
\label{e:dfsinh}
\eqe
 is needed. With $\Delta{\otc{\alpha_\mrp}}$ evaluated, the elastic angle strain and stress can  then be updated from
 \eqb{lll}
  \phi_\mrp^{n+1} \is  \phi^{n}_\mrp + h^{\mathrm{trial}}\,\Delta{\otc{\alpha_\mrp}}~,\\[4mm]
 \phi_\mre^{n+1} \is  \phi^{n+1}  - \phi_\mrp^{n+1} ,\\[4mm] 
 q^{n+1} \is q^n +    \Delta{\otc{\alpha_\mrp}}~,\\[4mm]
\tau^{n+1} \is \mu_\mrf\, \phi_\mre^{n+1}~.
\label{e:elastic_strain_plastic}
\eqe

From Eq.~(\ref{e:elastic_strain_plastic}.2) follows the elastic angle variation
\eqb{lll}
  \delta\phi_\mre \is   \delta\phi^{n+1}_\mre =  \delta\phi^{n+1} = \delta\theta_{12} ^{n+1} =  g_{12}^{\alpha\beta}\,\delta\auab~,
  \label{e:delta_phie_elastic}
\eqe  
the stress components  \eqref{e:tauab_eps_phi} 
\eqb{lll}
\tauab_\mra = 2\, {\tau}\,g_{12}^{\alpha\beta}~,
\label{e:alg_stress_plastic}
\eqe
and the corresponding material tangent \eqref{e:cdef}
\eqb{lll}
c_\mra^{\alpha\beta\gamma\delta}  = 4\, \big( \mu_\mrf + \mu_\mrf^2/g' \big)\,g_{12}^{\alpha\beta}\,g_{12}^{\gamma\delta} ~ + ~ 4\, \tau\,g_{12}^{\alpha\beta\gamma\delta}~.
\label{e:alg_tangent_plastic}
\eqe
The resulting predictor-corrector algorithm is summarized in Tab.~\ref{t:Summary_alg}.
\begin{table}[!htp]
\small
\begin{center}
\def\arraystretch{1.5}\tabcolsep=5.0pt
\begin{tabular}{|l| }
\hline

Given: current fiber angle $ \phi^{n+1}$, the previous values $ \phi^{n}$, $ \phi_\mrp^{n}$, $q^n$, shear stiffness  $\mu_\mrf$, and  \\[-0.2mm] 
 yield function $f_\mry$.\\
1. Trial step with the elastic \resc{assumption} $\phi^{\mathrm{trial}}_\mre = \phi^{n+1} - \phi^{n}_\mrp$, and   $\tau^{\mathrm{trial}} = \epsilon\,\phi^{\mathrm{trial}}_\mre~$ \\

 ~~$\bullet$  Compute trial yield function $f_\mry^{\mathrm{trial}}:=f_\mry(\tau^{\mathrm{trial}} , q^n)$ from Eq.~\eqref{e:alg_fytrial}. \\

2. If $f_\mry^{\mathrm{trial}} \leq 0$, then the step is indeed elastic. \\
 ~~$\bullet$ Set plastic strain $\Delta{\otc{\alpha_\mrp}}=0$.\\
 ~~$\bullet$ Compute elastic angle strain $\phi_\mre^{n+1}$ and elastic shear stress  $\tau$  from Eq.~\eqref{e:elastic_stress_strain_trial}.\\
 ~~$\bullet$ Compute stress $\tauab_\mra$ and corresponding tangents $c_\mra^{\alpha\beta\gamma\delta}$ from Eq.~\eqref{e:alg_stress_elastic} and \eqref{e:alg_tangent_elastic}.\\

3. If $f_\mry^{\mathrm{trial}} > 0$, then the step is plastic: \\ 
 ~~$\bullet$ Compute plastic strain $\Delta{\otc{\alpha_\mrp}}$ using Newton's method with residual \eqref{e:fsinh} and gradient  \eqref{e:dfsinh}.\\
 ~~$\bullet$ Compute elastic angle strain $\phi_\mre^{n+1}$ and elastic shear stress  $\tau$  from Eq.~\eqref{e:elastic_strain_plastic}.\\
 
 ~~$\bullet$ Compute stress $\tauab_\mra$ and corresponding tangents $c_\mra^{\alpha\beta\gamma\delta}$ from Eq.~\eqref{e:alg_stress_plastic} and \eqref{e:alg_tangent_plastic}.\\

4. Update $\phi_\mrp$ and $q$ from Eq.~\eqref{e:timedisWF}\\
 
 \hline
\end{tabular}
\end{center}
\caption{Summary of the predictor-corrector algorithm.}
\label{t:Summary_alg}
\end{table}

\remark{\label{rm:dphi_da_ab}As seen from Eq.~\eqref{e:delta_phie_trial} and \eqref{e:delta_phie_elastic}, the change of elastic angle $\phi_\mre$ with respect to the metric in the presented plasticity model is simply
\eqb{lll}
\ds\pa{\phi_\mre}{\auab} = g_{12}^{\alpha\beta}
\label{e:derive_phi_e}
\eqe
for both elastic and plastic states.
}

\section{\res{Analytical solution of angle plasticity in the picture~frame~test}}{\label{s:ana_picF}}

This section presents the analytical solution for the picture frame test  shown in \res{Fig.~\ref{f:Pic_HKUST} and Fig.~\ref{f:picframe_ana}a-b.} 
The aim is to find an analytical expression for the stress based on the angle plasticity model proposed in Sec.~\ref{s:modelelastoplasticity}.  Our solution is applicable to multiple loading (or unloading) cycles. To this end, we employ a so-called  expanding yield surface approach: It solves for the shear stress within a given loading or unloading interval using the initial conditions provided from the solution at the end of the previous  interval. \res{A supplementary Matlab code for this is provided at {\url{https://github.com/xuanthangduong/textile-picture-frame-test.git}}}

%

\subsection{\res{Various angle measures in the picture frame test}}

The initial configuration and fiber directions in the picture frame test are   seen in Fig.~\ref{f:picframe_ana}a. In this case the angle measures from Eq.~\eqref{e:defineGamma_0} and \eqref{e:defineGamma_current} become 
 \begin{empheq}[ ]{align}
\Theta_{12} &:=  \bL_1\cdot\bL_1 = \cos\Theta =  0 ~,\label{e:picF_defineGamma_0}\\[2mm]
\theta_{12}  &:=  \bell_1\cdot \bell_2 = \cos\theta~.
\label{e:picF_defineGamma_current}
\end{empheq}
 For the plotting of the testing results, it is convenient to use the shear angle 
\eqb{ll}
\gamma :=  90^\circ -  \theta~\quad$[deg]$~.
\eqe
Consider a loading (or unloading)  interval, denoted $\mathscr{C}$, where $\theta_{12} \in [\theta_{12}^0~,\theta_{12}^{\mathrm{max}}]$.  
  We define the relative angles $\bar\phi$ similarly to  Eq.~\eqref{e:definePhi1}-\eqref{e:definePhi3} but now with respect to the configuration at the start of interval $\mathscr{C}$. Here, the bar indicates an interval-wise quantity. This means that within interval $[\theta_{12}^0~,\theta_{12}^{\mathrm{max}}]$ the configuration at $\theta_{12}$ is identical to the current configuration $\sS$.  But the configuration at  $\theta_{12}^0$ now becomes the reference configuration and is denoted by $\bar{\sS}_0$. Between $\bar{\sS}_0$ and $\sS$, the intermediate configuration is denoted by $\bar{\hat\sS}$ -- generally different from $\hat\sS$. The \res{directions of two fiber pairs in} configurations $\bar{\sS}_0$ and $\bar{\hat\sS}$ are denoted by $\bell_\res{I}^0$ and  $\bar{\bellhat}_\res{I}$, respectively. Similarly to Eqs.~\eqref{e:defineGamma_0} and \eqref{e:defineGamma_hat},  fiber directions $\bell_\res{I}^0$ and  $\bar{\bellhat}_\res{I}$ constitute  the relative fiber angles, denoted $\theta_{12}^0$ and $\bar{\thetahat}_{12}$, respectively. 
  
 With these, the relative angles within interval $\mathscr{C}$ are defined as
 \begin{empheq}[ ]{align}
\bar\phi  &:=    \theta_{12} -   \theta_{12}^0~,\label{e:picFdefinePhi1}\\[2mm]
\bar\phi_{\mre} &:=  \theta_{12}  - \bar{\thetahat}_{12}~,\label{e:picFdefinePhi2}\\[2mm]
\bar\phi_{\mrp} &:= \bar \phi - \bar\phi_{\mre} ~.
\label{e:picFdefinePhi3}
\end{empheq}
Consequently $\bar\phi$ always ranges from zero to $\phi^{\mathrm{max}}:= \theta_{12}^{\mathrm{max}}-   \theta_{12}^0$. Note that $\phi^{\mathrm{max}}$ can  take a negative value.


\subsection{\res{Shear stress in a loading interval }}

Similar to the interval-wise angle measures defined in Eq.~\eqref{e:picFdefinePhi1}-\eqref{e:picFdefinePhi3}, the  shear stress can be expressed in an interval-wise manner as follows.

Consider  a loading or unloading interval  $\mathscr{C}$, which proceeds from the previous interval, say $\mathscr{C}^0$. Given that $\bar\phi$ ranges from  $0$ to $\phi^{\mathrm{max}}$ and the fabric deforms from elastic phase $\mathscr{C}_\mre$ to plastic phase $\mathscr{C}_\mrp$, 
the total shear stress $\tau$  can be computed  from the shear stress at the end of interval $\mathscr{C}^0$ plus an additional stress. That is,
\eqb{lll}
\tau =   \tau^0 + \bar{\tau}~,
\label{e:tauc_all}
\eqe
where  $\tau^0$  and $\bar\tau$ denote the shear stress due to the change of fiber angles in interval $\mathscr{C}^0$ and $\mathscr{C}$, respectively.  Here, $\bar\tau$ follows from \eqref{e:tau_eps_phi} with \eqref{e:picFdefinePhi3} as
\eqb{lll}
 \bar\tau := \mu_\mrf\,\bar\phi_\mre =   \mu_\mrf \,(\bar\phi - \bar\phi_\mrp)~.
\label{e:tauc}
\eqe
Here, the internal variable $\bar\phi_\mrp $ is still unknown and can be solved by the following approach, which we term the \textit{expanding yield surface approach}.

\subsection{\res{Analytical solution by an \textit{expanding yield surface approach}}}

We now aim at solving for $\bar{\phi}_\mrp$ to compute shear stress  $\bar\tau$. To this end,  performing time integration over  plastic phase  $\mathscr{C}_\mrp$ on both sides of the evolution equations \eqref{e:flowWF} and \eqref{e:evoldotWF} gives
 \begin{empheq}[ ]{align}
 h\,(\bar\phi_\mrp  +  \otc{\bar{c}_1}) &= {\otc{\alpha_\mrp}} +  \otc{\bar{c}_2}~, \label{e:time_int_lam}\\[2mm]
Q:= {\otc{\alpha_\mrp}}  +  \otc{\bar{c}_2} &= q   +  \otc{\bar{c}_3}~, \label{e:time_int_lam_q}
\end{empheq}


where $\bar\phi_\mrp$ is the  plastic angle defined in Eq.~\eqref{e:picFdefinePhi3}, $Q$ denotes \otc{an offset} of $q$ (and also ${\otc{\alpha_\mrp}}$) over $\mathscr{C}$, 
and parameters  $\otc{\bar{c}_1}$  $\otc{\bar{c}_2}$, and  $\otc{\bar{c}_3}$  are  constants to be determined from the  initial conditions, see Eq.~\eqref{e:constant_computed}. Inserting Eq.~\eqref{e:time_int_lam} into \eqref{e:time_int_lam_q} gives the explicit expression of the hardening variable  $q$  in terms of  $\bar\phi_\mrp$ as\footnote{Here we keep the constants separate for the sake of comparison with the general numerical procedure in Sec.~\ref{s:predictor_corrector}. Alternatively, one can combine the constants and eliminate variable ${\otc{\alpha_\mrp}}$.}
\eqb{lll}
q = {\otc{\alpha_\mrp}} + \otc{\bar{c}_2}  - \otc{\bar{c}_3} =  h\,(\bar\phi_\mrp + \otc{\bar{c}_1}) - \otc{\bar{c}_3} =: \tilde{q}(\bar\phi_\mrp )~.
\label{e:initialcon}
\eqe
With this and  angle $\bar\phi$ given, we can solve for $\bar\phi_\mrp $  
\eqb{lll}
f_\mry (\tau,\tilde q) = 0
 \label{e:fy_textile_ana_eq}
\eqe
in a plastic state. Here, 
\eqb{lll}
f_\mry = \bar{\tau}\,h - (\tau_\mry +  |\tau^0|)  - A \,\mathrm{asinh}(a\,\tilde{q}) - B \,\mathrm{tanh}(b\,\tilde{q}) -  C \,\tilde{q}^c~
 \label{e:fy_textile_ana}
\eqe
follows from \eqref{e:fy_textile} when inserting \eqref{e:tauc_all}, and $\bar\tau$ is expressed by  Eq.~\eqref{e:tauc} with \eqref{e:initialcon}. Therefore,
$f_\mry$ from Eq.~\eqref{e:fy_textile_ana} becomes a function of sole (scalar) variable $\bar\phi_\mrp$, and Eq.~\eqref{e:fy_textile_ana_eq} can be solved for $\bar\phi_\mrp$ by Newton's method using the derivative
\eqb{lll}
f'_\mry= - \ds h\, \mu_\mrf -\frac{A\,a\,h}{\sqrt{1 + a^2\,\tilde{q}^2}}  - B\,b\,h\,\mathrm{sech}^2\big(b\,\tilde{q}\big)- C\,c\,h\,\tilde{q}^{c-1} ~.
\label{e:picFdfsinh}
\eqe
As seen from Eq.~\eqref{e:fy_textile_ana},  the yield stress is shifted by $|\tau^0|$  since the current interval starts from the last point of the previous interval. In other words, the yield \otc{surface} expands by $|\tau^0|$.  

Further, the elastic phase in  interval $\mathscr{C}$ can be defined from function \eqref{e:fy_textile_ana} with the condition 
\eqb{lll}
f_\mry(\bar\tau) <0~. 
 \label{e:fy_textile_ana_fs_elas}
\eqe
Note that in this phase,  hardening variable $\tilde q$ does not evolve. That is,  it remains constant at the value  from the previous loading interval, say $\tilde q=\otc{q^0}$, and $\bar\tau = \mu_\mrf\,\bar\phi$ from Eq.~\eqref{e:tauc} since $\bar\phi_\mrp=0$, and function~\eqref{e:fy_textile_ana} becomes
\eqb{lll}
f_\mry = \mu_\mrf\, |\bar\phi|- (\tau_\mry +  |\tau^0|)  - A \,\mathrm{asinh}(a\,\otc{q^0}) - B \,\mathrm{tanh}(b\,\otc{q^0}) -  C \,(\otc{q^0})^c~.
 \label{e:fy_textile_ana_fs}
\eqe
With this, condition~\eqref{e:fy_textile_ana_fs_elas}  gives the definition of the elastic zone as
\eqb{lll}
 |\bar\phi| < \phi_\mry~,
\eqe
where we have defined the so-called yield angle $\phi_\mry$ as
\eqb{lll}
\phi_\mry := \ds\frac{1}{\mu_\mrf} \,\Big[ \tau_\mry +  |\tau^0|+  A \,\mathrm{asinh}(a\,\otc{q^0}) + B \,\mathrm{tanh}(b\,\otc{q^0})+  C \,(\otc{q^0})^c \Big]~.
 \label{e:fy_textile_ana_phis}
\eqe
In particular for the initial loading interval,  setting $\tau^0=0$, and $\otc{q^0}=0$ in Eq.~\eqref{e:fy_textile_ana_phis} gives $ \phi_\mry  = \tau_\mry /\mu_\mrf$.

\subsection{\res{Analytical solution procedure}}

At the start of interval $\scrC$,  the values of shear stress $\tau$, accumulated plastic angle ${\otc{\alpha_\mrp}}$,   hardening variable $q$, and accumulation $Q$ in Eq.~\eqref{e:time_int_lam_q}  are taken over from (the end of) the previous  interval $\scrC_0$ as  
\eqb{lllll}
\tau \is \tau^0~,\\[2mm]
{\otc{\alpha_\mrp}} \is {\otc{\alpha_\mrp^0}}~,\\[2mm]
q \is q^0~\\[2mm]
Q \is Q^0~,
\label{e:initi_0}
\eqe
where $Q^0 = \tau^0 = {\otc{\alpha_\mrp^0}}=\tau^0=q^0 = 0$ for the initial loading interval. With this, we can compute the yield angle  $\phi_\mry$ from Eq.~\eqref{e:fy_textile_ana_phis}.


For any given $\bar\phi\in [0, \phi^{\mathrm{max}}]$, if $\bar\phi$ falls in the elastic phase $\scrC_\mre$, i.e.~$|\bar\phi|\leq\phi_\mry$,  the shear stress $\eqref{e:tauc_all}$ becomes
\eqb{lll}
\tau = \tau^0 + \mu_\mrf\,\bar\phi~,
\label{e:tau_elas}
\eqe
since $\bar\phi_\mrp=0$. 

On the other hand, if $\bar\phi$ falls in the plastic phase $\scrC_\mrp$, i.e.~$|\bar\phi|>\phi_\mry$, we compute $\bar\phi_\mrp$  by solving Eq.~\eqref{e:fy_textile_ana_eq} with $f_\mry$ defined by Eq.~\eqref{e:fy_textile_ana}, where  the three constants appearing in Eq.~\eqref{e:time_int_lam} and \eqref{e:time_int_lam_q} are determined from the initial conditions at the start of plastic phase $\scrC_\mrp$. This gives
\eqb{lllll}
\otc{\bar{c}_3} \is  Q^0 - q^0,\\[1.5mm]
\otc{\bar{c}_2} \is  Q^0- {\otc{\alpha_\mrp^0}}~,\\[1.5mm]
\otc{\bar{c}_1} \is h^\mre\,Q^0~,\\[1.5mm]
h^\mre \is \sign \tau_\mre~.
\label{e:constant_computed}
\eqe
Here, $\tau_\mre$ denotes the shear stress \eqref{e:tau_elas} at the end of  the elastic phase $\scrC_\mre$ (i.e.~the start of plastic phase $\scrC_\mrp$), and we have used the condition $\bar\phi_\mrp=0$ at the start of plastic phase $\scrC_\mrp$, and the fact that $Q$, $q$, and ${\otc{\alpha_\mrp}}$ remain unchanged in $\scrC_\mre$ as \otc{$q$} and ${\otc{\alpha_\mrp}}$ evolve only in  plastic phases.

 With $\bar\phi_\mrp$ computed, apart from stress $\tau$ from  $\eqref{e:tauc_all}$, variables $q$, $Q$ and ${\otc{\alpha_\mrp}}$ from Eq.~\eqref{e:initialcon} and \eqref{e:time_int_lam} can be computed at the end of the current interval, so that they are available for the initial conditions of the next loading interval.

\section{\res{Model verification using the picture frame test}}{\label{s:model_verification}}


This section first verifies the preceding finite element formulation using the analytical solution of the proposed angle plasticity model for the picture frame test given in Sec.~\ref{s:ana_picF}. Second, the picture frame test is used to  investigate the influence of the model parameters. 
In these results,   shear forces are normalized by frame length $L_0$ and \res{stress measure} $\mu_0$. 

\subsection{\res{Finite element versus analytical solutions of the picture frame test}}
\begin{figure}[H] 
\begin{center} \unitlength1cm
\begin{picture}(0,6.6)
\put(-7.5,0.1){\includegraphics[width=0.44\textwidth]{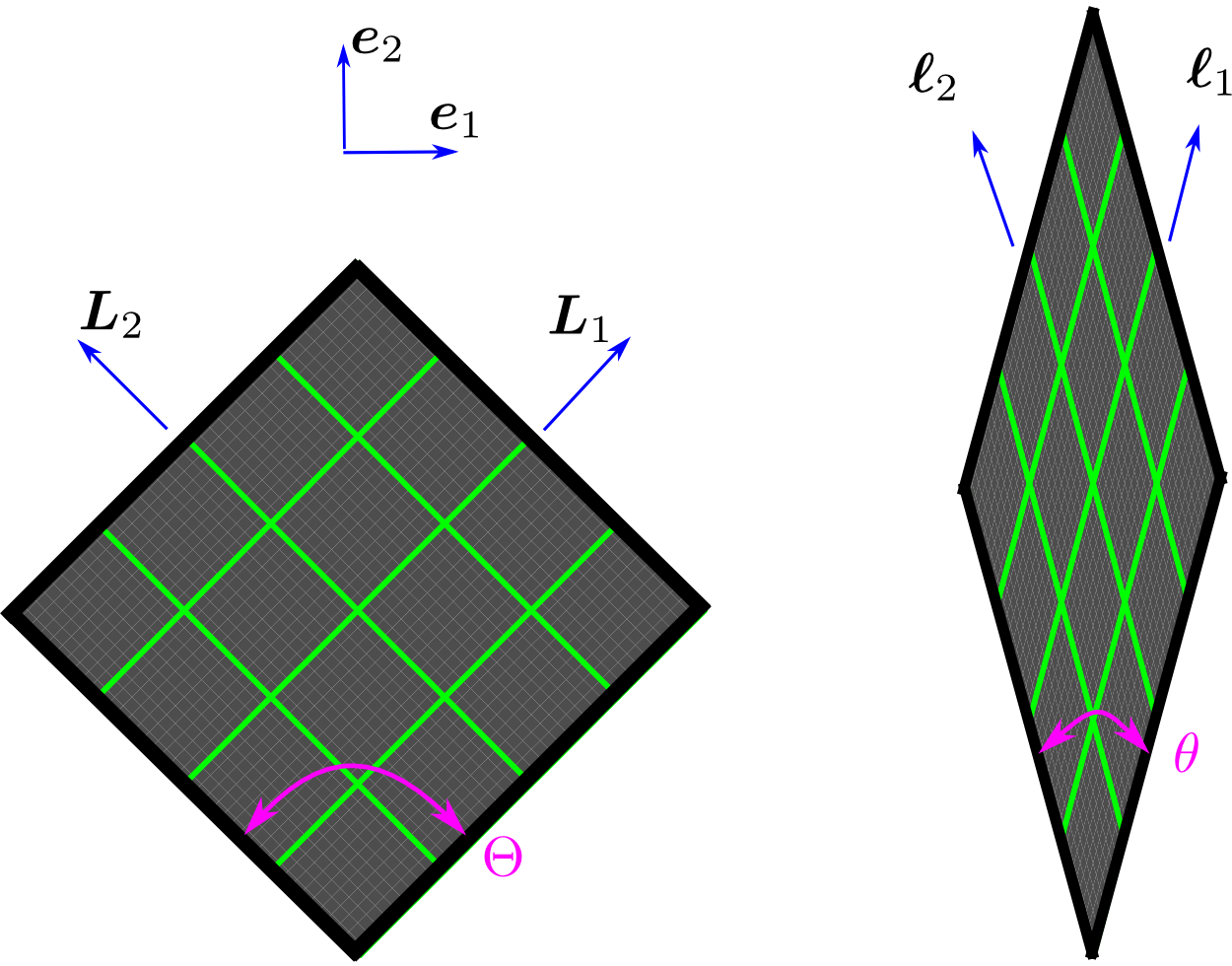}}
\put(0.5,0.05){\includegraphics[width=0.50\textwidth]{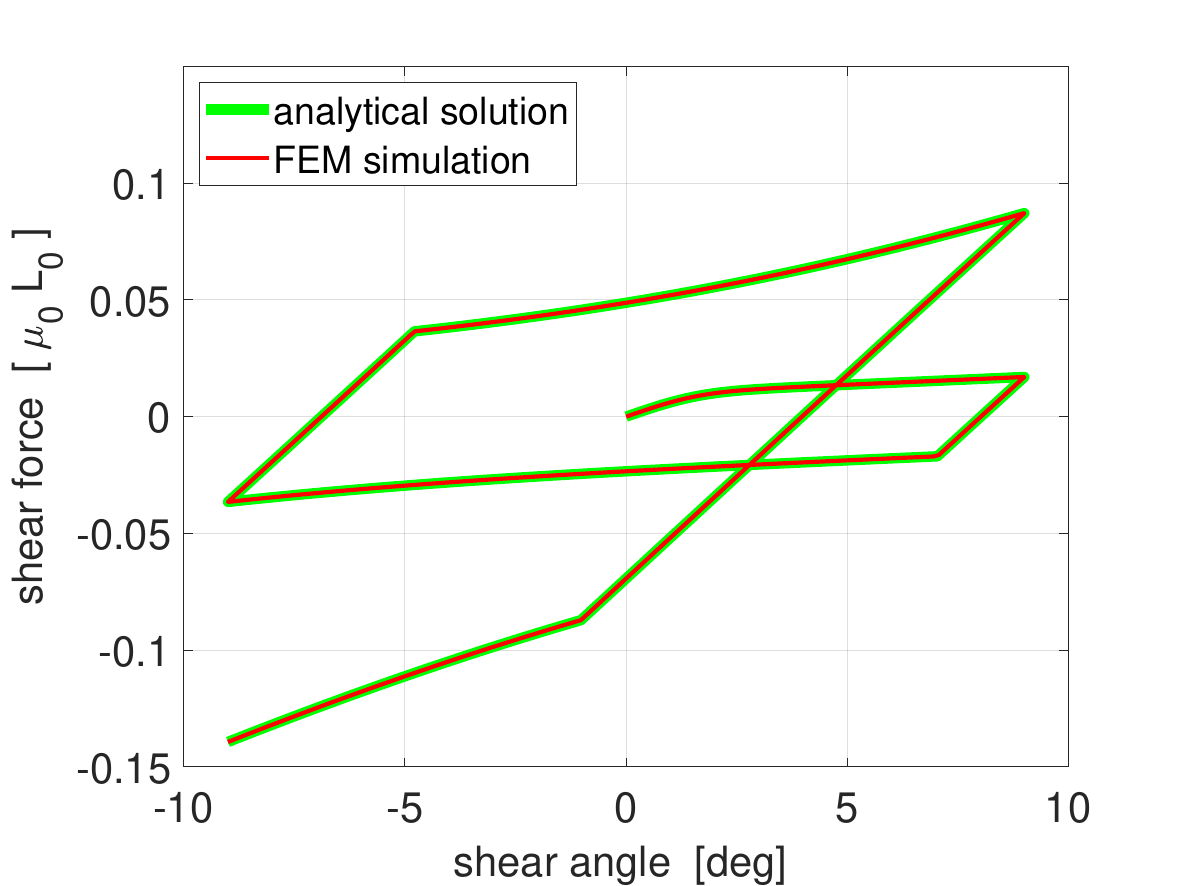}}

\put(-8.0,0.1){{\small{(a)}}}
\put(-2.8,0.1){{\small{(b)}}}
\put(0.6,0.1){{\small{(c)}}}

\end{picture}
\caption{Verification of the proposed angle plasticity model using the  picture frame test: (a)~Initial and (b)~deformed configurations with two fiber families (marked in green). (c) Comparison of FEM results with analytical solution for multiple loading and unloading cycles. Here, the following parameters are used: $\tau_\mry=0$, $\mu_\mrf= \mu_0$, $A=0.05\, \mu_0$, $a=  1$, $B= 0.01\, \mu_0$, $b=   55$, $C= 0.7\, \mu_0$ and $c= 5$. \res{A supplementary Matlab code for this is provided at {\url{https://github.com/xuanthangduong/textile-picture-frame-test.git}}}}
\label{f:picframe_ana}
\end{center}
\end{figure}

The picture frame test consists of  a square sheet of dimension $L_0\times L_0$ with  two embedded fiber families that is deformed into a rhombus as seen Fig.~\ref{f:picframe_ana}a-b. The deformation is applied by the Dirichlet boundary condition 
\eqb{lll}
\bar\bx (\varphi,\bar\bX) = \sqrt{2}\,\big(\cos\varphi\,\be_1\otimes\be_1 + \sin\varphi\,\be_2\otimes\be_2\big)\,\bar\bX~.
\eqe
 Here, $\varphi:= (\pi - \theta)/2$, and $\bar\bx$ and $\bar\bX$ denote the current and initial position of the boundary points, respectively. The plasticity model from Sec.~\ref{s:constitute} is used in the test. Assuming no wrinkling occurs, the problem can be solved analytically, as is shown in \res{Sec.}~\ref{s:ana_picF}. Fig.~\ref{f:picframe_ana}c shows that the finite element solution and analytical solution of the shear forces are identical (within machine precision). This verifies the implemented finite element formulation.

\subsection{Parameter study based on the picture frame test}
In the following,  the picture frame test is used to study the influence  of the parameters appearing in the angle plasticity \resc{model \eqref{e:fy_textile} on} the shear response. Additionally, the agreement  between  simulation and  exact solution is confirmed in all cases.

\subsubsection{Influence of tensile stiffness,  initial shear modulus and yield stress}

\begin{figure}[H] 
\begin{center} \unitlength1cm
\begin{picture}(0,6.5)

\put(-8.1,0){\includegraphics[width=0.50\textwidth]{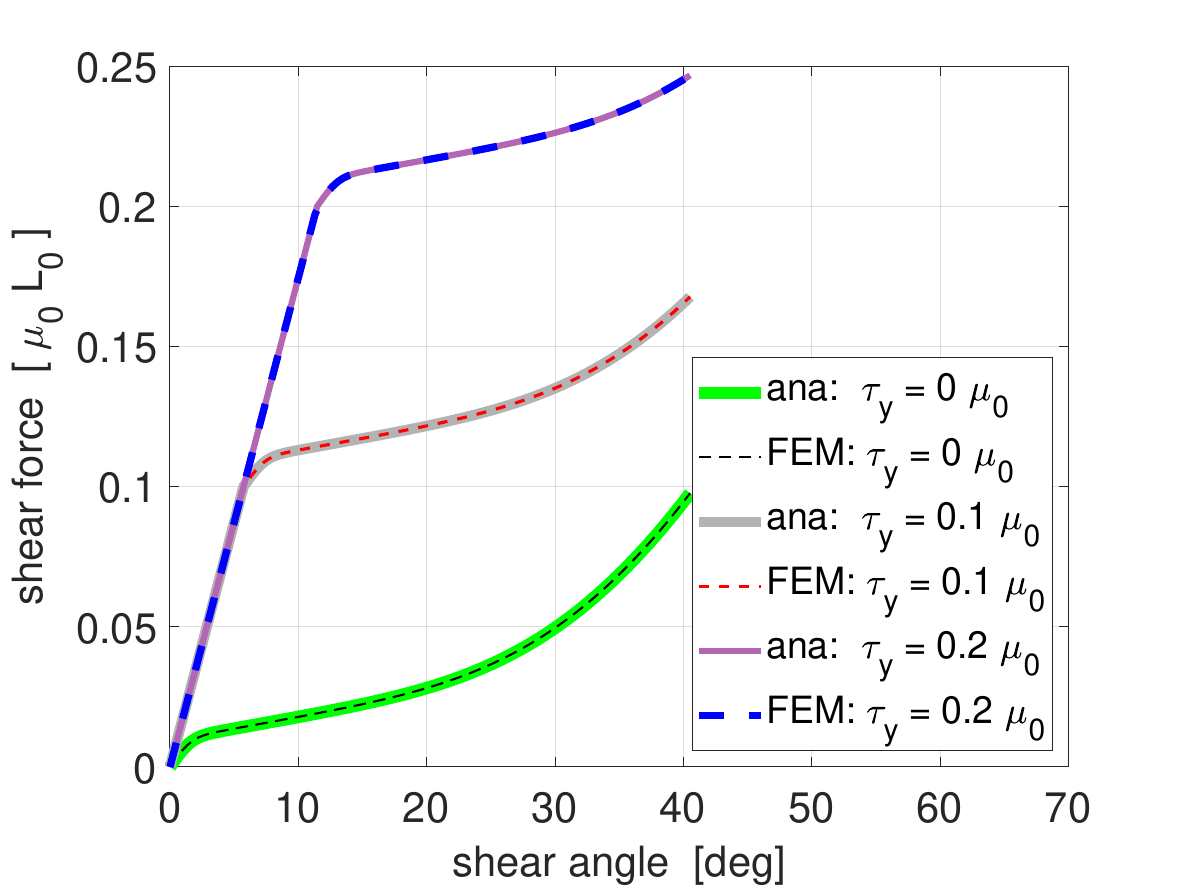}}
\put(0.1,0){\includegraphics[width=0.50\textwidth]{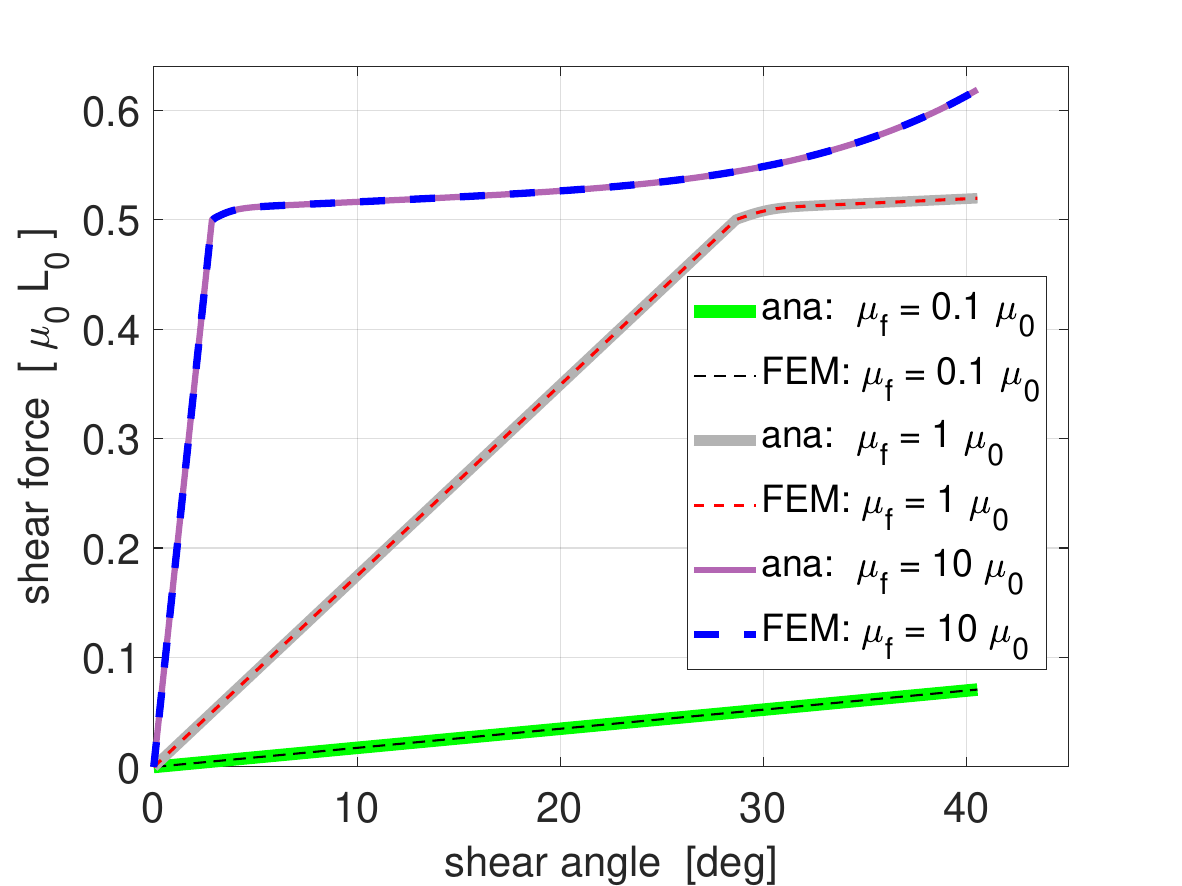}}

\put(-8.0,0.1){{\small{(a)}}}
\put(0.2, 0.1){{\small{(b) }}}

\end{picture}
\caption{Parameter study based on the picture frame test during phase I: influence of (a)~yield stress $\tau_\mry$ and  (b)~initial shear modulus $\mu_\mrf$ of model \eqref{e:fy_textile}. Here, unless otherwise varied and specified in the figure legends, the following parameters are used: $\tau_\mry=0.5\,\mu_0$, $\mu_\mrf= \mu_0$, $A=0.05\, \mu_0$, $a=  1$, $B= 0.01\, \mu_0$, $b=   55$, $C= 0.7\, \mu_0$ and $c= 5$.}
\label{f:picframe_ana_par}
\end{center}
\end{figure}

Tensile fiber stiffness $\epsilon_\mrL$ has no influence on the shear response of fabrics when it is sufficiently large.  
Fig.~\ref{f:picframe_ana_par}a-b shows the influence of yield stress $\tau_\mry$ and initial shear modulus $\mu_\mrf$ on to the elastic shear response of fabrics  (phase I, see \res{Sec.}~\ref{r:model_motive}). As expected, parameter $\tau_\mry$  offsets the curve vertically, while $\mu_\mrf$ controls its  slope  within the elastic range.

\subsubsection{Influence of angle plasticity parameters}
\begin{figure}[H]
\begin{center} \unitlength1cm
\begin{picture}(0,12.5)
\put(-8.1,6.5){\includegraphics[width=0.50\textwidth]{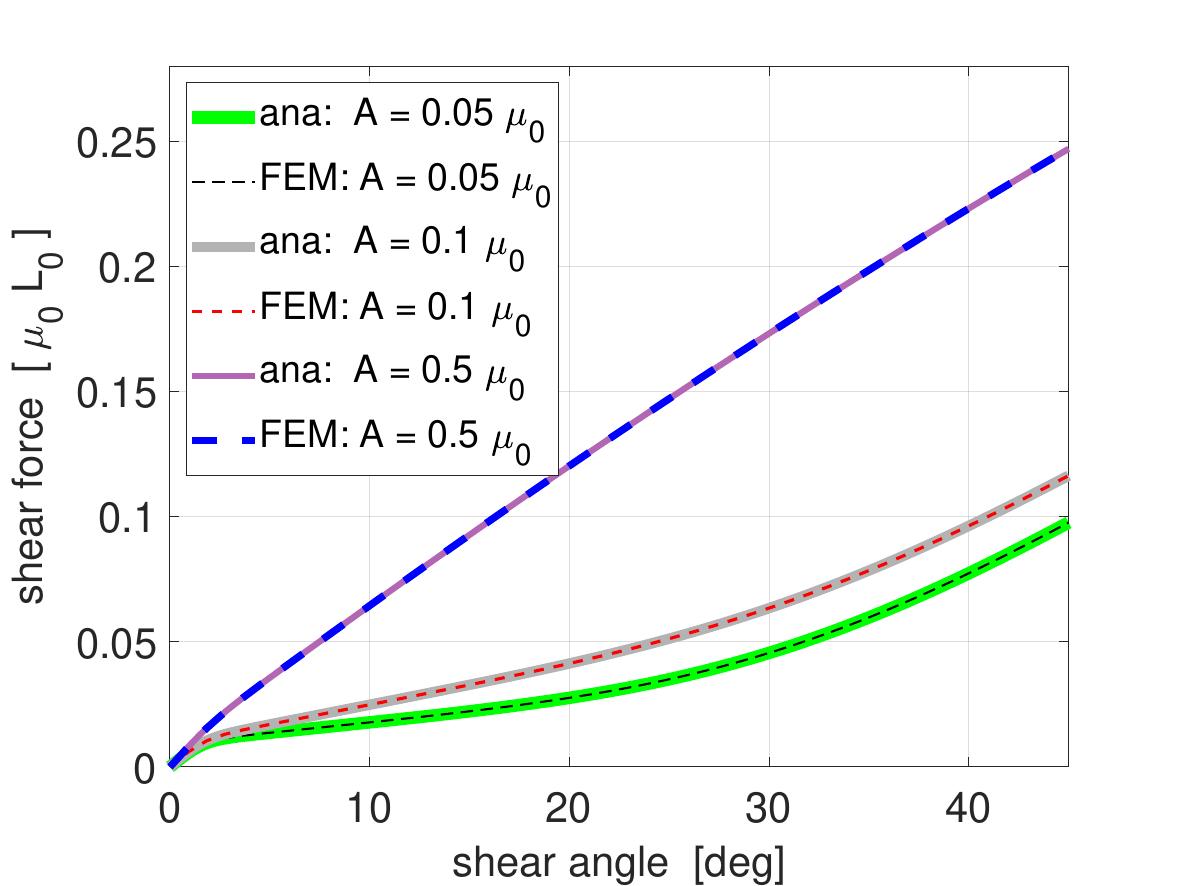}}
\put(0.1,6.5){\includegraphics[width=0.50\textwidth]{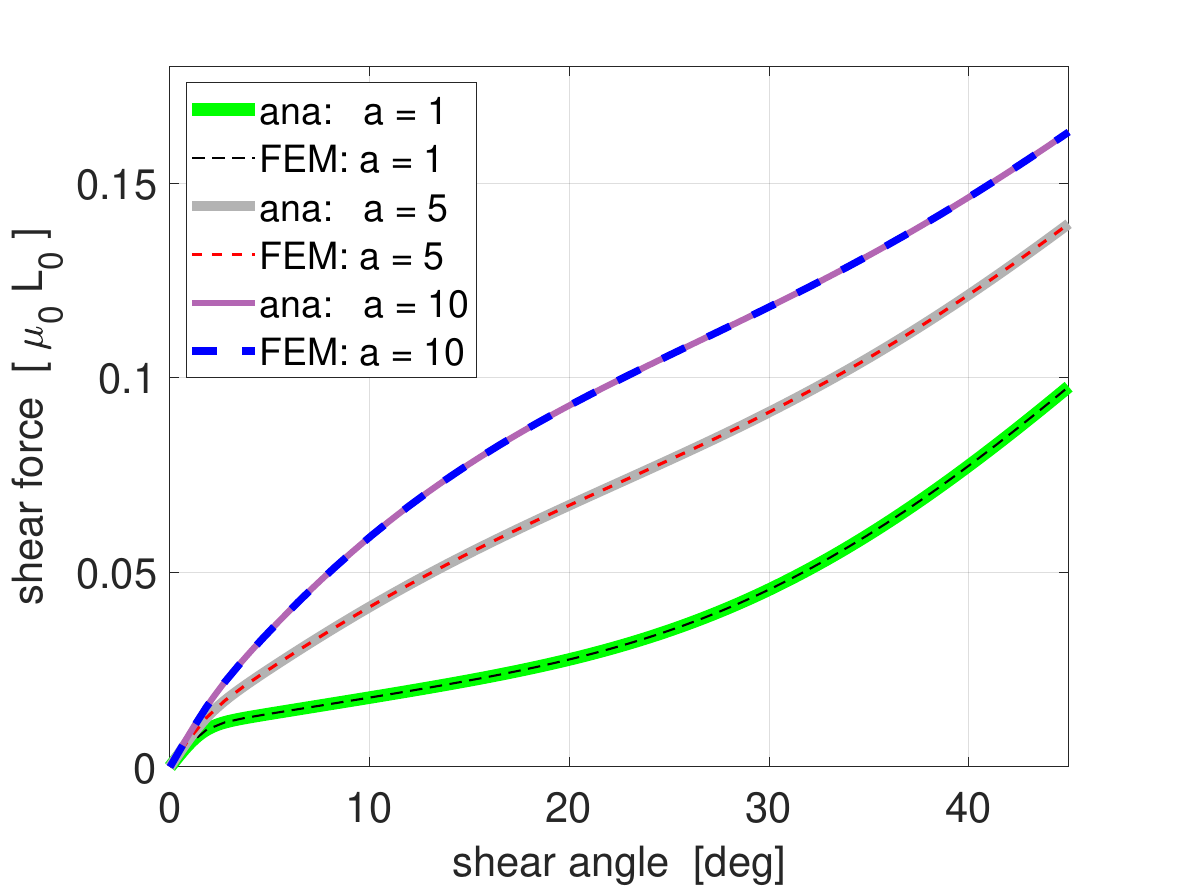}}
\put(-8.0,6.45){{\small{(a)}}}
\put(0.2,6.45){{\small{(b)}}}

\put(-8.1,-0.15){\includegraphics[width=0.50\textwidth]{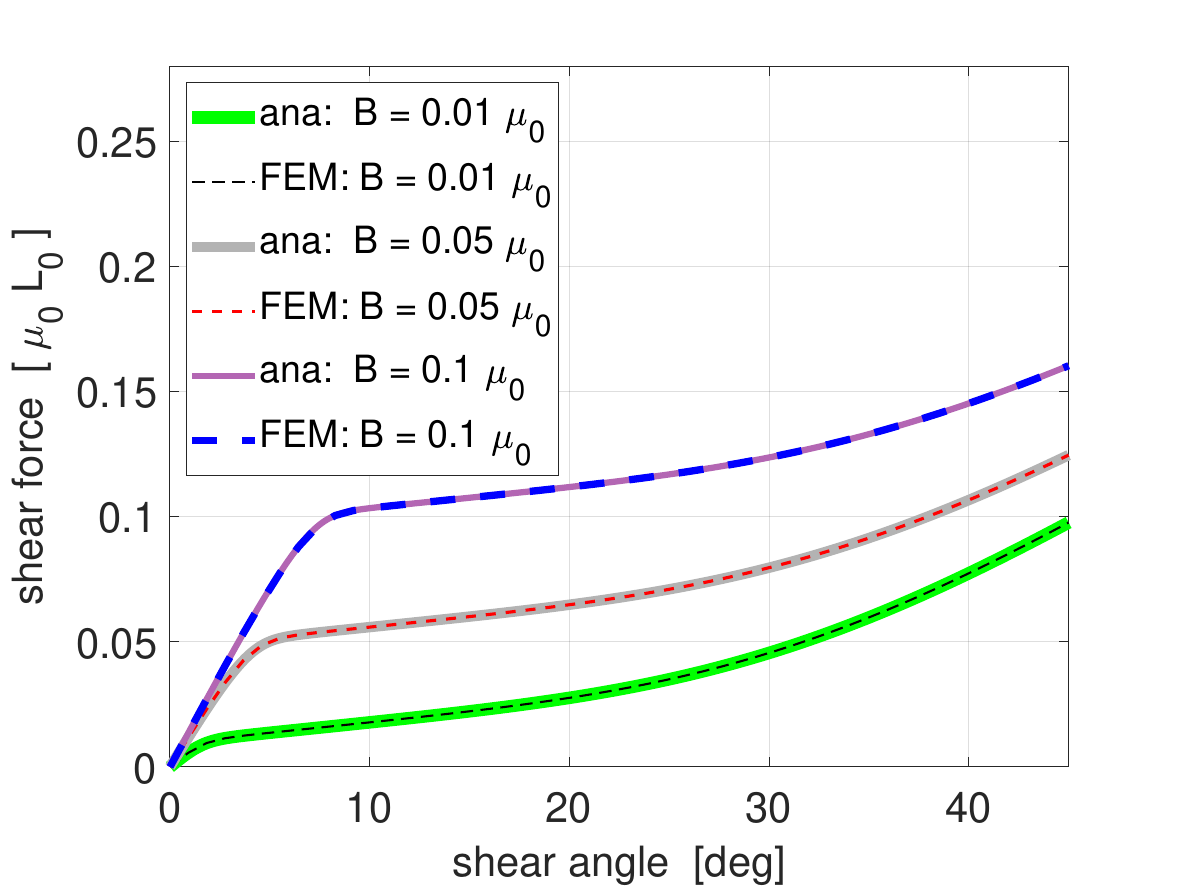}}
\put(0.1,-0.15){\includegraphics[width=0.50\textwidth]{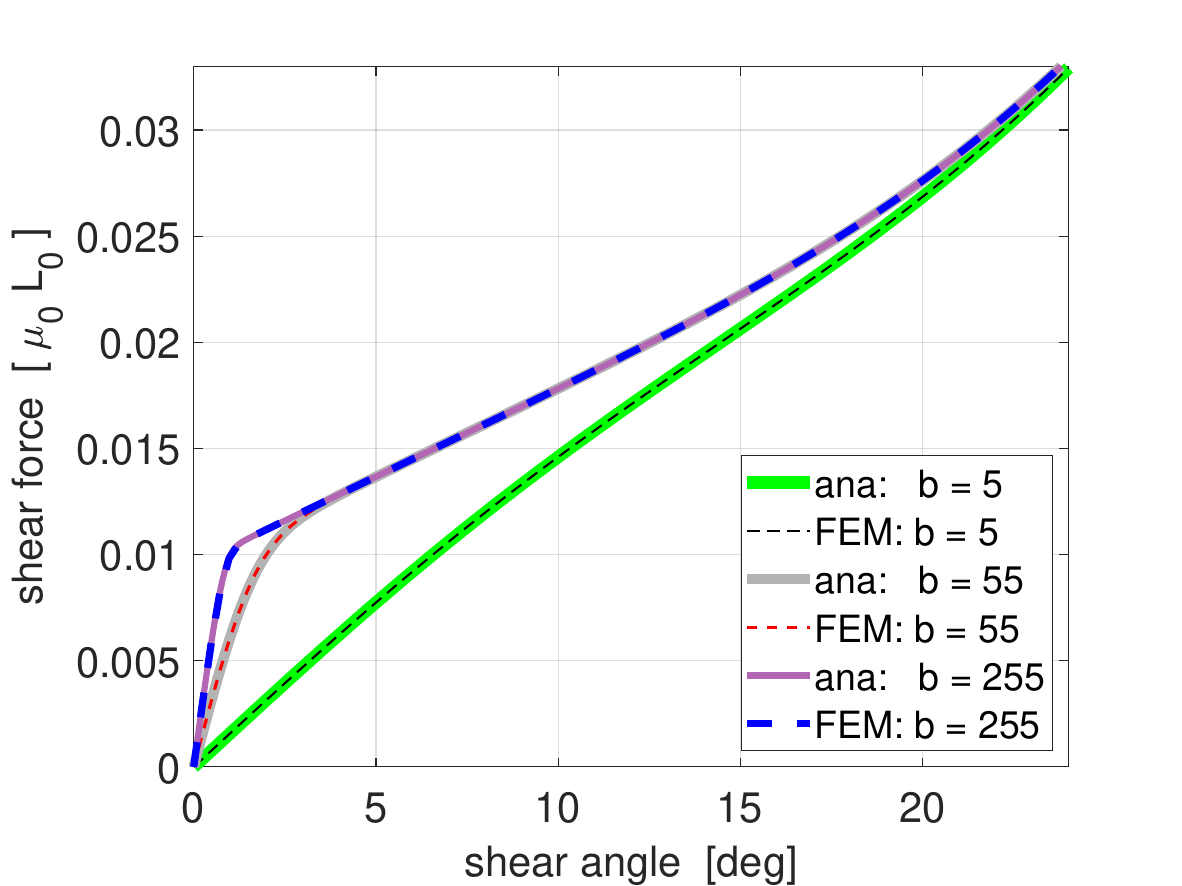}}
\put(-8.0,-0.1){{\small{(c)}}}
\put(0.2,-0.1){{\small{(d)}}}
\end{picture}
\caption[caption]{Parameter study based on the picture frame test \resc{during phase} II: (a)-(d)~influence of parameter $A$, $a$, $B$ and $b$, respectively, of model \eqref{e:fy_textile}. These parameters  govern the  low plastic resistance phase (phase II). Here, unless otherwise varied and specified in the figures, the following parameters are used: $\tau_\mry=0$, $\mu_\mrf= \mu_0$, $A=0.05\,\ \mu_0$, $a=  1$, $B= 0.01 \,\mu_0$, $b=   55$, $C= 0.7\,\ \mu_0$ and $c= 5$.}
 \label{f:eg_study_shear}
\end{center}
%
\begin{center} \unitlength1cm
\begin{picture}(0,7.0)

\put(-8.1,-0.15){\includegraphics[width=0.50\textwidth]{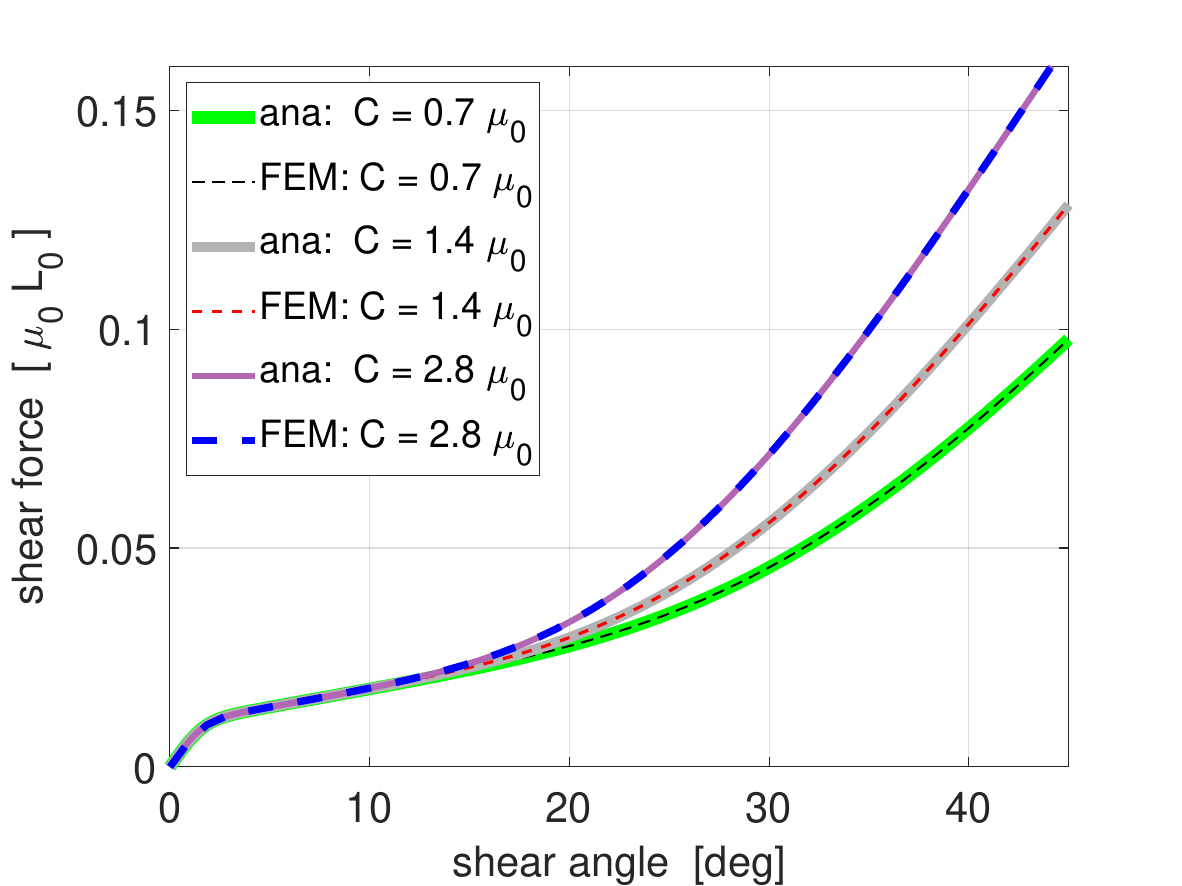}}
\put(0.1,-0.15){\includegraphics[width=0.50\textwidth]{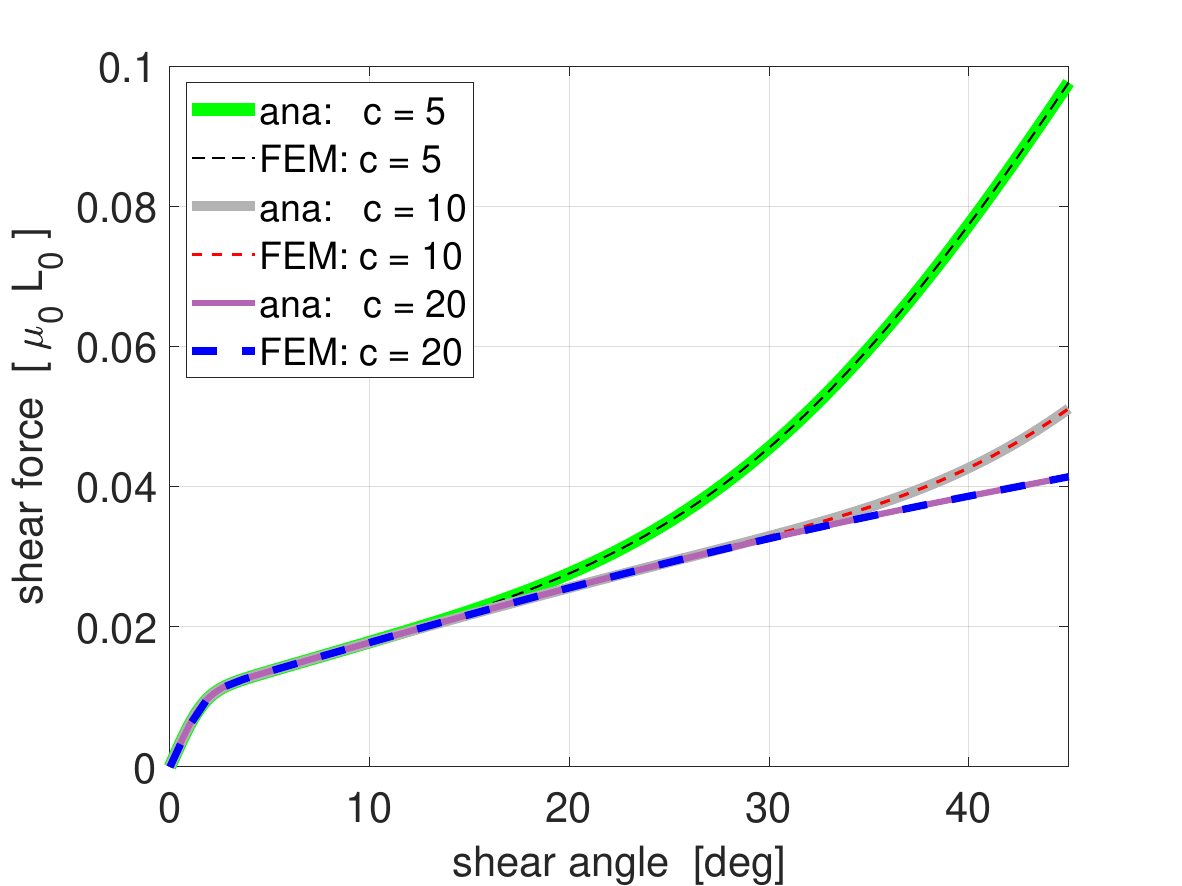}}
\put(-8.0,-0.1){{\small{(a)}}}
\put(0.2,-0.1){{\small{(b)}}}
\end{picture}
\caption[caption]{Parameter study based on the picture frame test during phase III: (a)-(b)~influence of parameter $C$ and $c$, respectively, of model \eqref{e:fy_textile}. These parameters  govern the increased hardening phase (phase III). Here, $\tau_\mry=0$, $\mu_\mrf= \mu_0$, $A=0.05\, \mu_0$, $a=  1$, $B= 0.01\, \mu_0$, $b=   55$ are used.}
 \label{f:eg_study_shear_Cc}
\end{center}
\end{figure}

Fig.~\ref{f:eg_study_shear}a-d show the  influence of parameters $A$, $a$, $B$, and $b$ in the proposed plasticity \resc{model \eqref{e:fy_textile}.} This set of parameters  govern the  low plastic resistance phase (phase II, see \res{Sec.}~\ref{r:model_motive}).
Note that $\tau_\mry$ is set to zero here, so that no elasticity is present.  
Parameters $A$ and $a$ change the global slope and offset of the curve as seen in Fig.~\ref{f:eg_study_shear}a-b, while parameters $B$ and $b$  control the slope and offset at small shear angles as shown in Fig.~\ref{f:eg_study_shear}c-d. 

On the other hand, parameters $C$ and $c$ govern the shear response during phase III (see \res{Sec.}~\ref{r:model_motive}) in the fabric. They thus   only alter the later part of the curve as seen in Fig.~\ref{f:eg_study_shear_Cc}. In theory, $\gamma=90^\circ$ is the limit of deformation as  the frame area reaches zero. Although the power law with parameters $C$ and $c$ does not strictly prevent this limit state, the computation becomes infeasible due to a \resc{vanishing} finite element Jacobian. In practice, due to a finite yarn thickness, the shear force usually changes rapidly for shear angles that are far from this limit state. For example, the data in Fig.~\ref{f:fitting_plastic_shear} shows that phase III already begins around $\gamma=35$-$45^\circ$.

%

\section{{Model calibration and validation}}{\label{s:model_verifi_cali_valid}}
The section presents the calibration and validation of the proposed angle plasticity model. The case of plain weave (glass) fabrics is considered using experimental data provided in the literature.
\subsection {Model calibration}
The following three subsections discuss the calibration of all the different material parameters. The results are summarized in Tab.~\ref{t:WFconstant}.

\begin{table}[!htp]
\small
\begin{center}
\def\arraystretch{1.5}\tabcolsep=3.5pt
\begin{tabular}{|c|l|l|l|l| }
\hline
 \parbox[t][][t]{0.9cm}{~para.} & \parbox[t]{1.0cm}{\,value}& \parbox[t]{1.2cm}{unit}   &\parbox[t]{3.5cm}{physical meaning}  &\parbox[t]{4.5cm}{calibrated from} \\ \hline\hline
$\mu_{\mrf}$ & $5.0$  & {{N/mm}}  & initial fabric shear modulus  & pic.\,frame (LSMP):  \cite{Cao2008a} \\ \hline
$\tau_\mry$ & $10^{-4}$  & {{N/mm}}  &initial fabric  yield stress  &   assumption \\ \hline
$A$ & $8.8 $ & {{N/mm}}  & yield function parameter \#1  & pic.\,frame (KUL): \cite{Cao2008a} \\ \hline
$a$ & $0.0024$  & {{-}}  &  yield function parameter  \#2  & pic.\,frame (KUL): \cite{Cao2008a} \\ \hline
$B$ &$ 0.0028 $ & {{N/mm}}  &  yield function parameter  \#3 & pic.\,frame (KUL): \cite{Cao2008a} \\ \hline
$b$ & $65.0 $ & {{-}}  &  yield function parameter  \#4  & pic.\,frame (KUL):  \cite{Cao2008a}\\ \hline
$C$ & $1.0$  & {{N/mm}}  & yield function parameter  \#5  & pic.\,frame (UN\,\&\,UML):  \cite{Cao2008a}\\ \hline
$c$ &$ 11.0 $& {{-}}  &  yield function parameter  \#6 &pic.\,frame (UN\,\&\,UML):  \cite{Cao2008a} \\ \hline
$\epsilon_{\mrL}$ & $110.0$  & {{N/mm}}  &  tensile fabric stiffness  & uniaxial tensile:  \cite{pengCao2005}\\ \hline
$\beta_{\mrn} $  & 3.023   & {N\,mm} & out-of-plane fabric bend. \!stiffness  & fiber cantilever: \cite{ITAdata2022} \\ \hline
$\beta_{\mrg} $ & 3.023   & {N\,mm} & in-plane fabric bend. \!stiffness  & assumption \\ \hline
$\beta_{\tau} $  & 3.023   & {N\,mm} & fiber torsion stiffness  &  assumption \\ \hline
\end{tabular}
\end{center}
\caption{Calibrated material parameters of fabric model \eqref{e:eg_Wsimple} suitable to  \resc{plain} weave fabrics of glass fibers. Here, we use the following data sources reported by \citet{pengCao2005,Cao2008a} and \citet{ITAdata2022} 
to fit the proposed plasticity model: Laboratoire de M{\'e}canique des Syst{\`e}mes et des Proc{\'e}d{\'e}s (LMSP),  Katholieke Universiteit Leuven in Belgium (KUL), University of Nottingham in UK (UN), 
and University of Massachusetts Lowell in USA (UML). 
}
\label{t:WFconstant}
\end{table}

\subsubsection{Calibration of the initial tensile and shear modulus of plain weave fabric}{\label{s:fitting_tensile_shere}}

\begin{figure}[H]
\begin{center} \unitlength1cm
\begin{picture}(0,5.6)

\put(-8.1,-0.15){\includegraphics[width=0.50\textwidth]{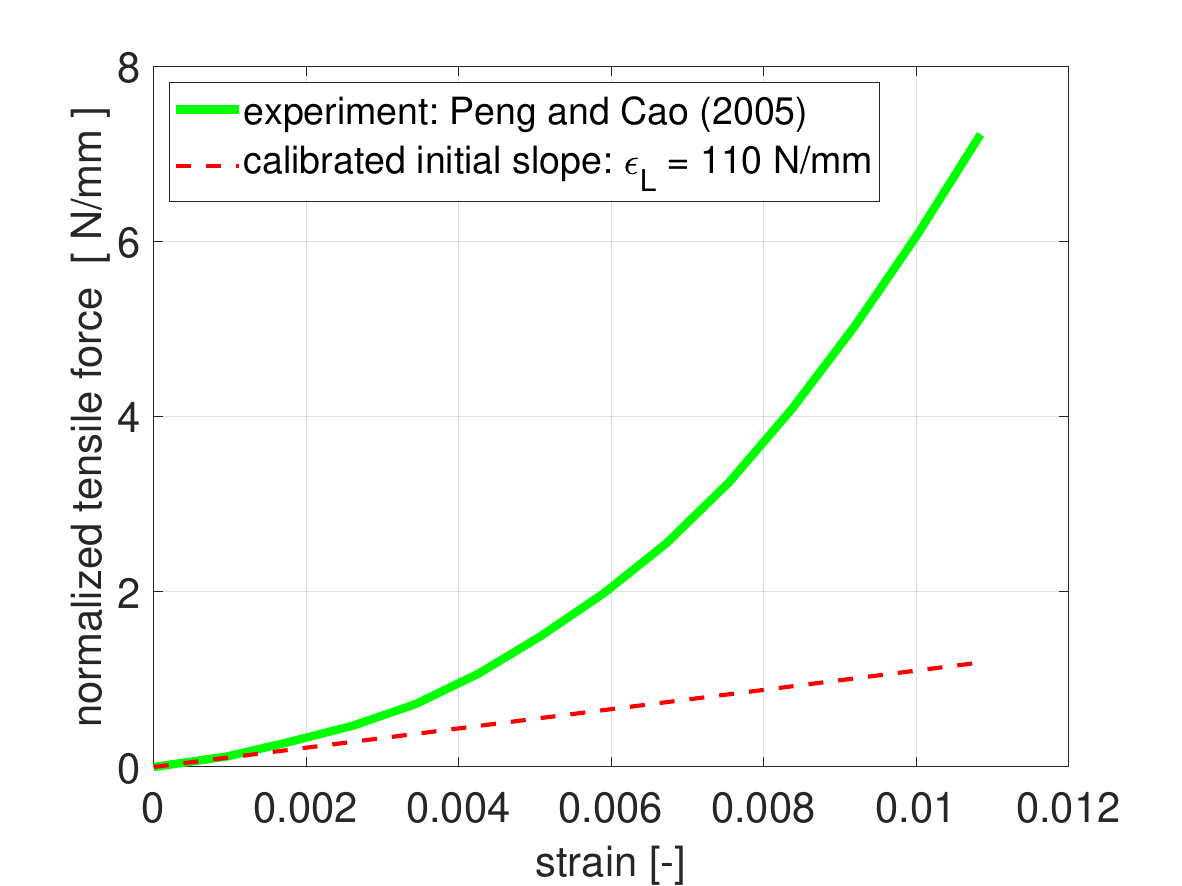}}
\put(-6.5,1.3){\includegraphics[width=0.055\textwidth]{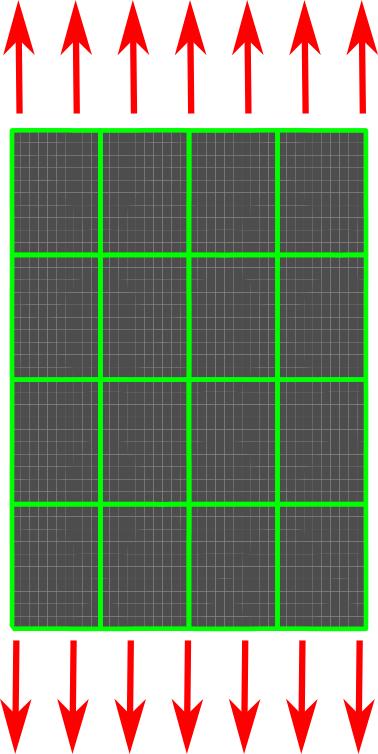}}

\put(0.1,-0.15){\includegraphics[width=0.50\textwidth]{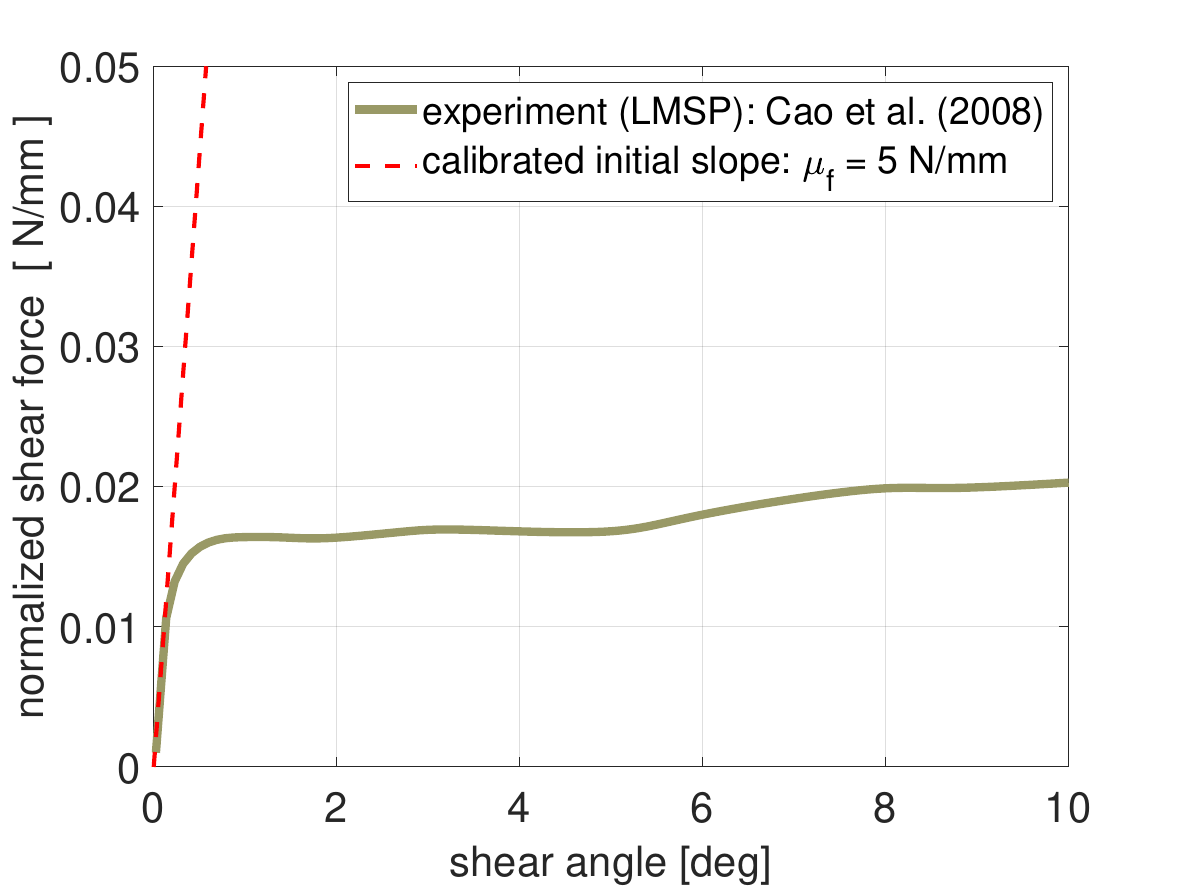}}
\put(1.8,2.4){\includegraphics[width=0.12\textwidth]{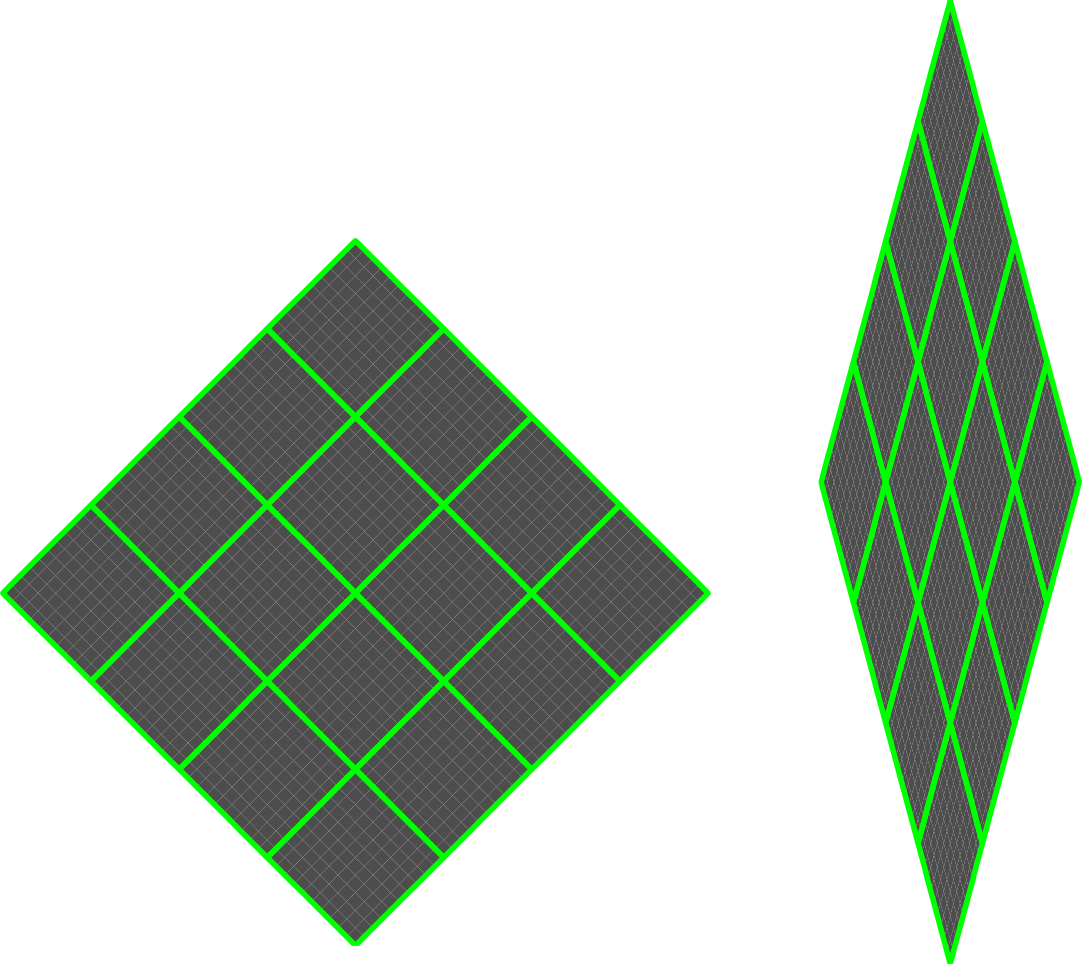}}

\put(-8.0,-0.1){{\small{(a)}}}
\put(0.2,-0.1){{\small{(b)}}}
\end{picture}
\caption[caption]{Model calibration of the initial tensile stiffness $\epsilon_\mrL$ and initial elastic shear modulus $\mu_\mrf$ for plain weave dry  fabric  -- see Eq.~(\ref{e:eg_W1}.1) and \eqref{e:tau_eps_phi} , respectively --  from  (a) uniaxial tensile test and  (b) picture frame test, respectively. The former uses the experimental data  of the uniaxial tensile test averaged over the four fabric samples provided by  \cite{pengCao2005}. The latter uses the data for the shear force normalized by the frame length, reported by Laboratoire de M{\'e}canique des Syst{\`e}mes et des Proc{\'e}d{\'e}s (LMSP) in \citet{Cao2008a}.  The  shear angle is defined as $\gamma:=90^\circ- \theta$.}
\label{f:fitting_tensile_shear}
\end{center}
\vspace{-0.5cm}
\end{figure}

Tensile stiffness $\epsilon_\mrL$  in Eq.~(\ref{e:eg_W1}.1) can be calibrated from experimental data of uniaxial tensile tests as shown in Fig.~\ref{f:fitting_tensile_shear}a. 
 Here,  in order to account for the undulation of yarns within the fabric, we use the tensile data provided by  \cite{pengCao2005} for fabric samples  instead of a single yarn, and our model is fitted to the data such that the initial slopes are matched.

Initial fabric shear modulus  $\mu_\mrf$  in Eq.~\eqref{e:tau_eps_phi}, on the other hand, needs to be calibrated from a shear test, such as the picture frame test, as shown in Fig.~\ref{f:fitting_tensile_shear}b.  Here, the data set from Laboratoire de M{\'e}canique des Syst{\`e}mes et des Proc{\'e}d{\'e}s (LMSP) \citep{Cao2008a} is used, 
 since it was obtained from the largest ratio of the fabric length ($240$mm, excluding the clamp-affected zone) to the frame length ($245$mm),  cf.~Tab.~3 in \cite{Cao2008a}. Such a setup maximizes the interaction between yarns in the initial loading phase, which may contribute to the largest initial slope among other testing setups as seen in Fig.~\ref{f:fitting_plastic_shear}a and Fig.~25 in \cite{Cao2008a}. 
 

\subsubsection{Calibration of the proposed  yield function}
The material parameters required to calibrate the proposed yield function \eqref{e:fy_textile} are $\tau_\mry$, $A$, $a$, $B$, $b$, $C$, and $c$ -- see also \res{Sec.}~\ref{r:model_motive}.   In principle, these seven parameters can be determined from the picture frame test. However there is no unloading data available for glass fibers in the literature. We therefore assume that the initial yield stress $\tau_\mry$ has a small value ($10^{-4}$ N/mm). This is justified since rotational sliding between warp and weft yarns is expected to start  for small \resc{shear angles.}

\begin{figure}[H]
\begin{center} \unitlength1cm
\begin{picture}(0,5.7)

\put(-8.1,-0.15){\includegraphics[width=0.50\textwidth]{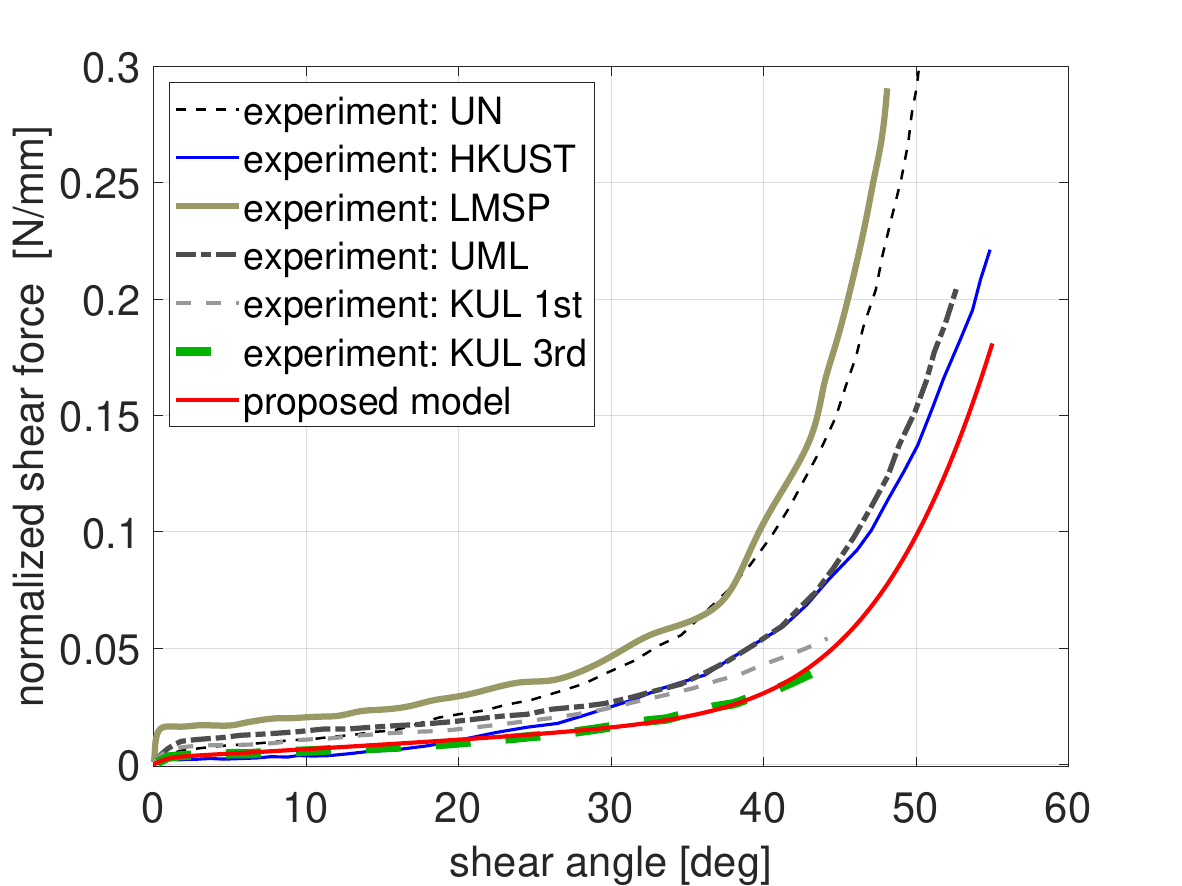}}
\put(-6.8,1.1){\includegraphics[width=0.12\textwidth]{figs/picture_frame_x_small.png}}

\put(0.1,-0.15){\includegraphics[width=0.50\textwidth]{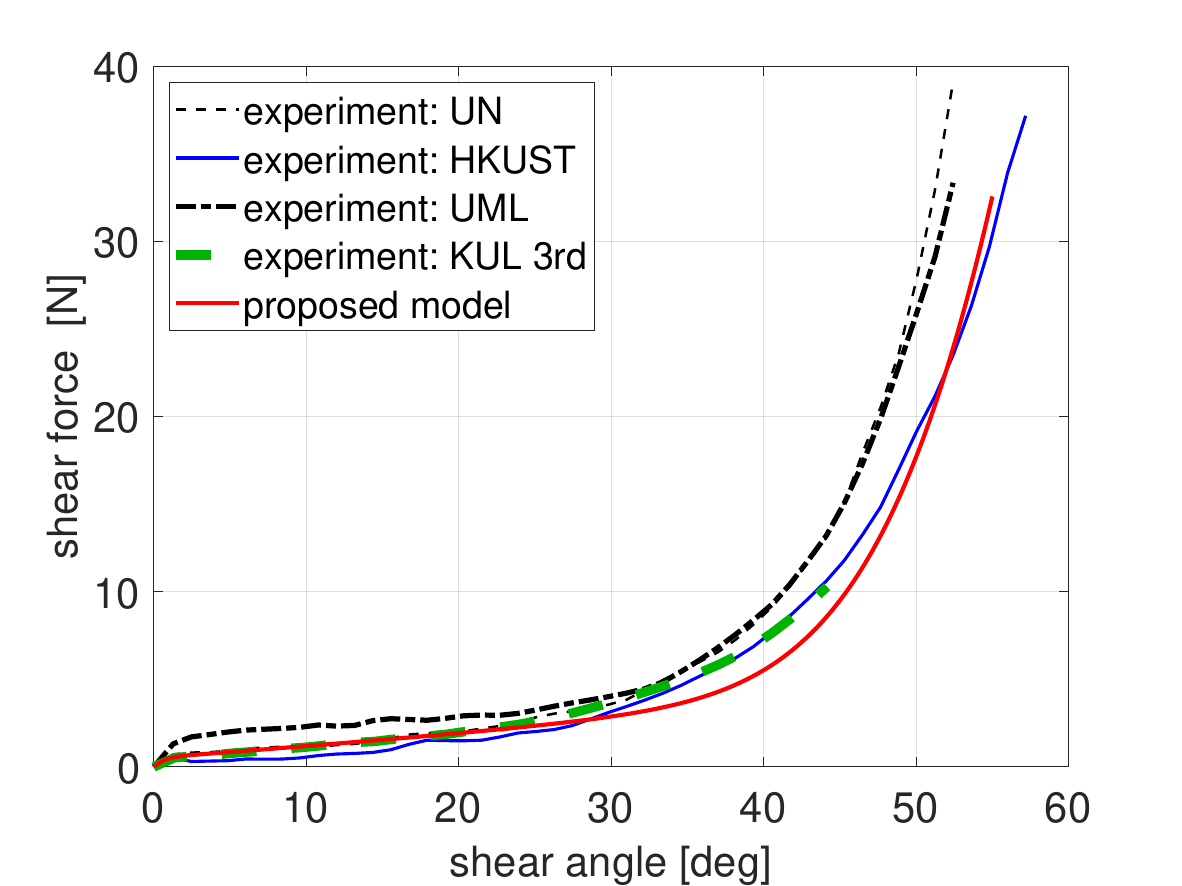}}
\put(1.8,1.3){\includegraphics[width=0.12\textwidth]{figs/picture_frame_x_small.png}}

\put(-8.0,-0.1){{\small{(a)}}}
\put(0.2,-0.1){{\small{(b)}}}
\end{picture}
\caption[caption]{Model calibration for the yield function parameters $A$, $a$, $B$, $b$, $C$, and $c$ -- see Eq.~\eqref{e:fy_textile} -- from the picture frame test  for (a) the shear force 
 normalized by the frame length,  and (b) the shear force  that has been scaled by the equivalent fabric size $L_0 = 180$mm. I.e.~$F_\mathrm{scaled}:= (L/L_0)\,F_\mathrm{reported}$, where $F_\mathrm{reported}$ denotes the shear force reported from experiments with fabric length $L$.  Here, the experimental curves for the normalized shear force and for  $F_\mathrm{reported}$ are taken from  Fig.~25 and  Fig.~22, respectively, in \cite{Cao2008a} for the various \resc{data sources.} 
 }
 \label{f:fitting_plastic_shear}
\end{center}
\vspace{-0.5cm}
\end{figure}

Apart from $\tau_\mry$, the four parameters $A$, $a$, $B$, $b$ -- governing the rotational sliding between yarns -- are fitted to  the experimental data from  \cite{Cao2008a}, as  shown in Fig.~\ref{f:fitting_plastic_shear}a. As seen, there is large scattering between the experimental data from different sources. In particular,  the  LMSP data --  while its first phase is useful in determining the initial shear modulus (see  Sec.~\ref{s:fitting_tensile_shere}) -- becomes unsuitable for the purpose of fitting  $A$, $a$, $B$, $b$, since its second phase contains  the shear resistance of the yarns at the sample edges due to the large fabric-frame length ratio used. The data from KUL,  on the other hand,  is more suitable (up to the point of the increased hardening phase) since it was obtained from three repetitions on a single sample. It was \resc{reported that} the second and the third repetitions gave similar data that was significantly lower than the first instance  (KUL 1st). It is noticeable in Fig.~\ref{f:fitting_plastic_shear}a that  the KUL 1st  curve  lies approximately at the mean of the data from UN, HKUST, and UML. This leads to our conjecture that the rotational sliding phase of  the KUL 3rd  curve is more accurate than the others, and we therefore use it to calibrate parameters  $A$, $a$, $B$, $b$ in our proposed model.

For the calibration of parameters $C$ and $c$ -- governing the increased hardening behavior (phase III), see \res{Sec.}~\ref{r:model_motive} --  the KUL data is insufficient as it does not go far enough as is seen from Fig.~\ref{f:fitting_plastic_shear}a.  We therefore calibrate  $C$ and $c$ such that the third phase of the curve from our model approximately matches the slope of the UN and UML curves. This is seen in Fig.~\ref{f:fitting_plastic_shear}a.

Further, we verify the applicability of the frame length normalization employed for our calibration by plotting the shear force versus shear angle in Fig.~\ref{f:fitting_plastic_shear}b. As seen, the curve for our fitted model is in good agreement with the KUL 3rd curve up to a shear angle of $25$ degrees. After that, our proposed curve deviates from the   KUL 3rd curve but it still generally follows the trend of the  UN and UML curves.

\subsubsection{Calibration of other parameters}{\label{s:fitting_beanding}}
Finally, we need to calibrate the bending stiffnesses  $\beta_\mrn$ and   $\beta_\mrg$  and torsional  stiffness $\beta_\tau$ for fiber bending model~(\ref{e:eg_W1}.\otc{2-3}). For $\beta_\mrn$ and   $\beta_\mrg$, we use the value of $3.023$ N/mm from the bending test of a single yarn reported in \cite{ITAdata2022}. To the best of our knowledge, no data for the torsion of yarns is available in the literature. We thus use $\beta_\tau=3.023$ N/mm in our simulations.

\subsection {Model validation using the bias extension test}
This section validates our calibrated plasticity model using the bias extension test for a rectangular sheet of plain weave fabric \citep{Cao2008a}. 
\begin{table}[!htp]
\small
\begin{center}
\def\arraystretch{1.4}\tabcolsep=5.0pt
\begin{tabular}{|c|l|l|l|}
\hline
 \parbox[t][][t]{1.65cm}{~sample} & \parbox[t]{2.5cm}{size [mm$^2$]}& \parbox[t]{2.6cm}{aspect~ratio}   &\parbox[t]{4.5cm}{\resc{experimental} data source}   \\ \hline\hline
\#1 & $100\times 200$  &  $1\!:\!2$& UN  \\ \hline
\#2 & $115\times 230$   & $1\!:\!2$ &HKUST,  INSA-NU  \\ \hline
\#3 & $150\times 300$   & $1\!:\!2$ & INSA-NU  \\ \hline
\#4  & $100\times 250$  & $1\!:\!2.5$  &UN \\ \hline
\#5 & $100\times 300$  &  $1\!:\!3$  & UN,  INSA-NU \\ \hline
\#6 & $150\times 450$   & $1\!:\!3$  & INSA-NU  \\ \hline
\end{tabular}
\end{center}
\caption{List of considered samples for the model verification using the bias extension test.  The experimental data -- here extracted from \cite{Cao2008a}  cf.~Fig.~38 therein -- originates from different institutions:  University of Nottingham in UK (UN), Hong Kong University of Science and Technology (HKUST),  INSA-Lyon in France and Northwestern University in USA  (INSA-NU).}
\label{t:data_bias_list}
\end{table}

We consider the six samples  listed in  Tab.~\ref{t:data_bias_list}. These samples vary in size, but can be grouped into three sample classes based on the appearing aspect ratios $1\!:\!2$, $1\!:\!2.5$ and $1\!:\!3$. 
The three sample classes are discretized by $32 \times 64$, $32 \times 80$ and $32 \times 96$  quadratic NURBS finite elements, respectively.  The calibrated material parameters from Tab.~\ref{t:WFconstant} are used in all subsequent computations. \res{Here, it is noted that, since we use identical material parameters for the two fiber families, and in-plane bending stiffness is relatively small compared to the tensile fabric stiffness, the influence of the in-plane bending on the following results appears insignificant. However, it is still needed for the convergence of shear bands as well as computational stability. This is shown in \cite{shelltextileIGA}.}
%

%
%
%

\begin{figure}[H]
\begin{center} \unitlength1cm
\begin{picture}(0,9.3)

\put(-8.6,-0.8){\includegraphics[height=0.60\textwidth]{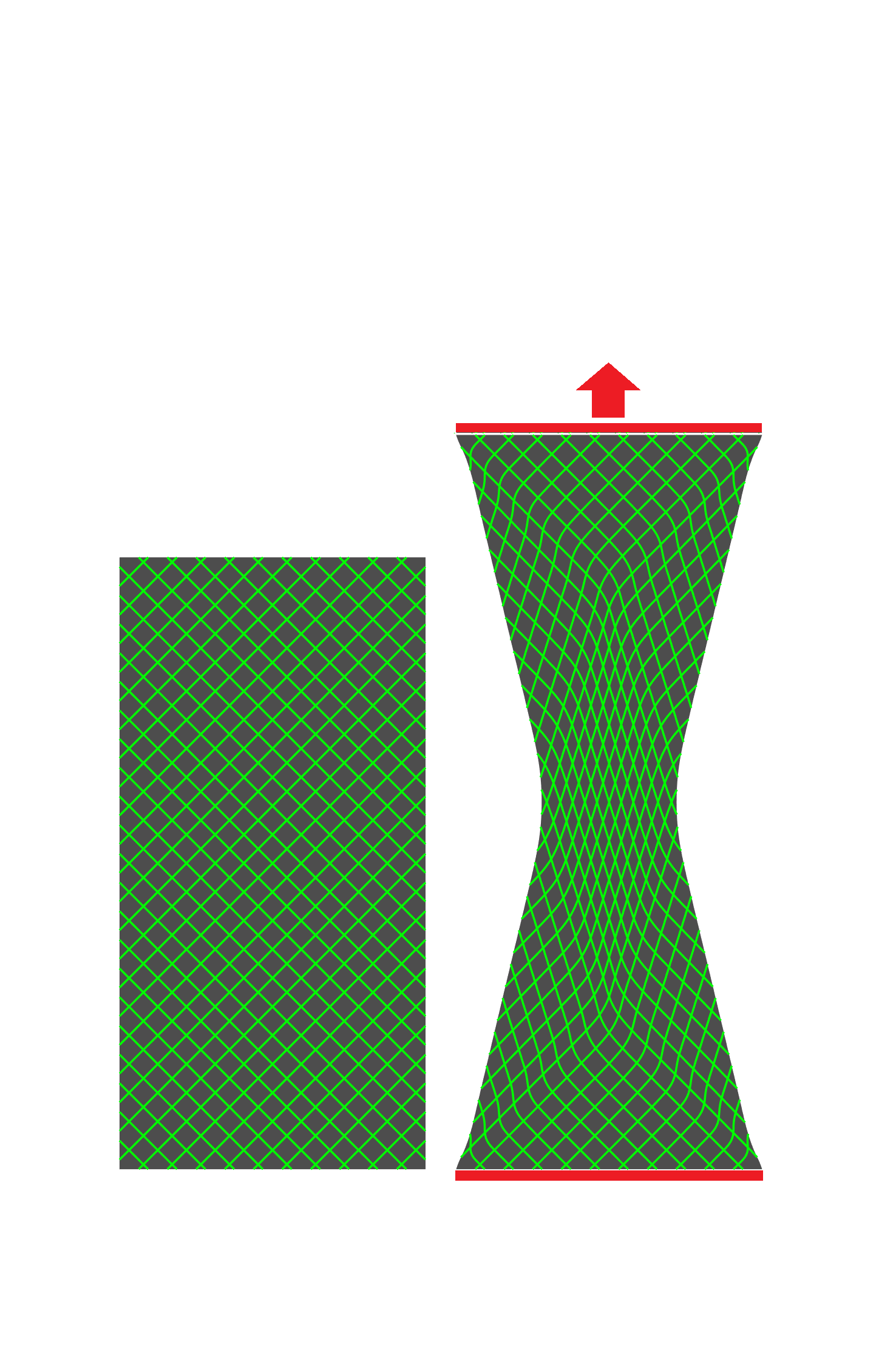}}
\put(-3.0,-0.8){\includegraphics[height=0.60\textwidth]{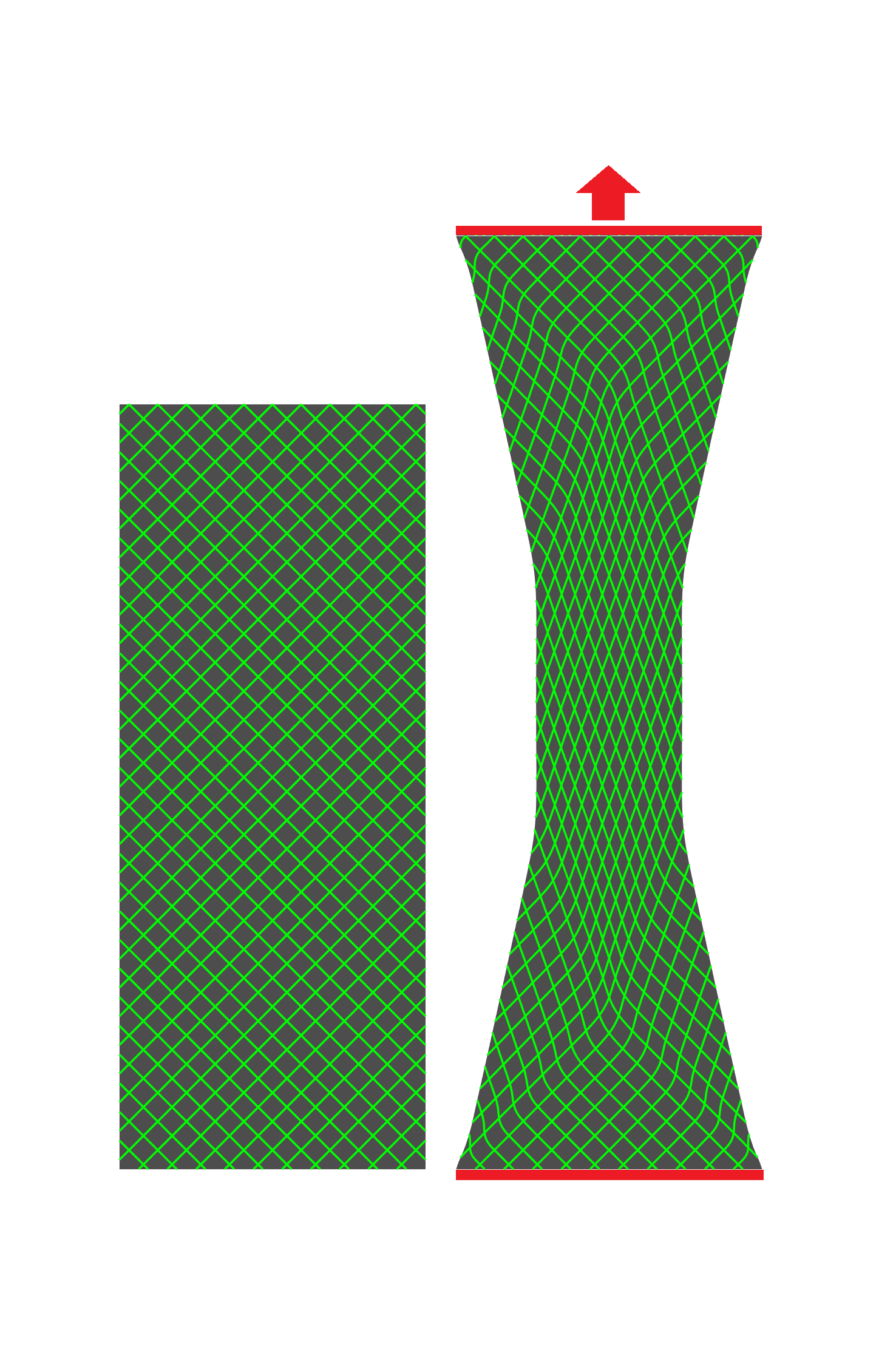}}
\put(2.6,-0.8){\includegraphics[height=0.63\textwidth]{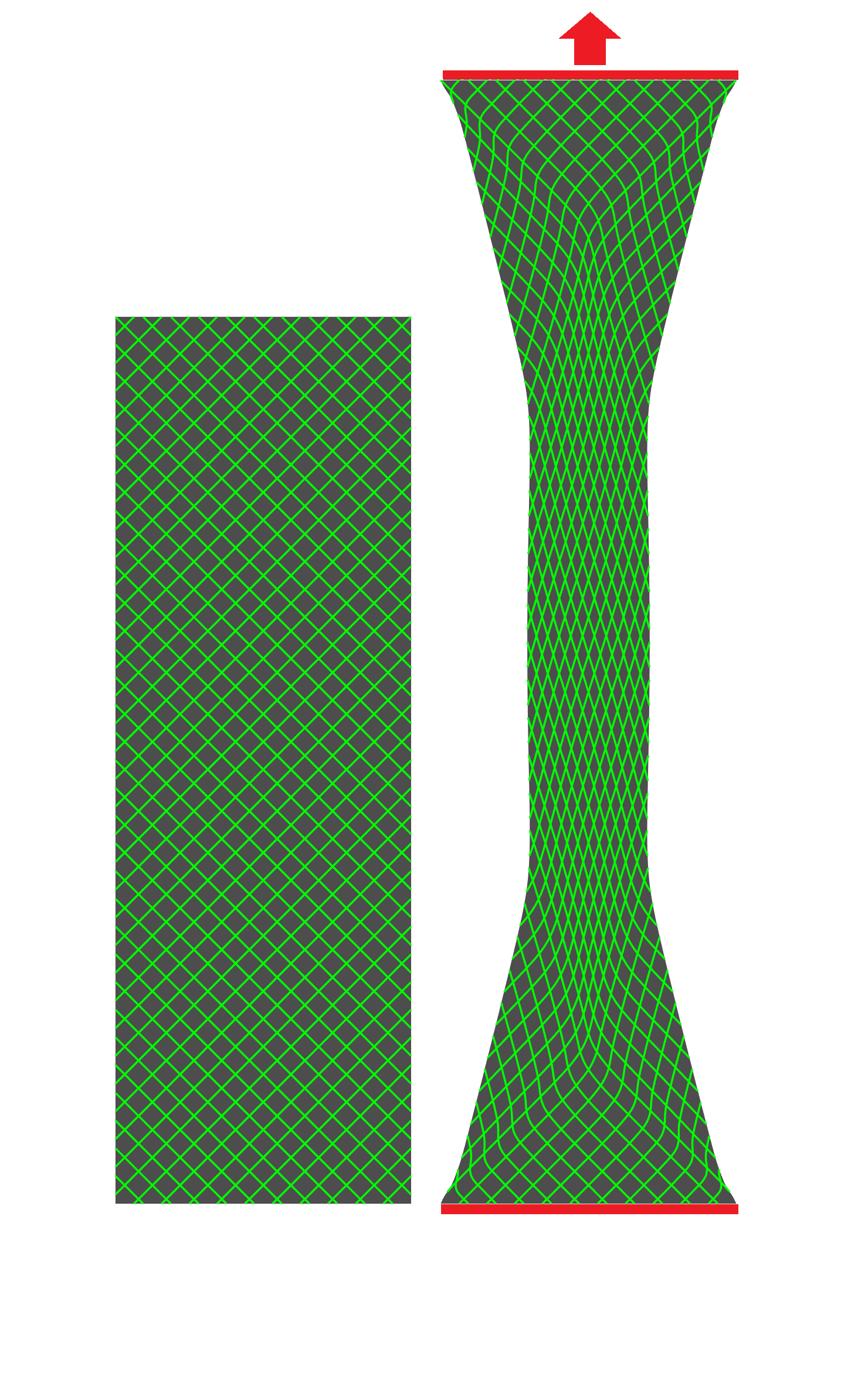}}
\put(-7.8,6.0){\tiny{Samples:}}
\put(-7.8,5.65){\tiny{\#1: $100\!\times\!200$mm$^2$}}
\put(-7.8,5.35){\tiny{\#2: $115\!\times\!230$mm$^2$}}
\put(-7.8,5.05){\tiny{\#3: $150\!\times\!300$mm$^2$}}

\put(-2.2,6.43){\tiny{Sample:}}
\put(-2.2,6.08){\tiny{\#4: $100\!\times\!250$mm$^2$}}

\put(3.4,7.8){\tiny{Samples:}}
\put(3.4,7.45){\tiny{\#5: $100\!\times\!300$mm$^2$}}
\put(3.4,7.15){\tiny{\#6: $150\!\times\!450$mm$^2$}}

\put(-7.4,0.1){{\small{(a)~Aspect ratio $1\!:\!2$}}}
\put(-1.8,0.1){{\small{(b)~Aspect ratio $1\!:\!2.5$}}}
\put(3.8,0.1){{\small{(c)~Aspect ratio $1\!:\!3$}}}
\end{picture}
\caption[caption]{Bias extension test for three \resc{sample} classes of plain weave fabrics: Initial and deformed configurations resulting from our computations. The deformed shapes with the same aspect ratio are almost identical regardless of size. The green lines illustrate the considered two fiber families, \res{while the red bars at top and bottom represent a clamped boundary with prescribed displacements.}}
\label{f:Bias_x0_xx}
\end{center}
\end{figure} 

Fig.~\ref{f:Bias_x0_xx} shows the initial and deformed configurations resulting from our simulations. Regardless of difference in size, almost identical shapes are expected for the same aspect ratio due to the  relatively small  bending stiffness used (
see Tab.~\ref{t:WFconstant}). This is confirmed in our computations.

\begin{figure}[H]
\begin{center} \unitlength1cm
\begin{picture}(0,10.1)
%
\put(-7.5,0){\includegraphics[height=0.55\textwidth]{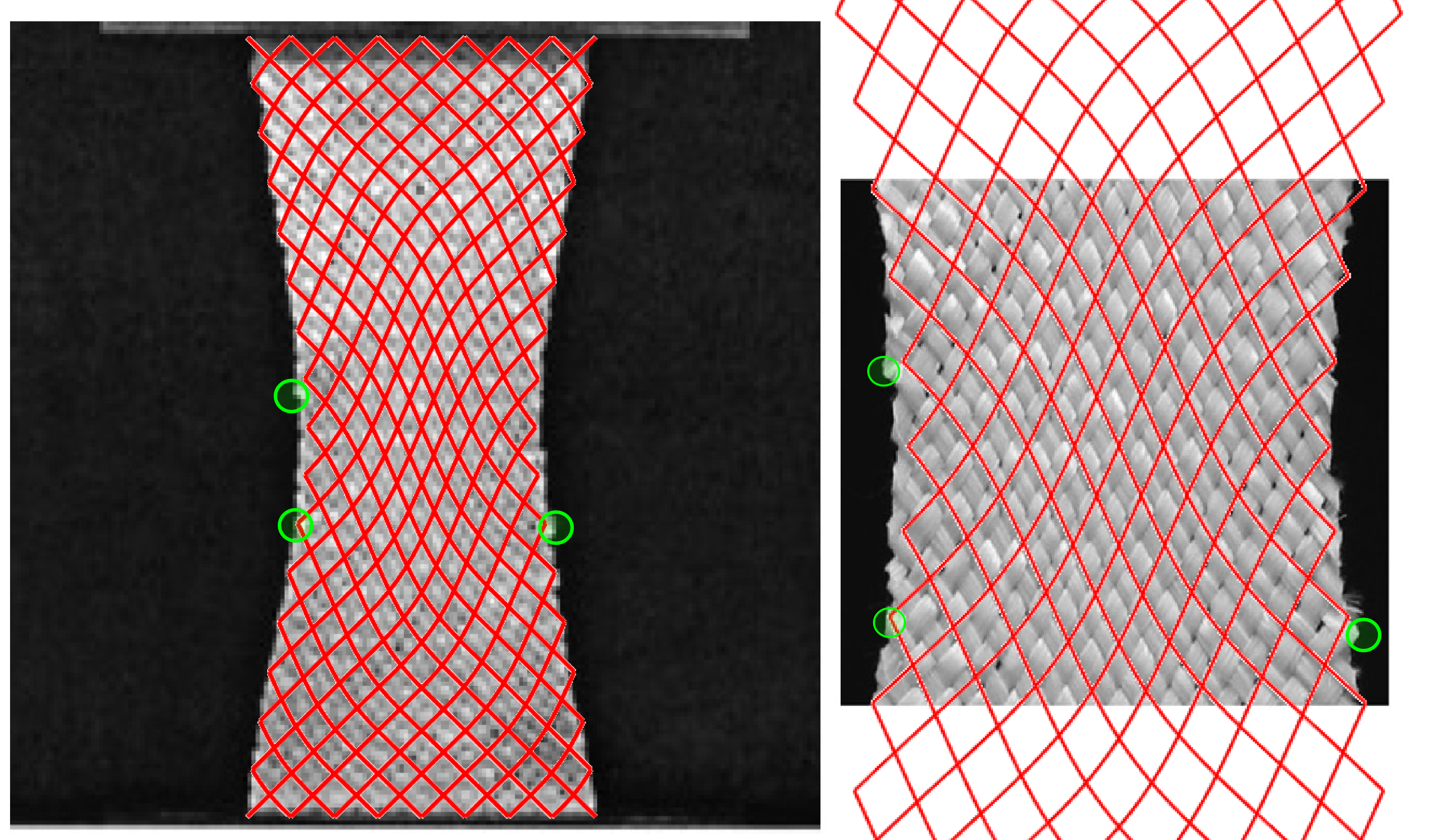}}

\end{picture}
\caption[caption]{\res{Bias extension test for sample \#2 at displacement $\bar{u}=30$mm: Overlaying our computational result (red \resc{curves}) on the experiment result for  the full view (left) and an enlarged view (right), where the green circles are corresponding points. The experimental results here are taken from  \cite{Zhu2007}, with permission from Elsevier.} }
\label{f:Bias_x0_xx_exp1}
\end{center}

\end{figure}

\res{Fig.~\ref{f:Bias_x0_xx_exp1} further compares the deformed shape of sample \#2 at displacement $30$mm (i.e.~before significant damage occurs) from our simulation to the experiment reported by \cite{Zhu2007}.  As the comparison shows, our  prediction is generally in  good agreement with the experimental results in terms of overall shape and fiber angles.}

\begin{figure}[H]
\begin{center} \unitlength1cm
\begin{picture}(0,22.0)
\put(-8,15.5){\includegraphics[height=0.45\textwidth]{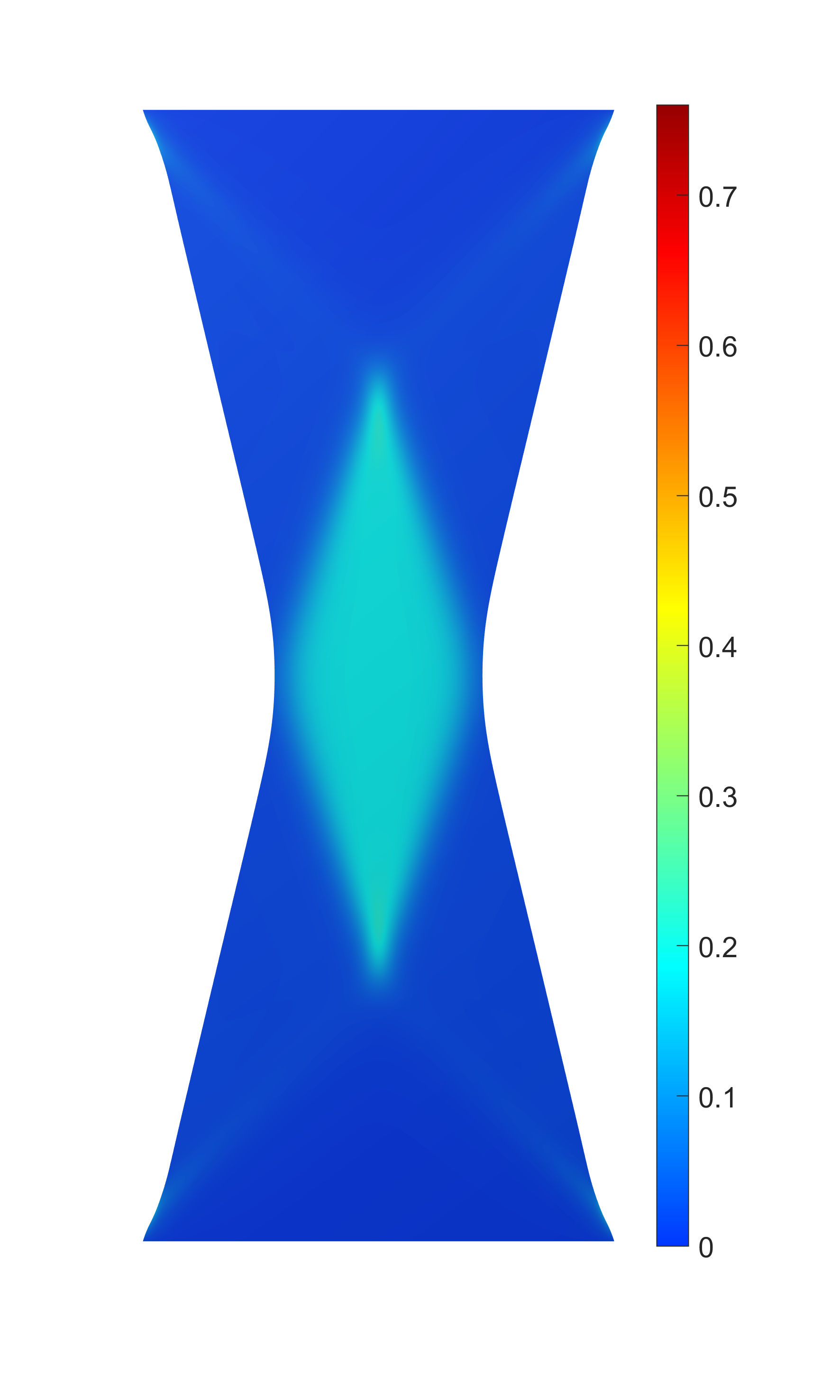}}
\put(-4.1,15.5){\includegraphics[height=0.45\textwidth]{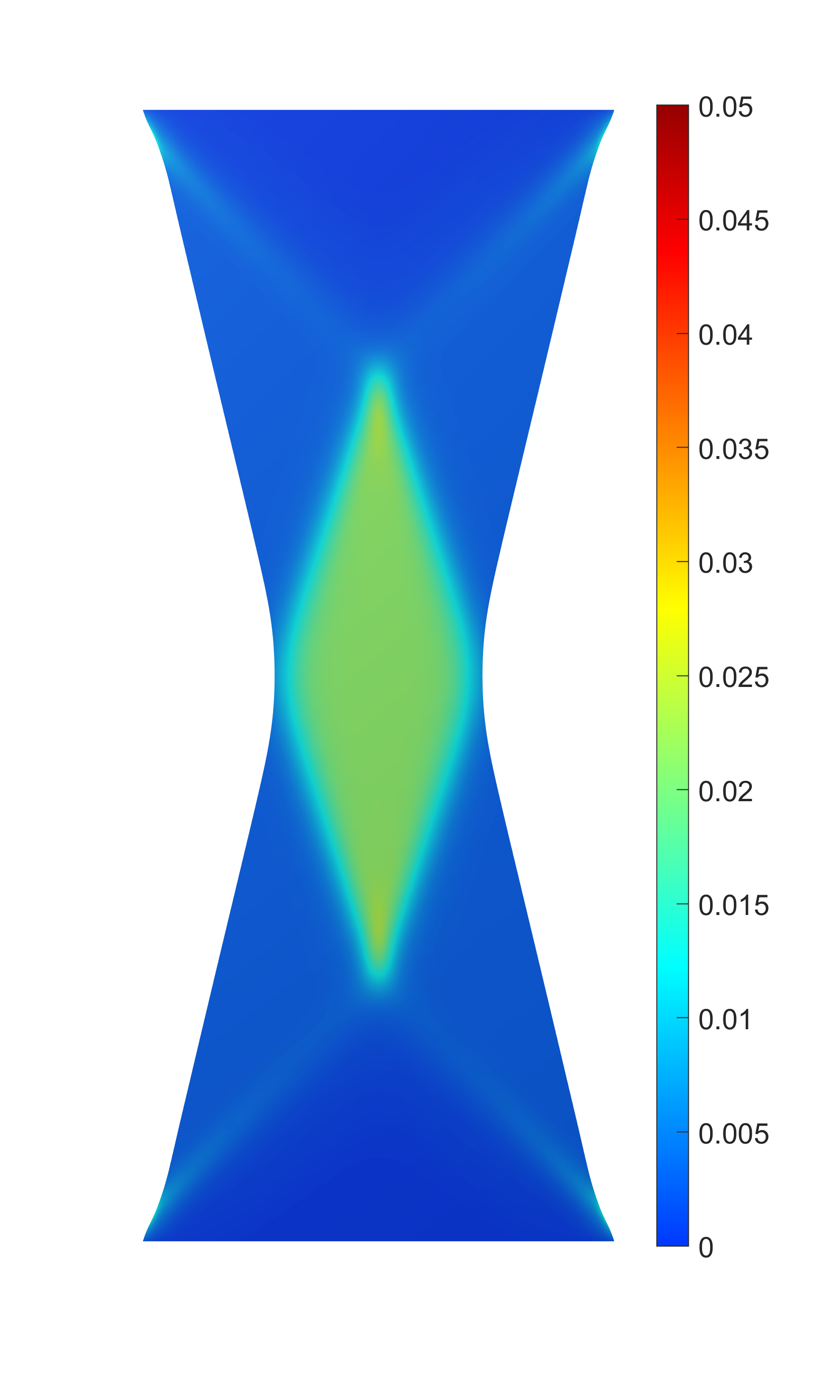}}
\put(-0.1,15.5){\includegraphics[height=0.45\textwidth]{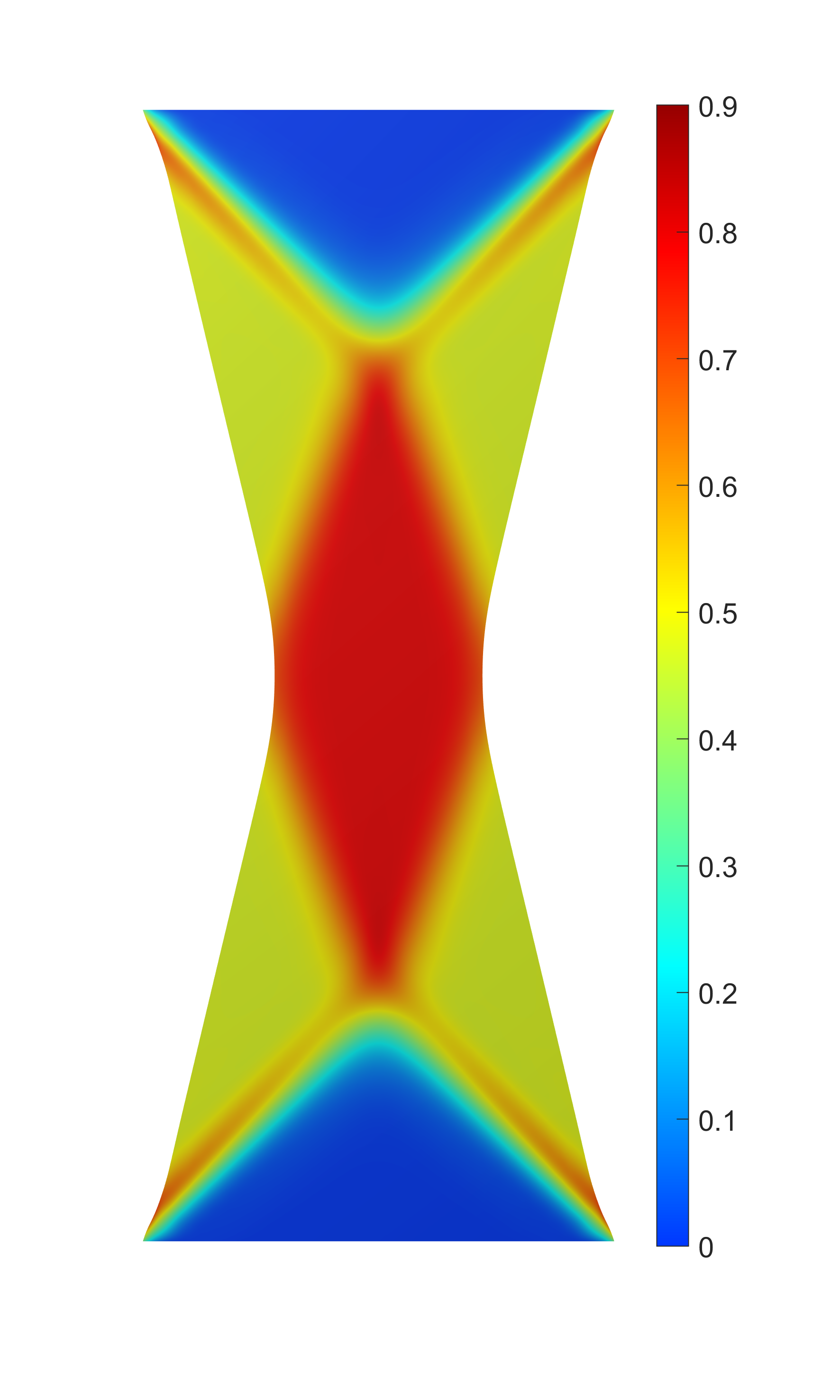}}
\put(3.8,15.5){\includegraphics[height=0.45\textwidth]{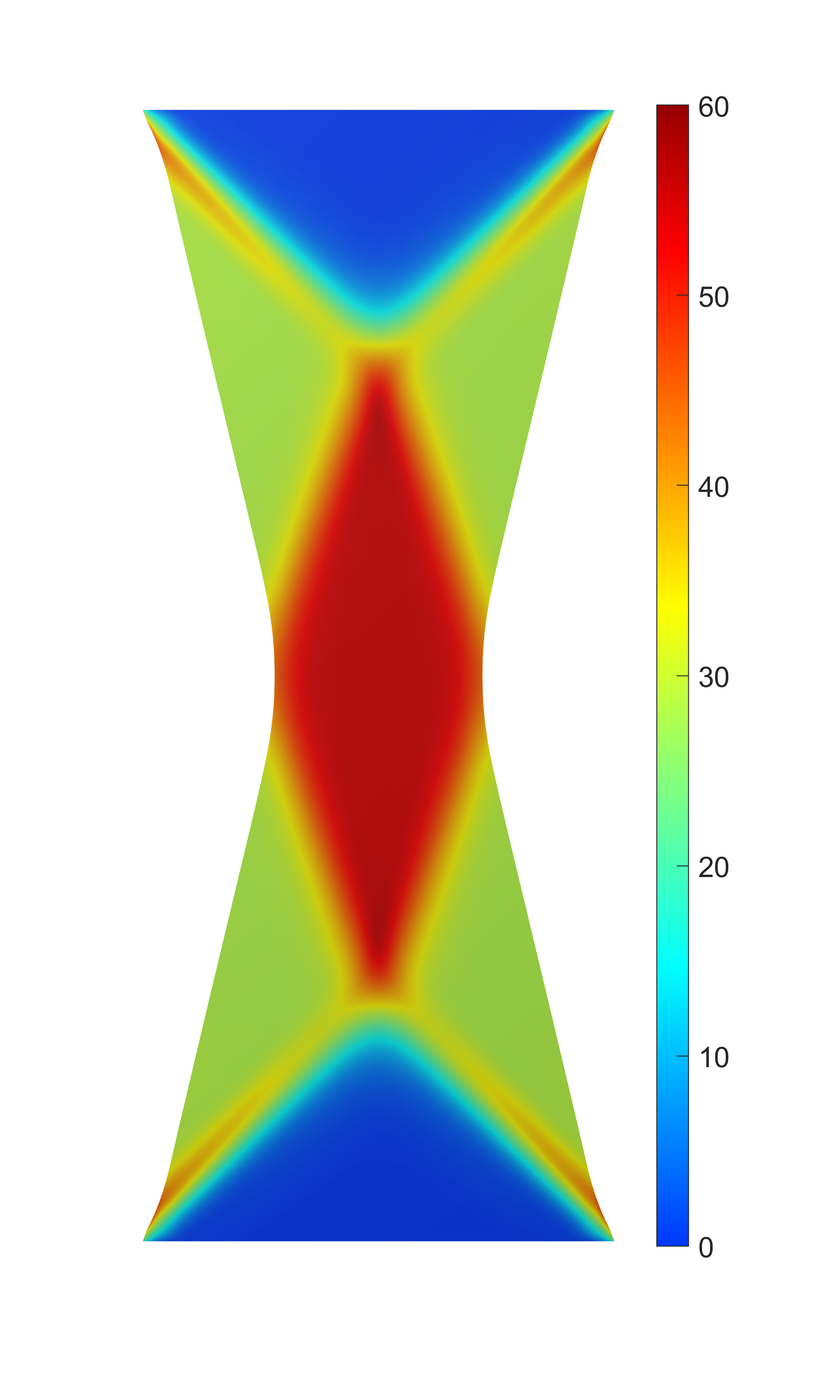}}

\put(-5.3, 22.4){ \quad \small{$\sigma_{12}$ }}

\put(-1.4, 22.4){~\quad \small{$\phi_{\mre}$ }}

\put(2.6, 22.4){~ \quad\small{ $\phi_{\mrp}$ }}

\put(6.5, 22.4){~\quad \small{$\gamma$}}

\put(-7.9, 15.7){\small{(a)~Aspect ratio $1\!\!:\!\!2$}}
\put(-8,7.8){\includegraphics[height=0.45\textwidth]{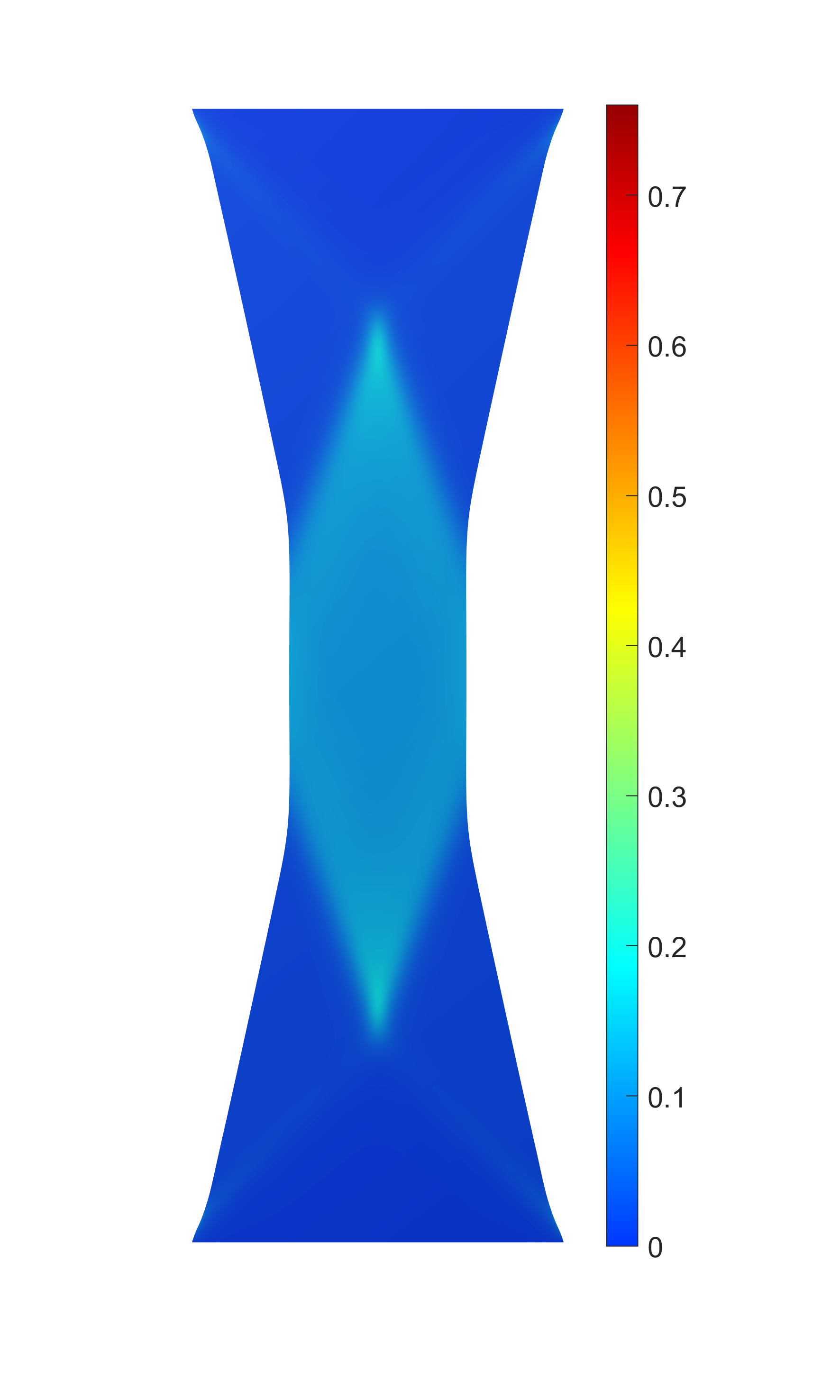}}
\put(-4.1,7.8){\includegraphics[height=0.45\textwidth]{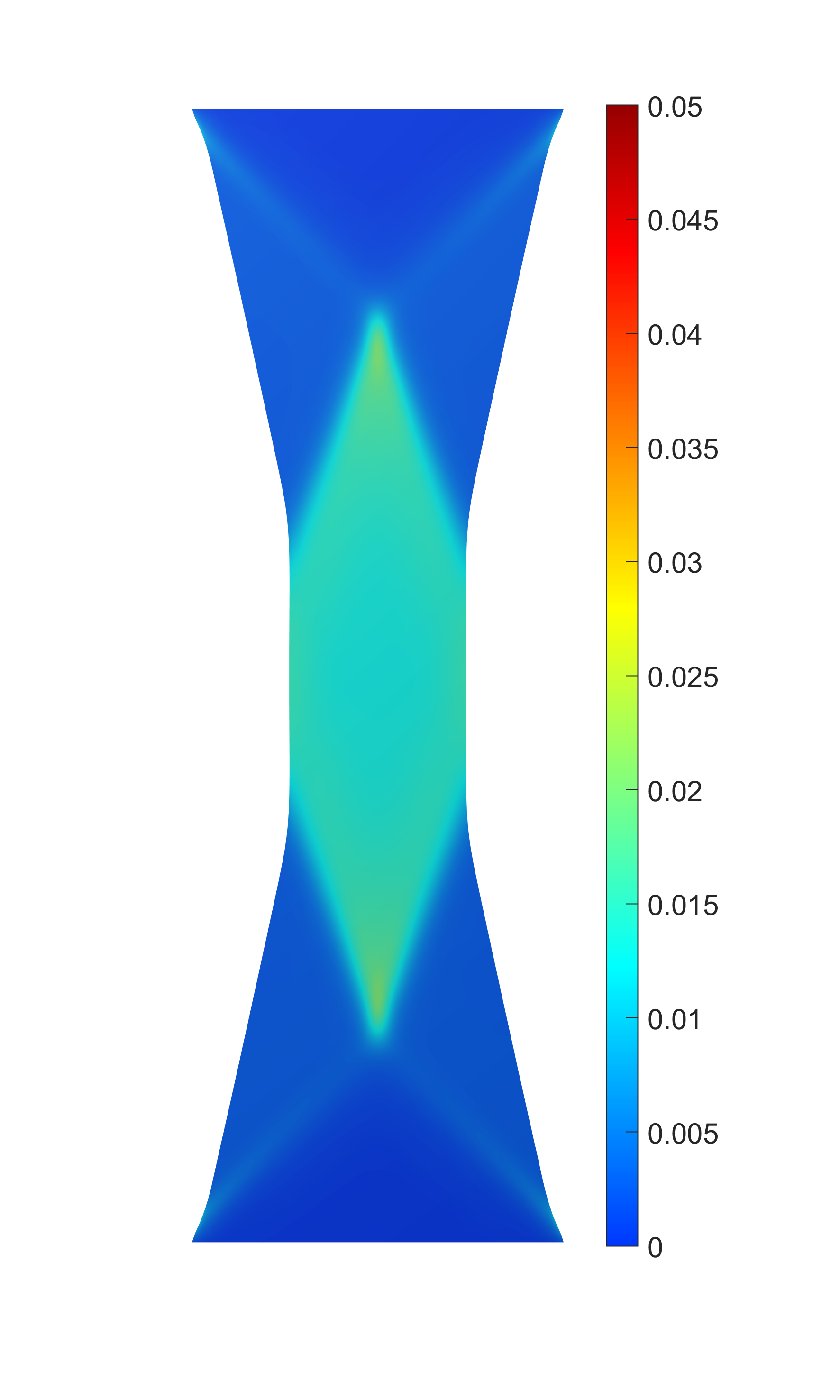}}
\put(-0.1,7.8){\includegraphics[height=0.45\textwidth]{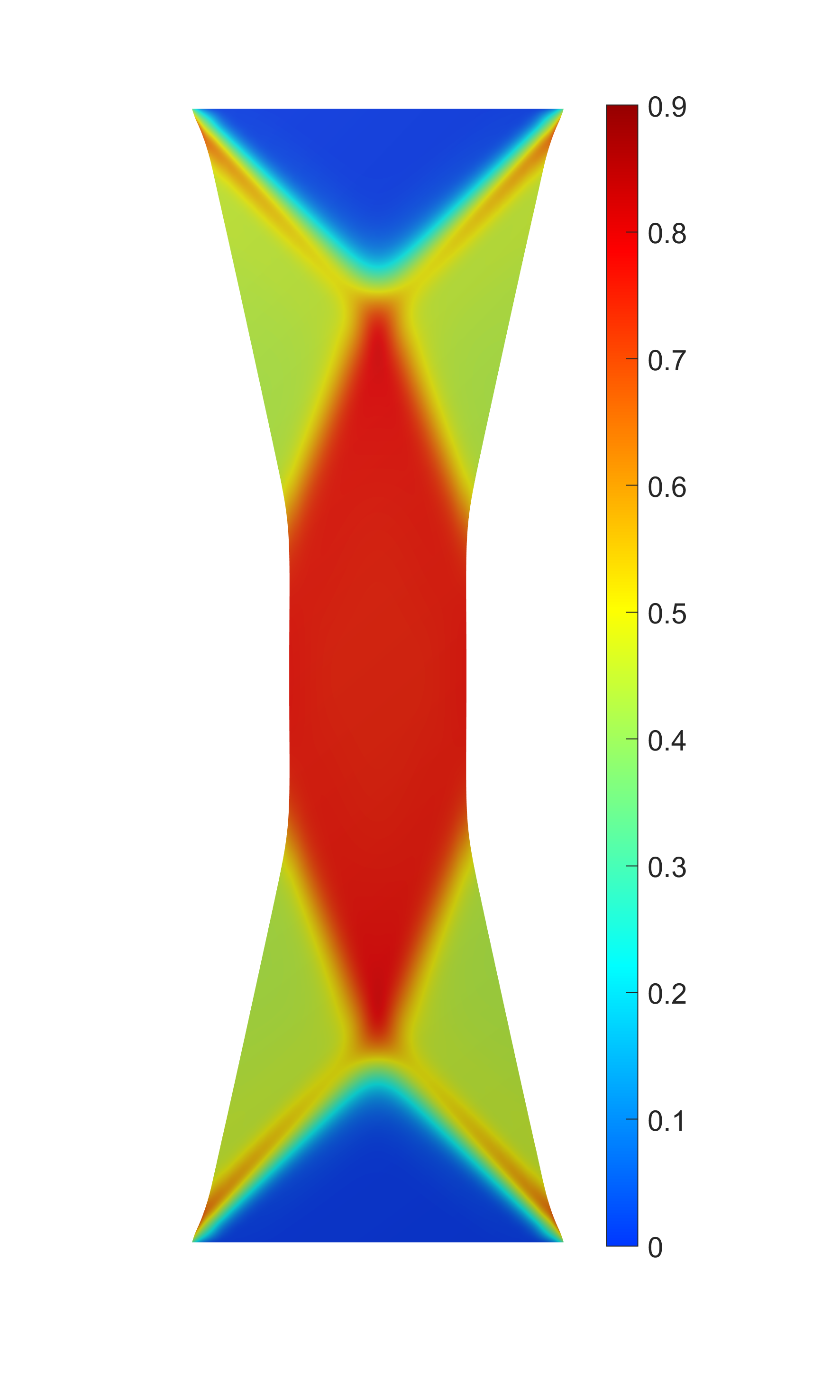}}
\put(3.8, 7.8){\includegraphics[height=0.45\textwidth]{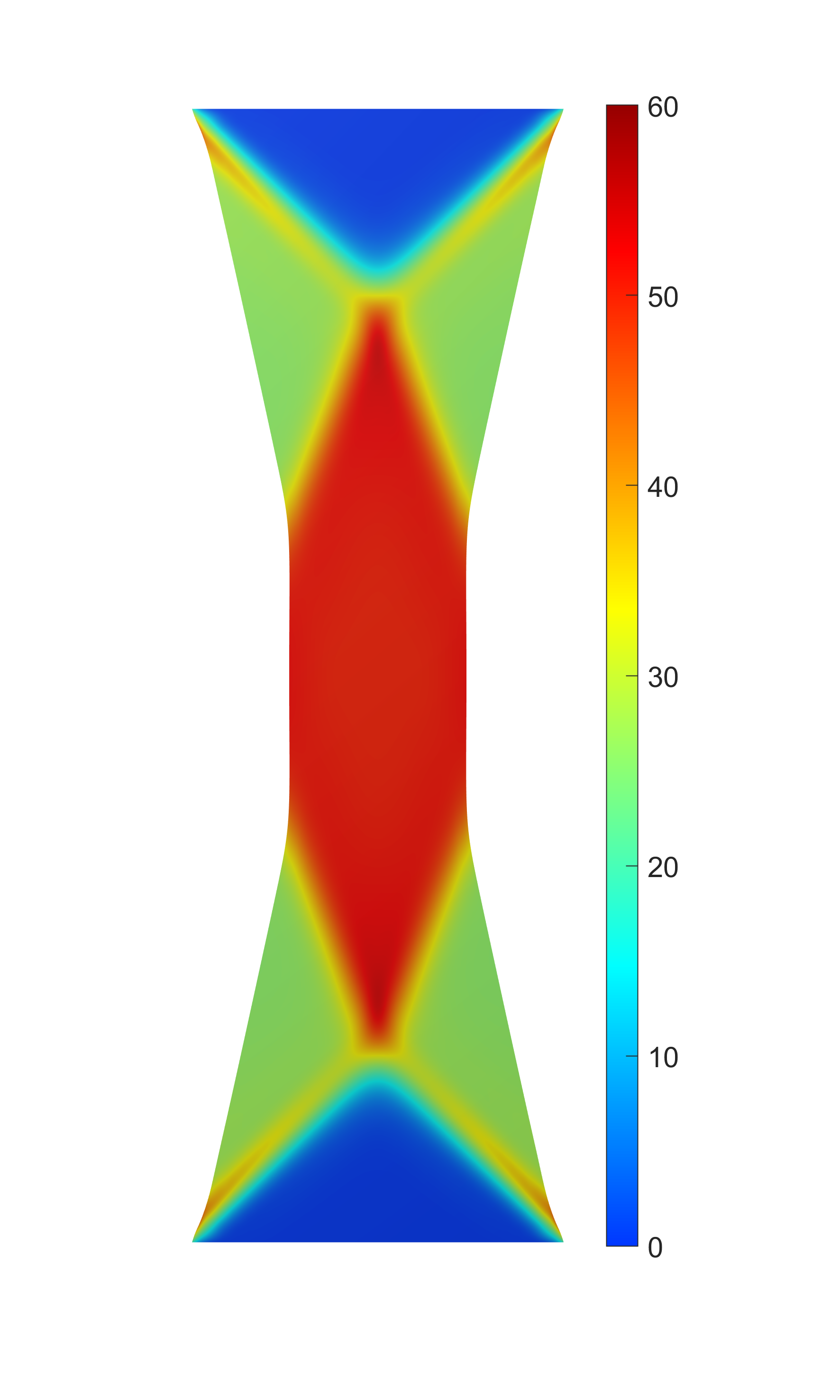}}

\put(-5.3, 14.7){~~\small{\!$\sigma_{12}$} }

\put(-1.4, 14.7){~~\small{\! $\phi_{\mre}$ }}

\put(2.6,14.7){~~\small{\! $\phi_{\mrp}$ }}

\put(6.5, 14.7){~~~\small{$\gamma$}}

\put(-7.9, 7.9){\small{(b)~Aspect ratio $1\!:\!2.5$}}
\put(-8, 0.0){\includegraphics[height=0.45\textwidth]{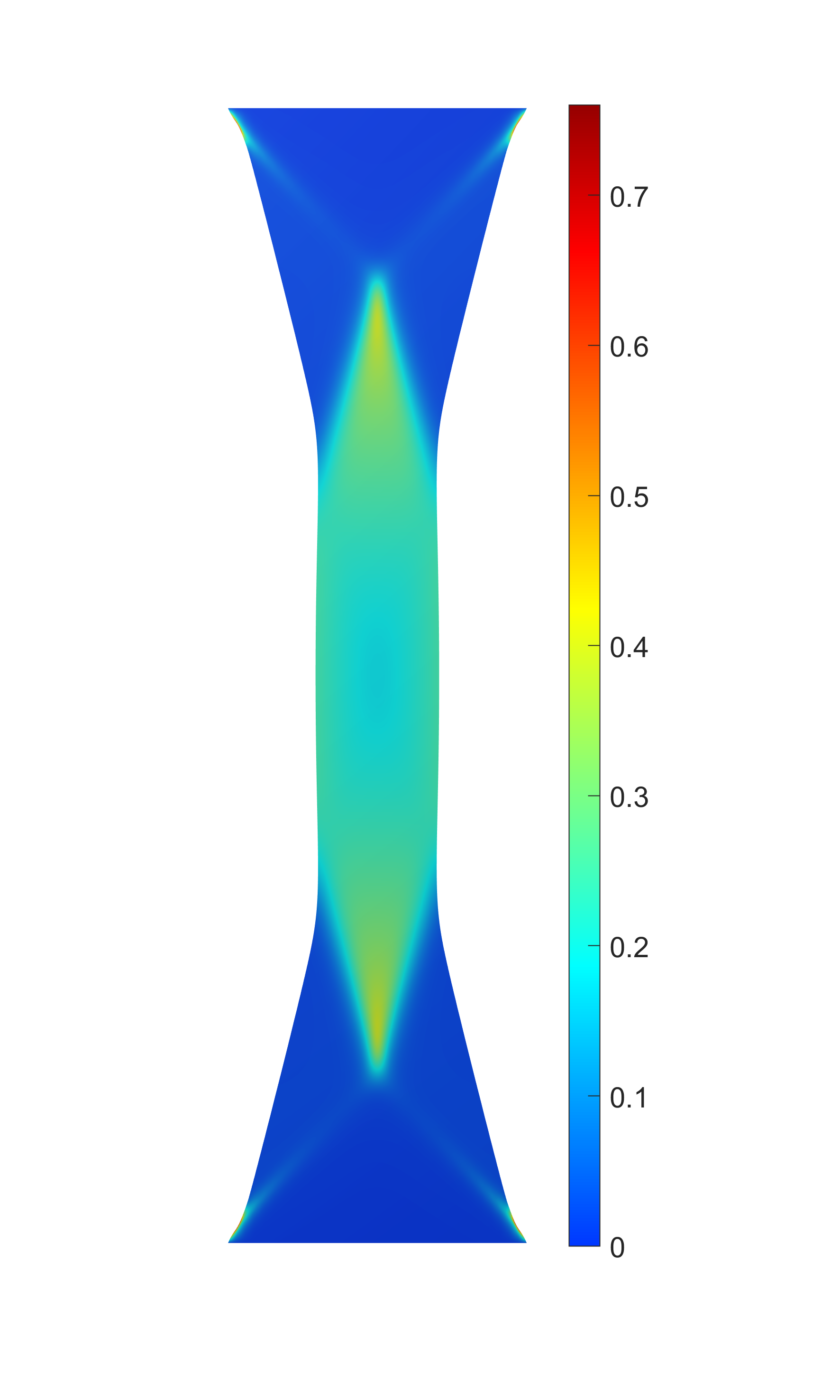}}
\put(-4.1,0.0){\includegraphics[height=0.45\textwidth]{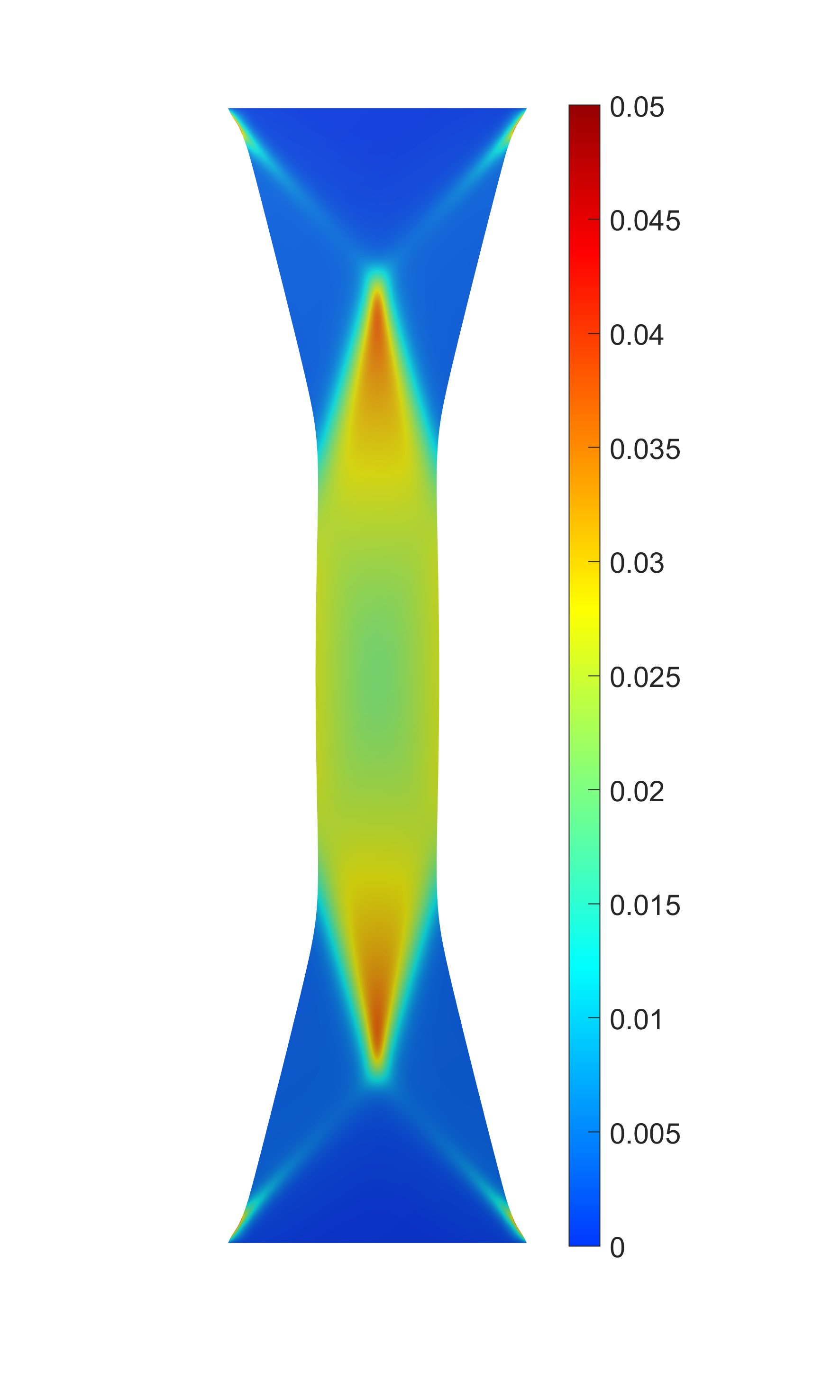}}
\put(-0.1,0.0){\includegraphics[height=0.45\textwidth]{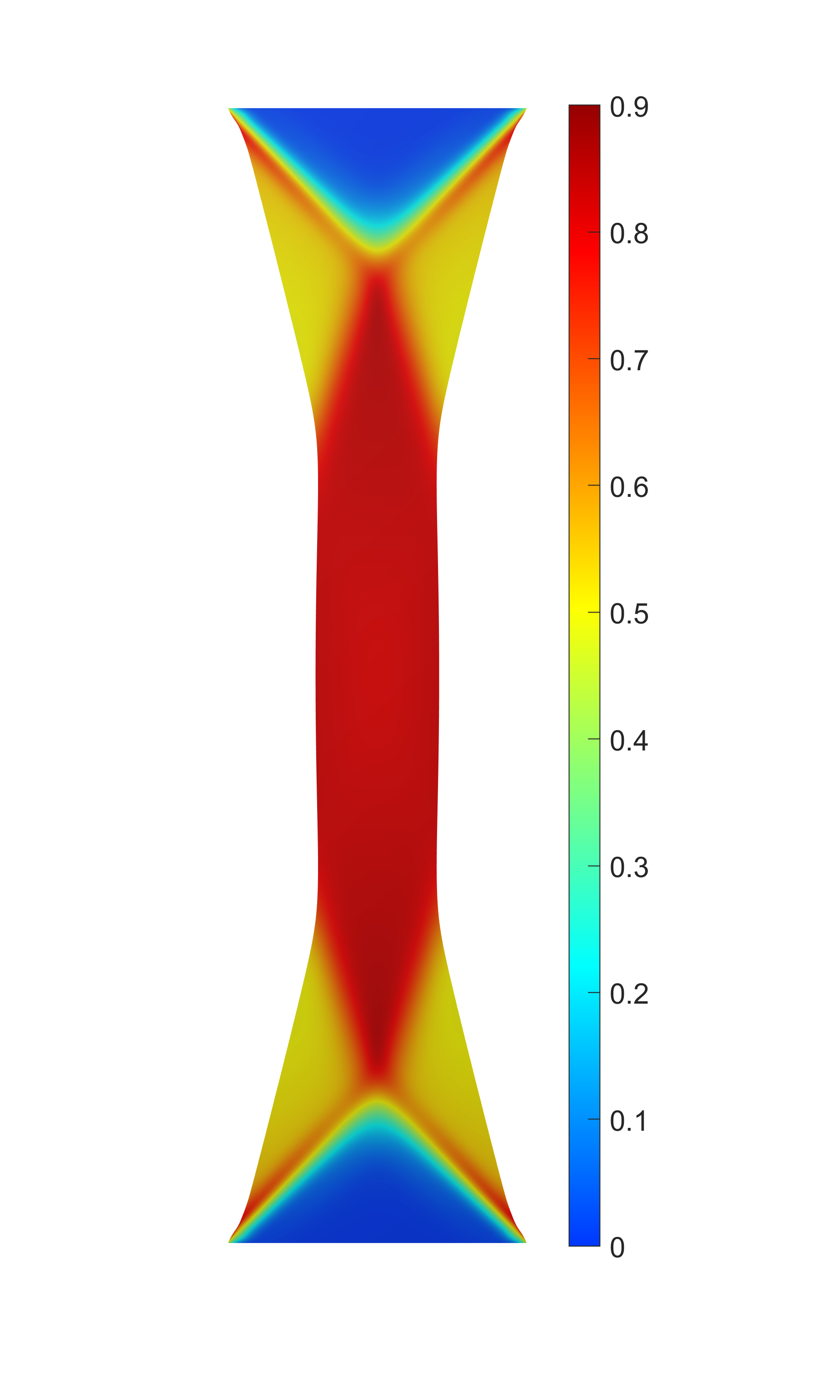}}
\put(3.8,0.0){\includegraphics[height=0.45\textwidth]{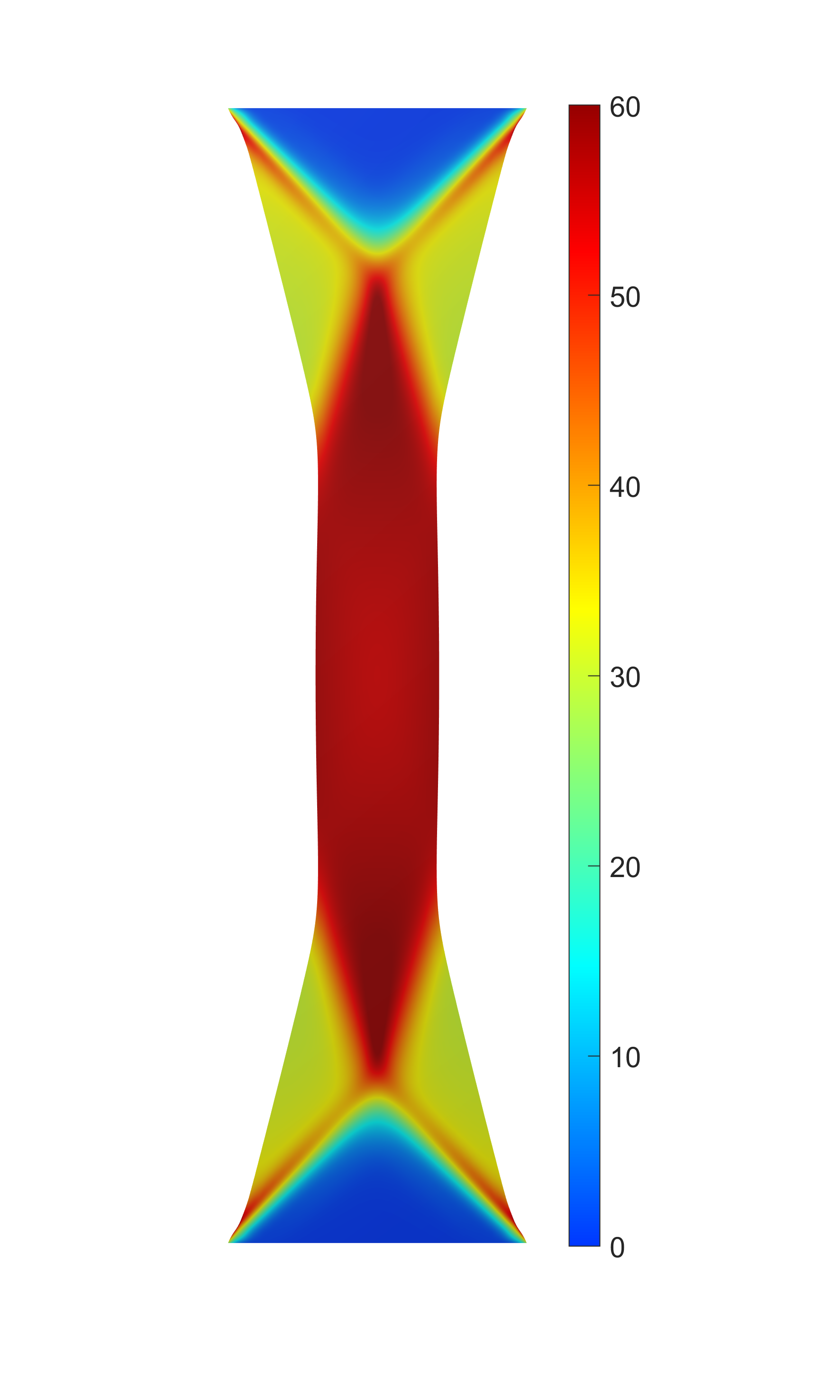}}

\put(-5.3, 6.9){\small{ $\sigma_{12}$ }}

\put(-1.4, 6.9){\small{ $\phi_{\mre}$ }}

\put(2.6,6.9){\small{ $\phi_{\mrp}$ }}

\put(6.5, 6.9){~~\small{$\gamma$}}

\put(-7.9, 0){\small{(c)~Aspect ratio $1\!:\!3$}}

\end{picture}
\caption[caption]{Bias extension test of plain weave fabrics:  From left to right: True shear stress $\sigma_{12}:=\tau/J$ [N/mm] according to Eq.~\eqref{e:tau_def}, elastic angle strain $\phi_{\mre}$, plastic angle strain $\phi_{\mrp}$,  and total shear angle defined by $\gamma:=90^\circ-\mathrm{arccos}(\theta)$ [deg] for the three sample classes with initial aspect ratio (a)~$1\!:\!2$, (b)~$1\!:\!2.5$, and (c)~$1\!:\!3$. }
\label{f:Bias_color_plots}
\end{center}
\end{figure}

Next, Fig.~\ref{f:Bias_color_plots} shows the distributions of shear stress $\sigma_{12}$, elastic angle strain $\phi_\mre$, plastic angle strain $\phi_\mrp$ and total shear angle $\gamma:=90^\circ-\mathrm{arccos}(\theta_{12} )$ resulting from our simulations for all samples. As seen from Fig.~\ref{f:Bias_color_plots}\,(2.\,\&\,3.\,column),  plastic angle $\phi_\mrp$ is much larger than  elastic angle $\phi_\mre$. This reflects the small yield stress $\tau_\mry$ and consequently leads to small $\phi_\mre$ and small stress $\sigma_{12}$. As is also seen from the plots of shear angle $\gamma$ in Fig.~\ref{f:Bias_color_plots}\,(4.\,column),  $\gamma$ is not only non-zero at the center of the samples, but also at the off-center. Note that $\gamma$ is observable from experiments and our computations are consistent with the optical measurement presented in \cite{Cao2008a} (cf.~Fig.~37 therein -- note that the shear angle  was defined there by $\mathrm{arccos}(\theta_{12} )$ in degrees). Our simulation results in Fig.~\ref{f:Bias_color_plots} (1\,\&\,2.\,column) further show that nonzero $\phi_\mre$ (and hence \resc{nonzero}  $\sigma_{12}$) mostly concentrate at the center and the shear bands of the samples. This is reasonable since the total shear angle $\gamma$ is also large there --  see Fig.~\ref{f:Bias_color_plots}\,(4.\,column) -- which corresponds to the increased hardening phase (phase III) (which already starts at around $\gamma= 35$-$45^\circ$, as seen from  Fig.~\ref{f:fitting_plastic_shear}a.). Meanwhile $\gamma$ is relatively small ($\gamma = 25^\circ$)  away from the center, which still allows fibers to freely rotate against each other there.

\begin{figure}[H]
\begin{center} \unitlength1cm
\begin{picture}(0,21.5)
\put(-8.1,14.6){\includegraphics[width=0.50\textwidth]{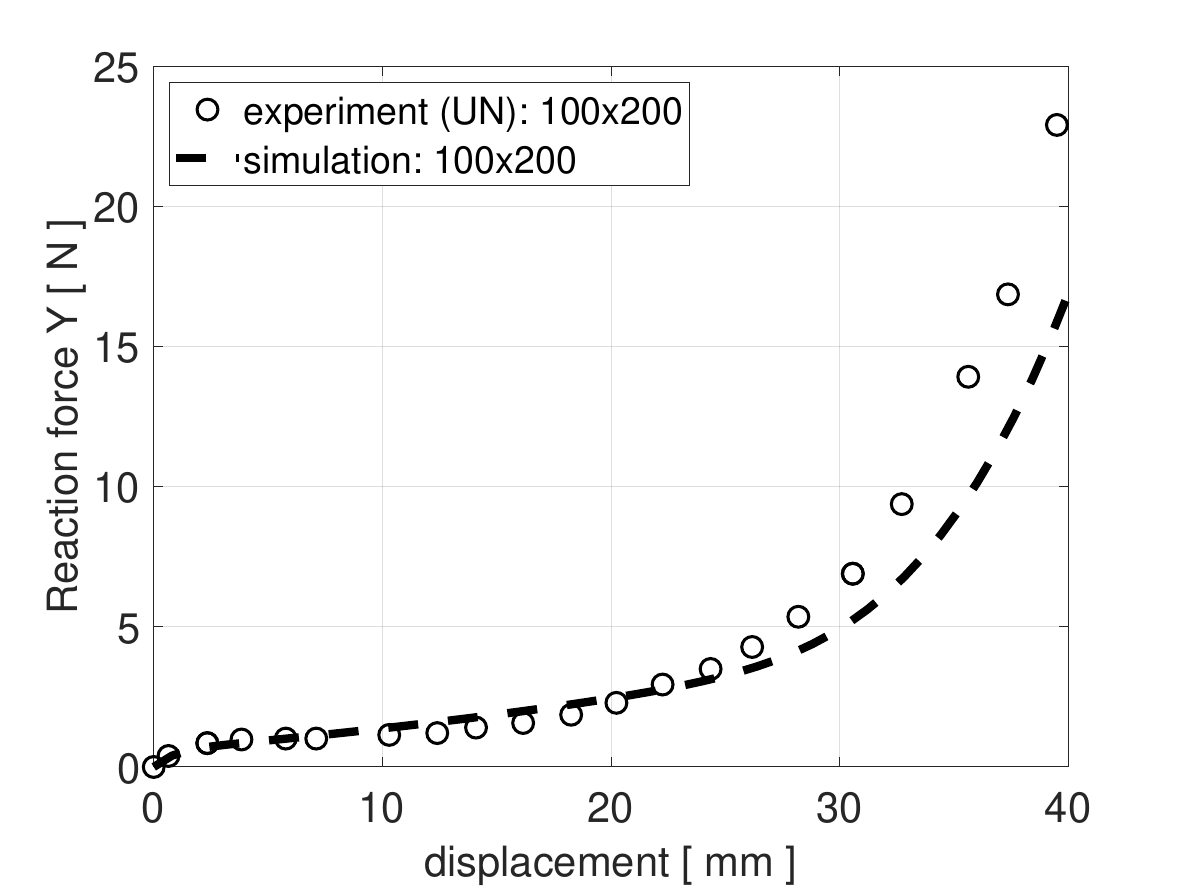}}
\put(0.5,14.6){\includegraphics[width=0.50\textwidth]{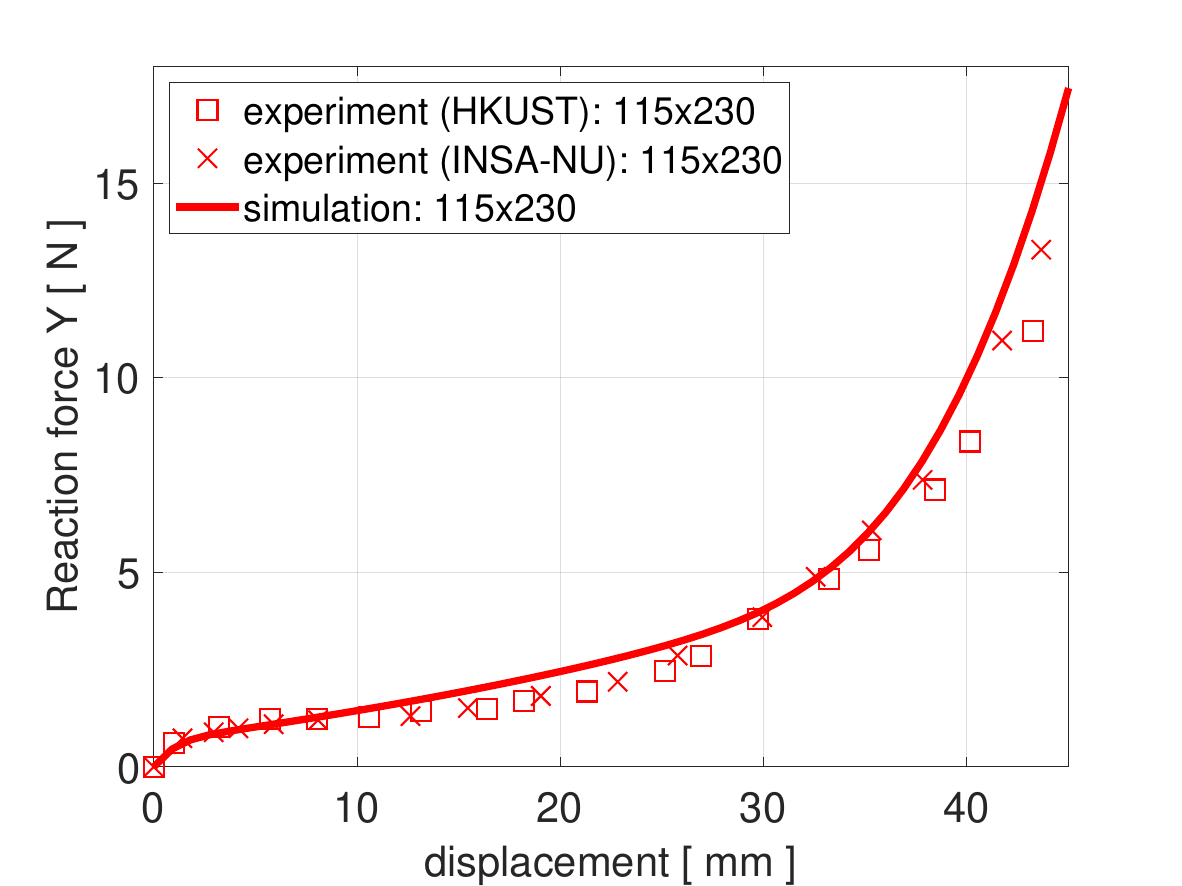}}

\put(-8.1,7.6){\includegraphics[width=0.50\textwidth]{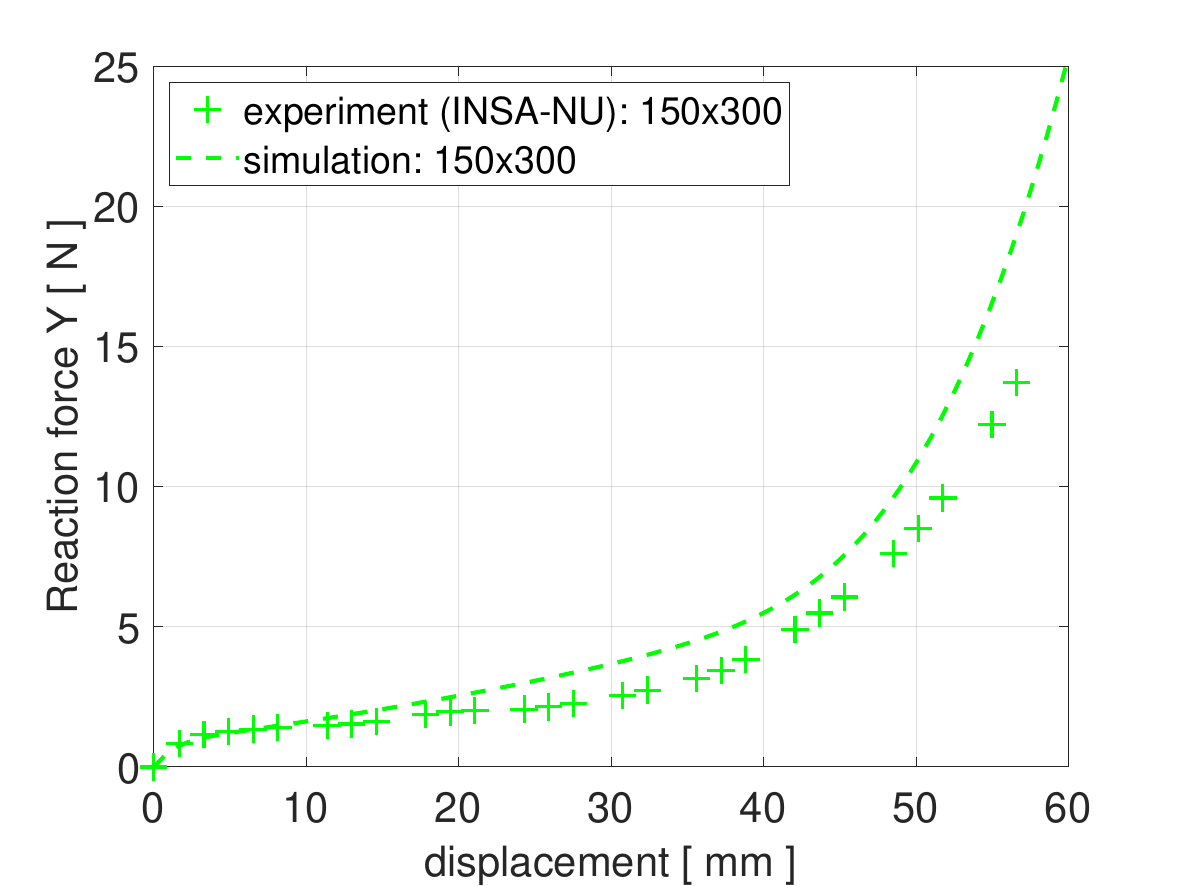}}
\put(0.5,7.6){\includegraphics[width=0.50\textwidth]{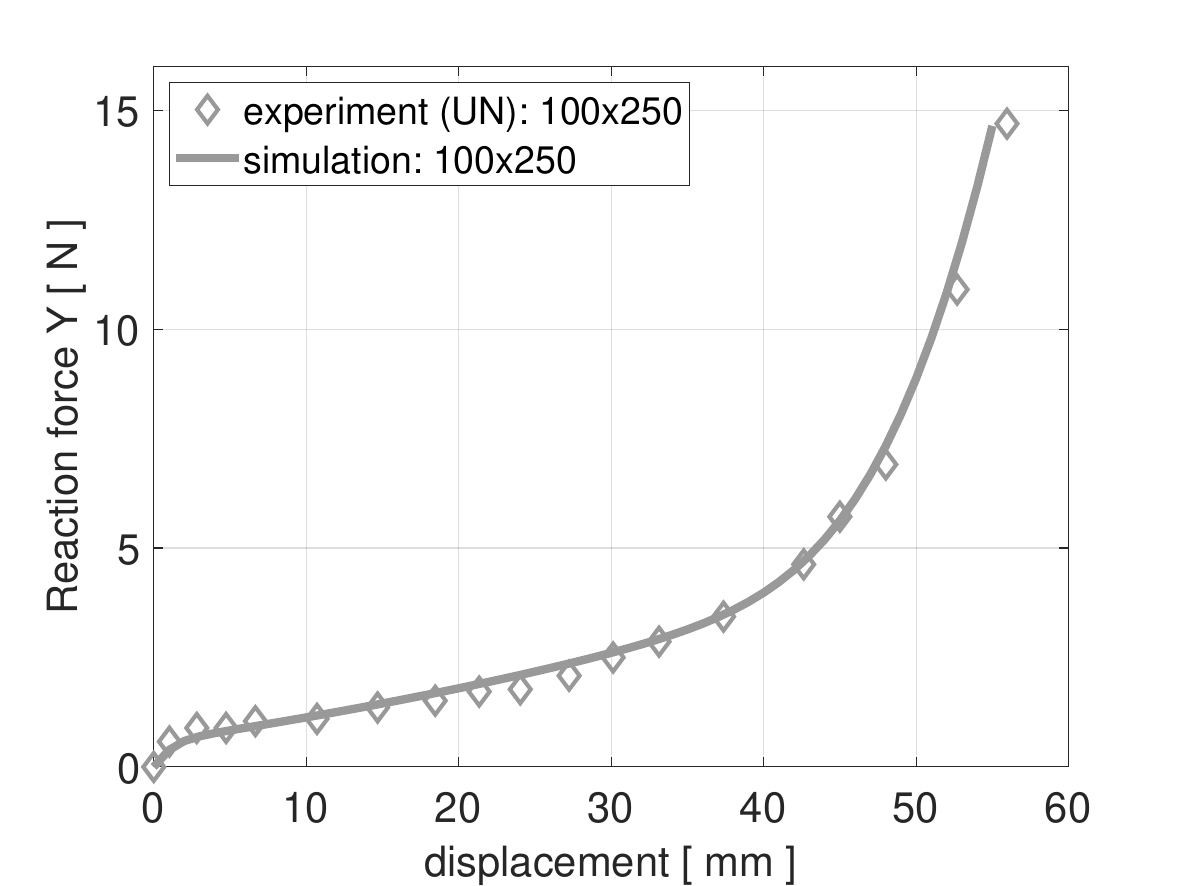}}

\put(-8.1,0.6){\includegraphics[width=0.50\textwidth]{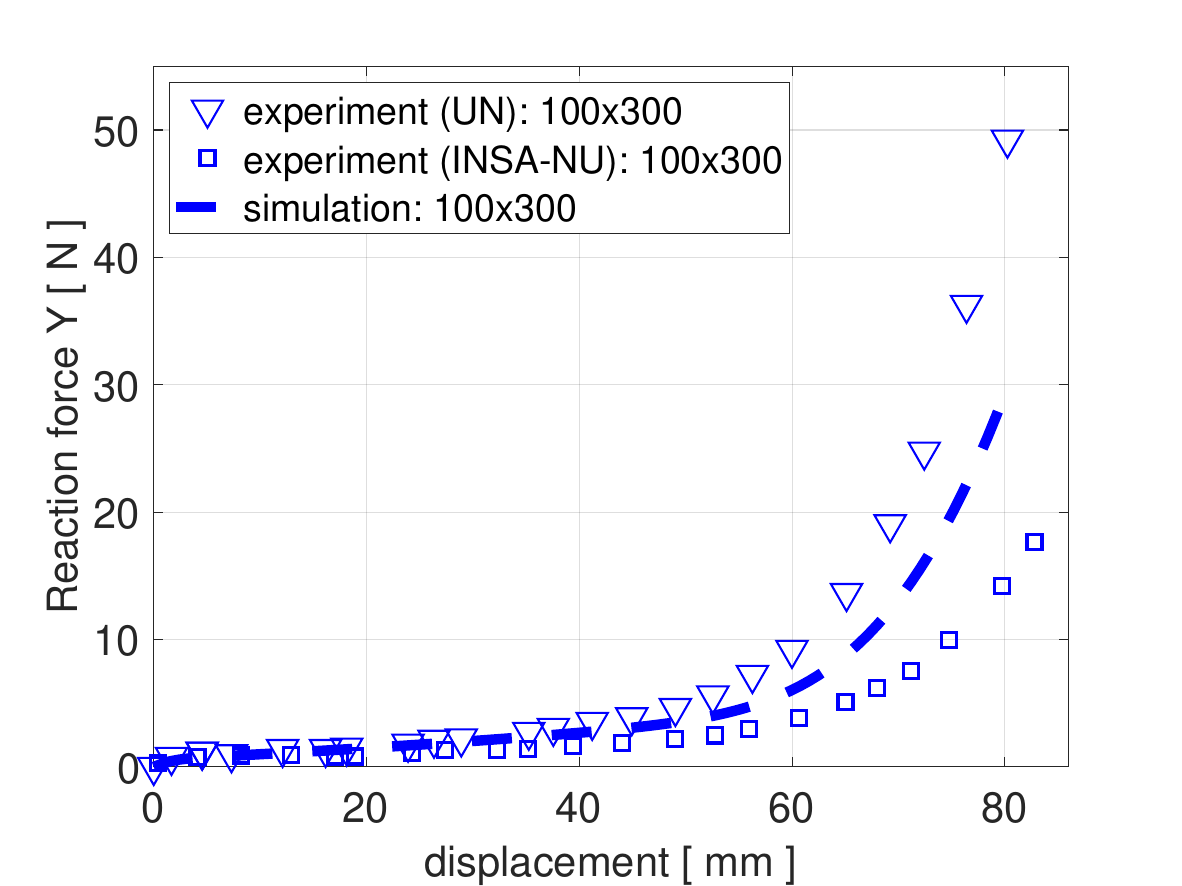}}
\put(0.5,0.6){\includegraphics[width=0.50\textwidth]{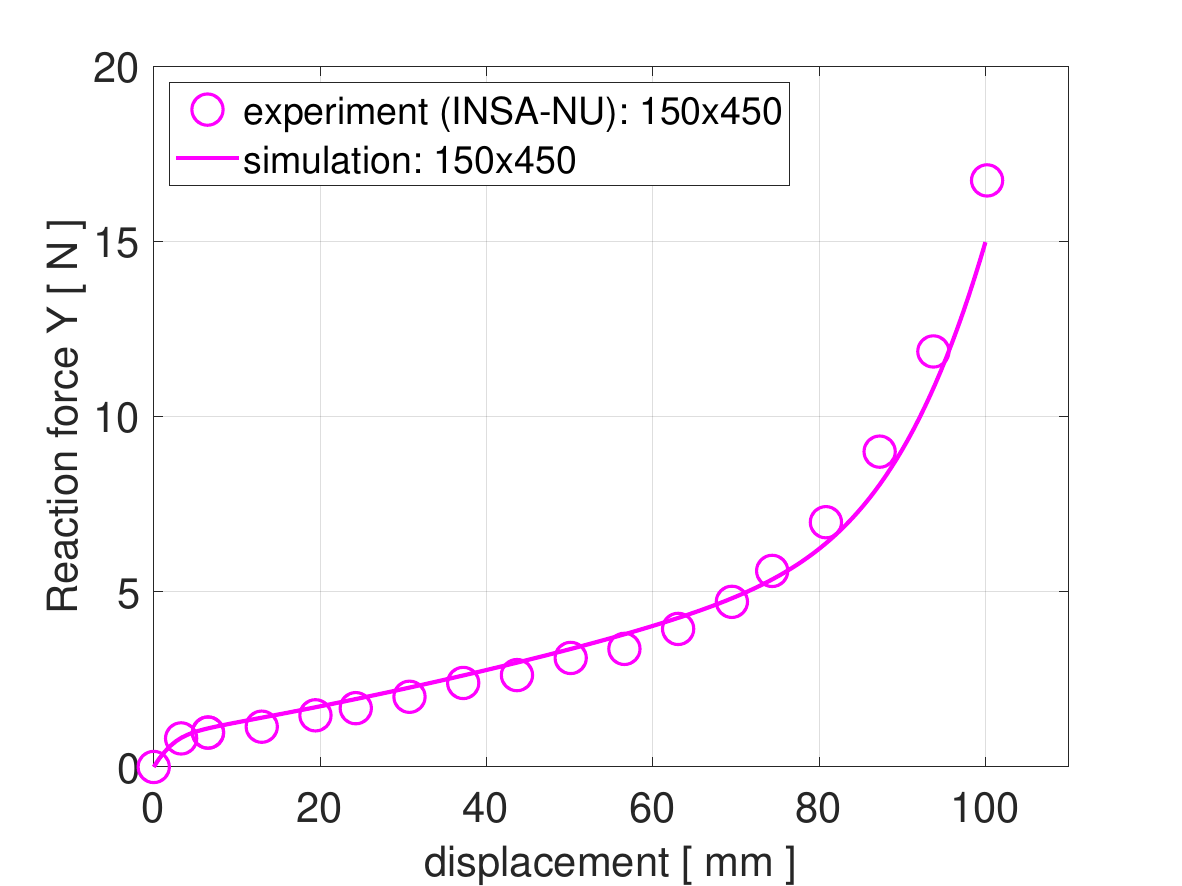}}

\put(-8,14.2){{\small{(a)~sample \#1}}}
\put(0.6, 14.2){{\small{(b)~sample  \#2}}}

\put(-8,7.2){{\small{(c)~ sample  \#3}}}
\put(0.6, 7.2){{\small{(d)~sample  \#4}}}

\put(-8,0.0){{\small{(e)~sample  \#5}}}
\put(0.6, 0.0){{\small{(f)~sample  \#6}}}

\end{picture}
\caption[caption]{Bias extension test of plain weave fabrics: Comparison of results from computation and experiment for reaction force versus displacement for all the six samples. Since these experimental results have not been used for calibration they validate our proposed model.
}
\label{f:Bias_Fy_compare_allsamples}
\end{center}
\end{figure}

Further, Fig.~\ref{f:Bias_Fy_compare_allsamples} compares our simulation results to the experimental results for the vertical reaction force  (pulling force) versus the vertical displacement. As seen, our model generally agrees well with the experiments, especially for samples \#2, \#4 and \#6. Since these experimental results have not been used for calibration they validate our proposed model. The results from sample \#5 show that the experimental results can vary strongly for different data sources. Our results lie in-between this variation.

\begin{figure}[H]
\begin{center} \unitlength1cm
\begin{picture}(0,7.5)

\put(-8.0,0.6){\includegraphics[width=0.50\textwidth]{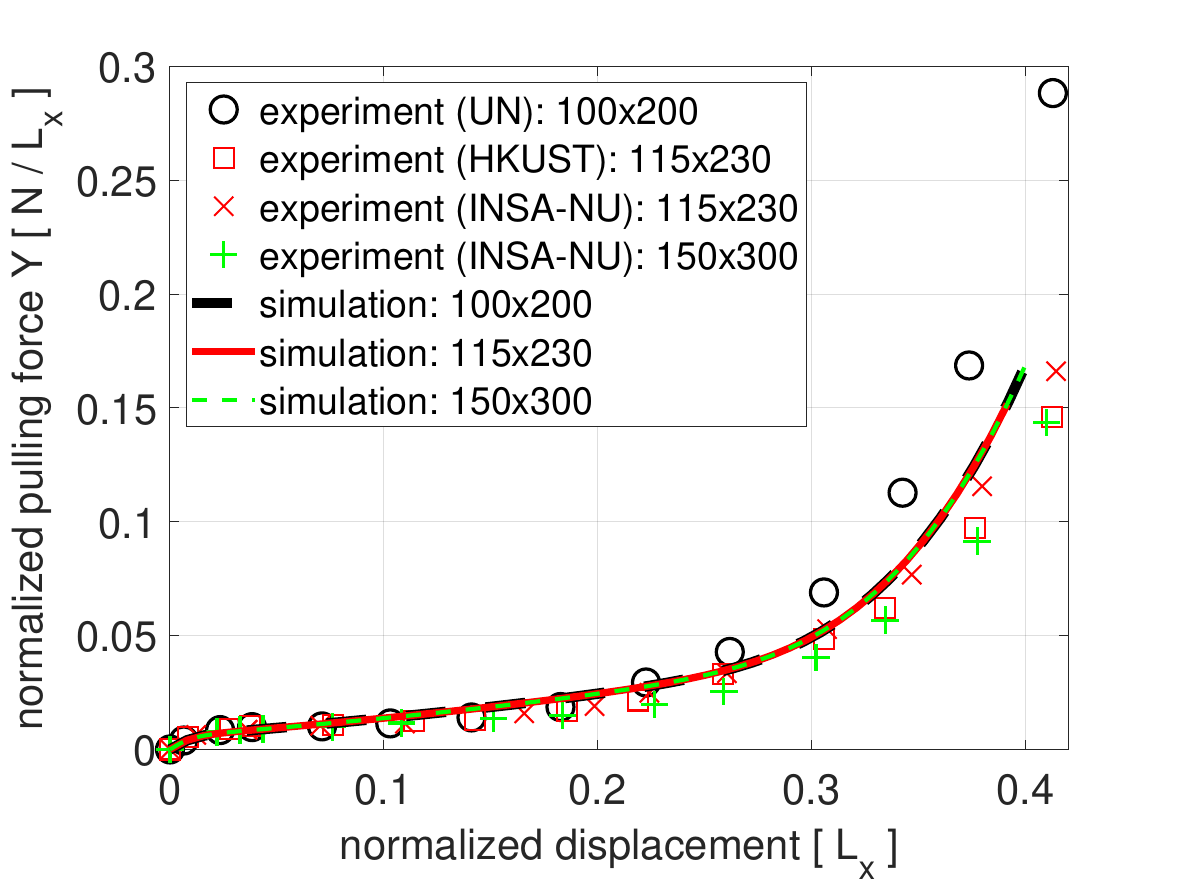}}
\put(0.5,0.6){\includegraphics[width=0.50\textwidth]{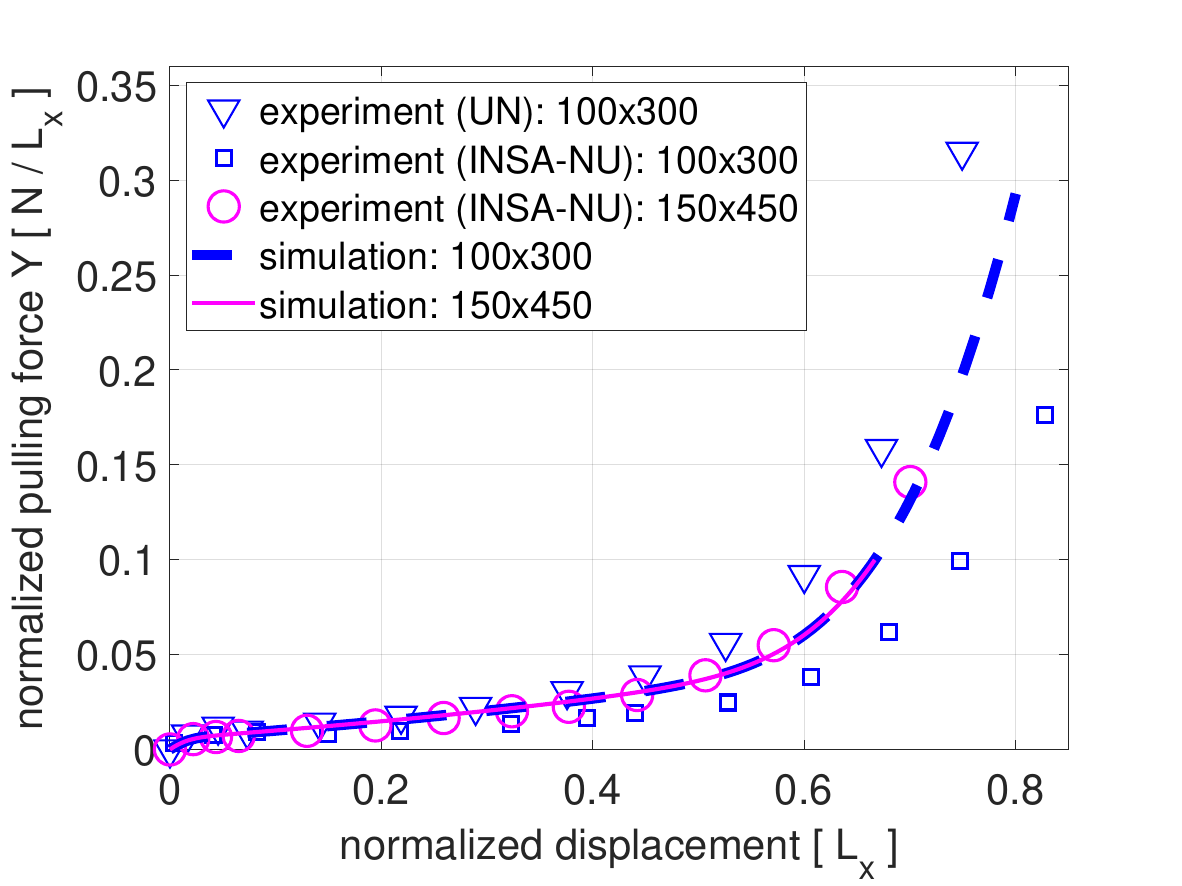}}

\put(-8.0,-0.0){{\small{(a)~Aspect ratio $1\!:\!2$}}}
\put(0.2,-0.0){{\small{(b)~Aspect ratio $1\!:\!3$}}}

\end{picture}
\caption[caption]{Bias extension test of plain weave fabrics: Comparison of results from computation and experiment for normalized reaction force versus displacement for the samples with equal aspect ratios. The  reaction force and the displacement are normalized by the corresponding sample widths. 
}
\label{f:Bias_Fy_compare_asratio}
\end{center}
\end{figure}

Finally, we check if the experimental data from different sources is consistent for different sample sizes. Therefore we plot the normalized reaction forces 
 in  Fig.~\ref{f:Bias_Fy_compare_asratio}. As seen, our simulation curves are almost identical for the same aspect ratio. The experimental data curves are generally consistent at low displacements, but scattered at large displacements. Our prediction lie approximately at the mean values of the experimental curves as Fig.~\ref{f:Bias_Fy_compare_asratio} show.

\section{Three-dimensional example} {\label{s:three_dim_eg}}

\begin{figure}[H]
\begin{center} \unitlength1cm
\begin{picture}(0,5.0)
\put(-8.0,-0.3){\includegraphics[height=0.36\textwidth]{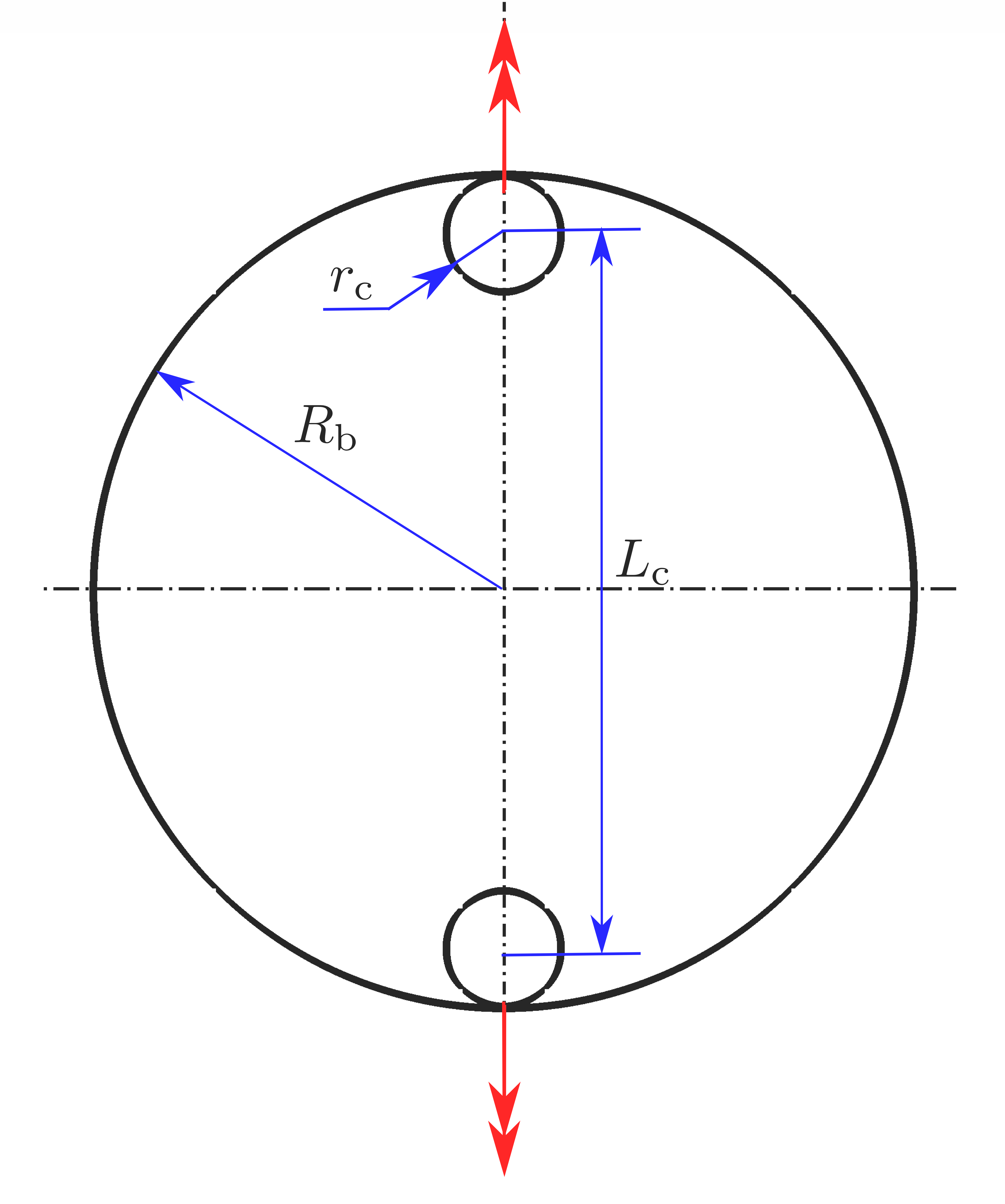}}
\put(-3.5,0.0){\includegraphics[height=0.30\textwidth]{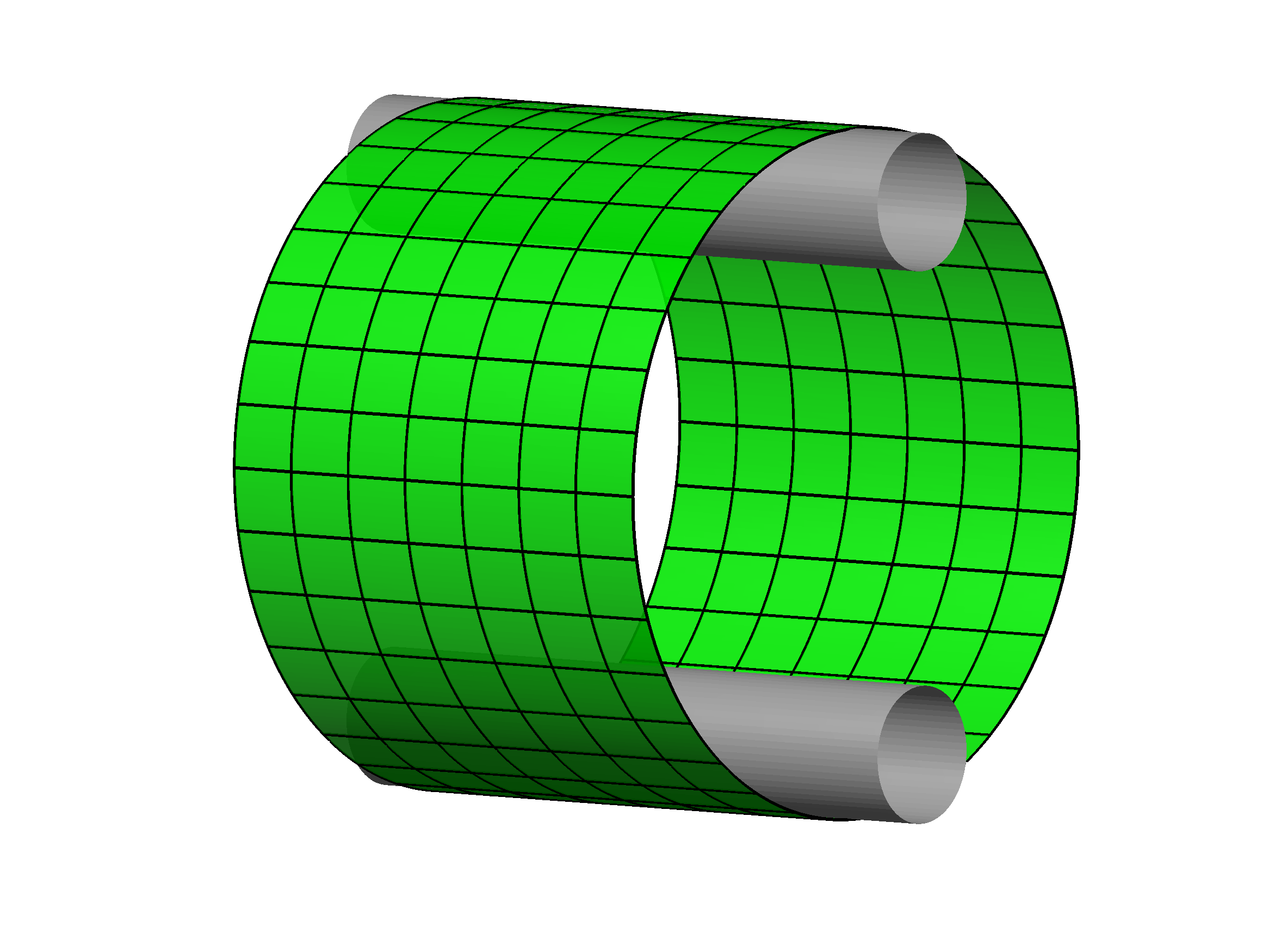}}
\put(2.3,0.0){\includegraphics[height=0.30\textwidth]{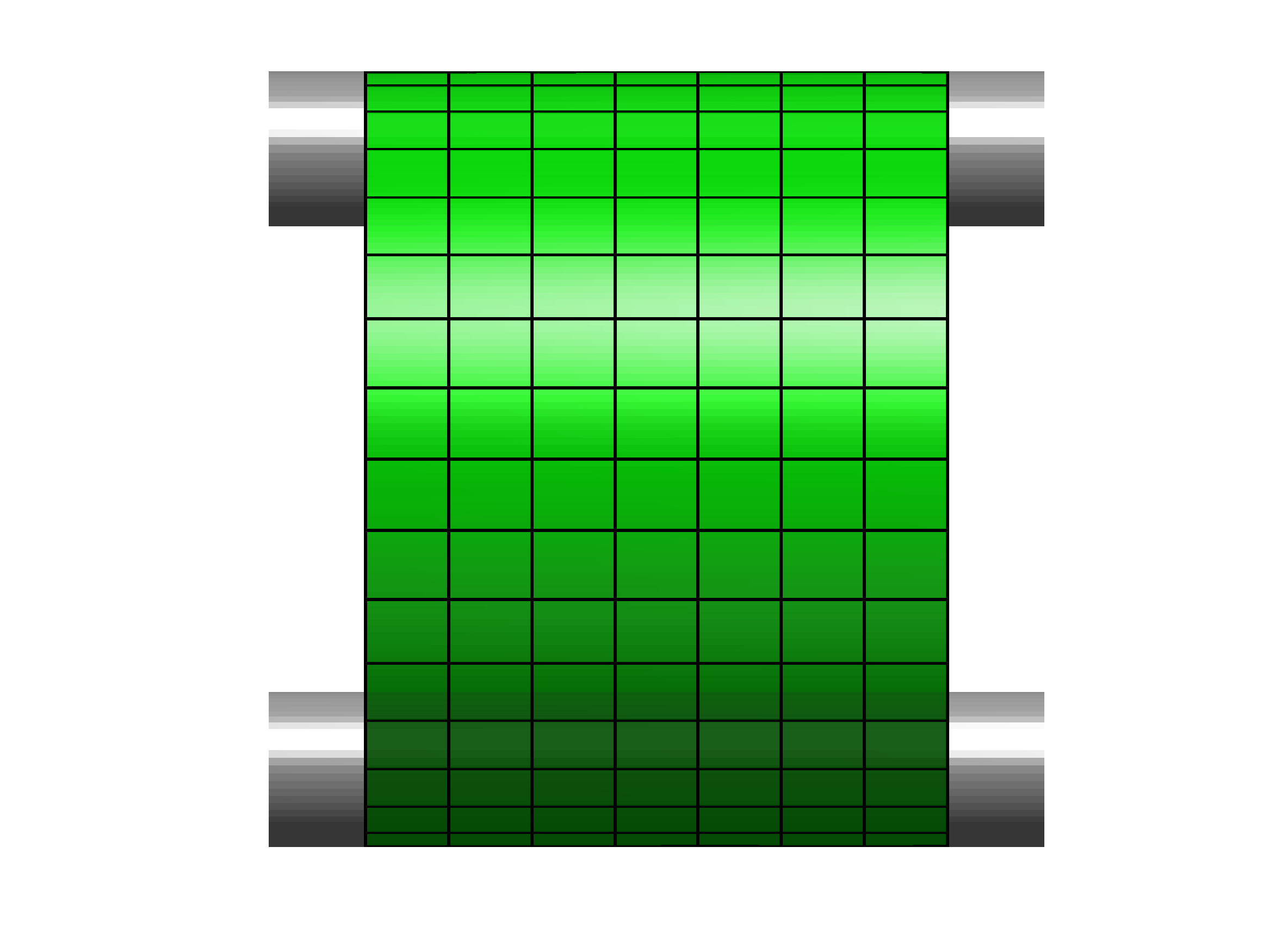}}
\end{picture}
\caption{Stretching of a textile band: Initial configuration at different views (from left to right): Side view, 3D view, and front view, respectively.  }
\label{f:rubber_x0}
\end{center}
\end{figure}

The last example demonstrates that the proposed angle plasticity model -- which is a 2D model within the tangent plane -- also works in the presence of large 3D deformations. Therefore, the following twisting example is considered: A cylindrical textile band 
with initial radius $R_{\mrb} = 45\,L_0$ and width $67.5\,L_0$, see Fig.~\ref{f:rubber_x0}, is placed in contact with two rigid cylindrical tubes with radius $r_{\!\mrc} = 9\,L_0$ and distance $L_\mrc =  72.0009\,L_0$.  This placement leaves a small initial penetration necessary for stabilizing  contact within the penalty formulation. The textile band is discretized by $7\times 34$  quadratic NURBS finite elements.  The material parameters from Tab.~\ref{t:WFconstant} are used for  the textile band.  For the subsequent simulation, the finite element nodes on the twisting axis are fixed in the horizontal directions in order to prevent rigid body motion of the textile band.

\begin{figure}[H]
\begin{center} \unitlength1cm
\begin{picture}(0,6.0)

\put(-8.1,-0.15){\includegraphics[width=0.50\textwidth]{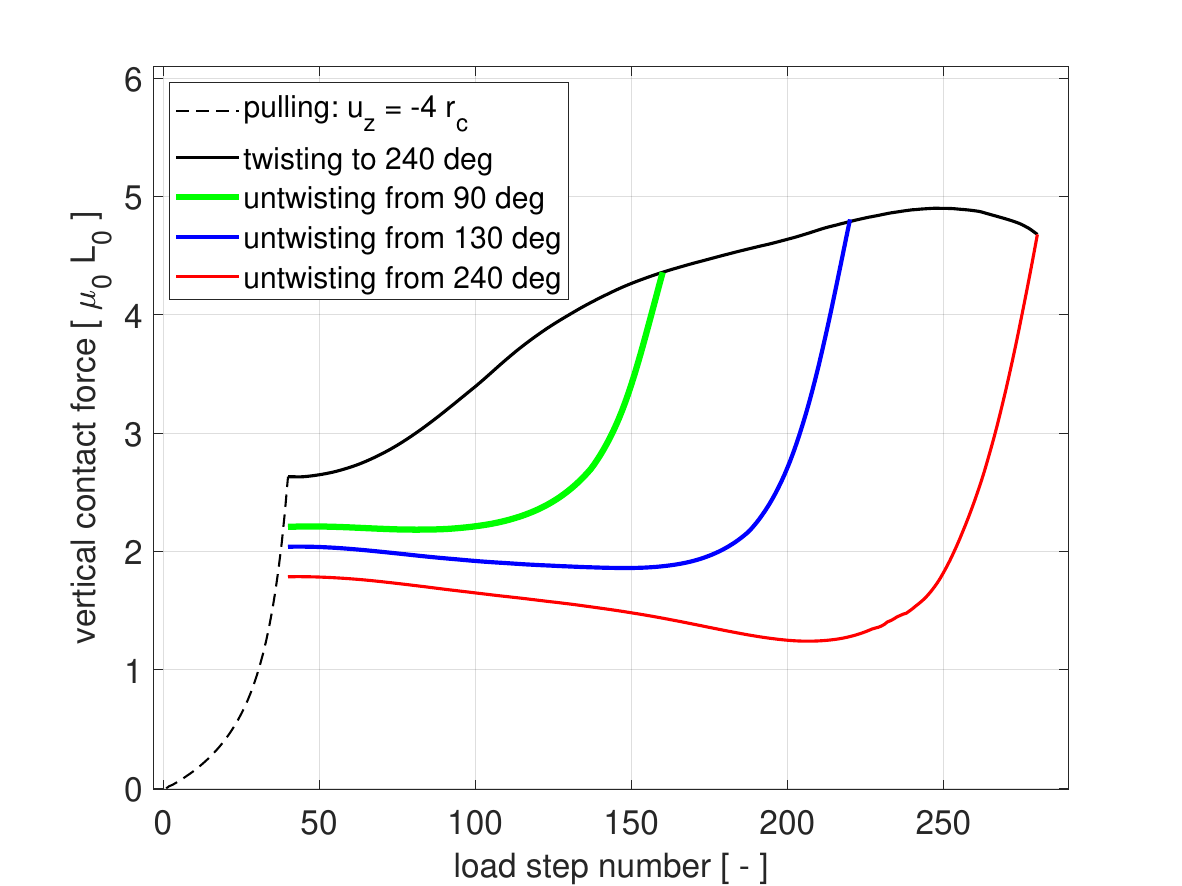}}
\put(0.6,-0.15){\includegraphics[width=0.50\textwidth]{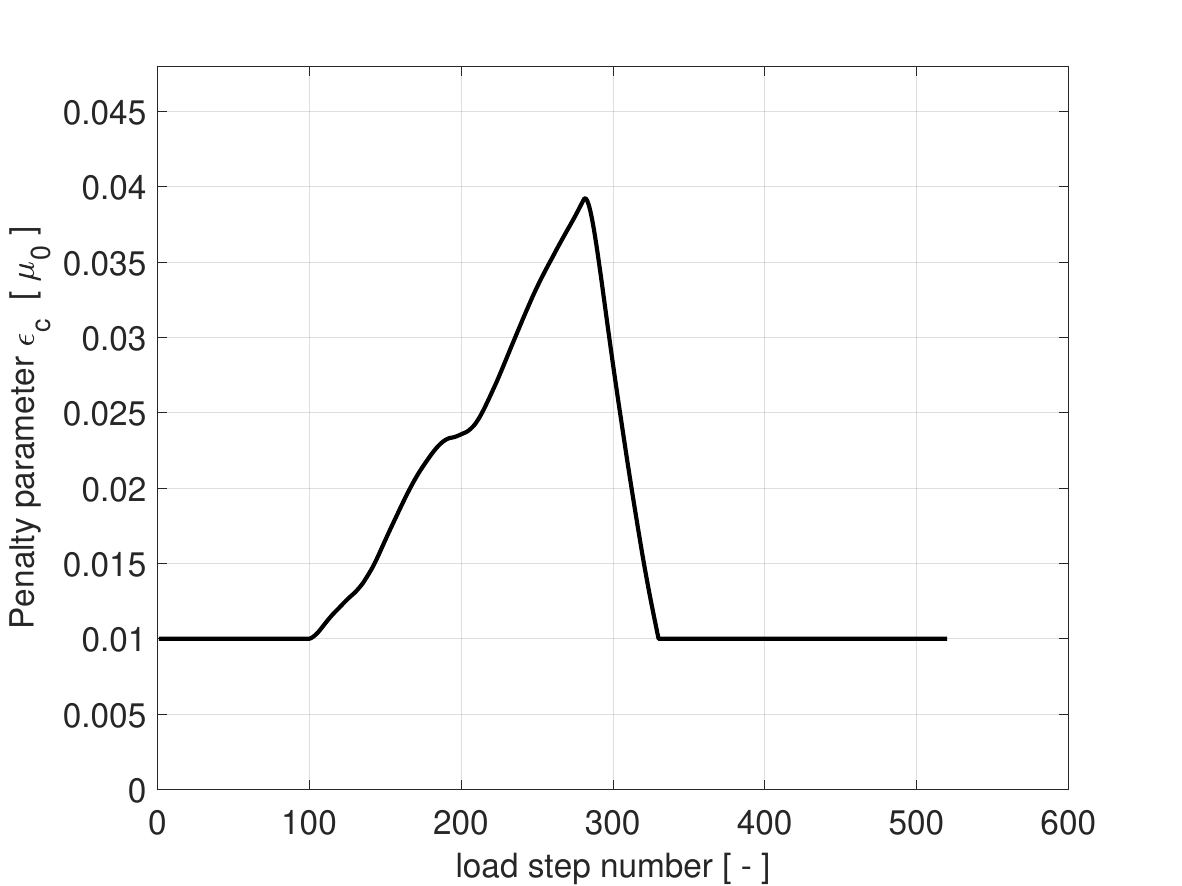}}
\put(-8.0,-0.1){{\small{(a)}}}
\put(0.4,-0.1){{\small{(b)}}}

\end{picture}
\caption{Twisting of a textile band: (a)~vertical reaction (i.e.~contact) force and (b) penalty parameter versus load step number. The load step number in (a) is reversed during unloading.}
\label{f:rubber_load_disp}
\end{center}
\end{figure}

\begin{figure}[H]
\begin{center} \unitlength1cm
\begin{picture}(0,22.2)
\put(-8.1,15.0){\includegraphics[height=0.44\textwidth]{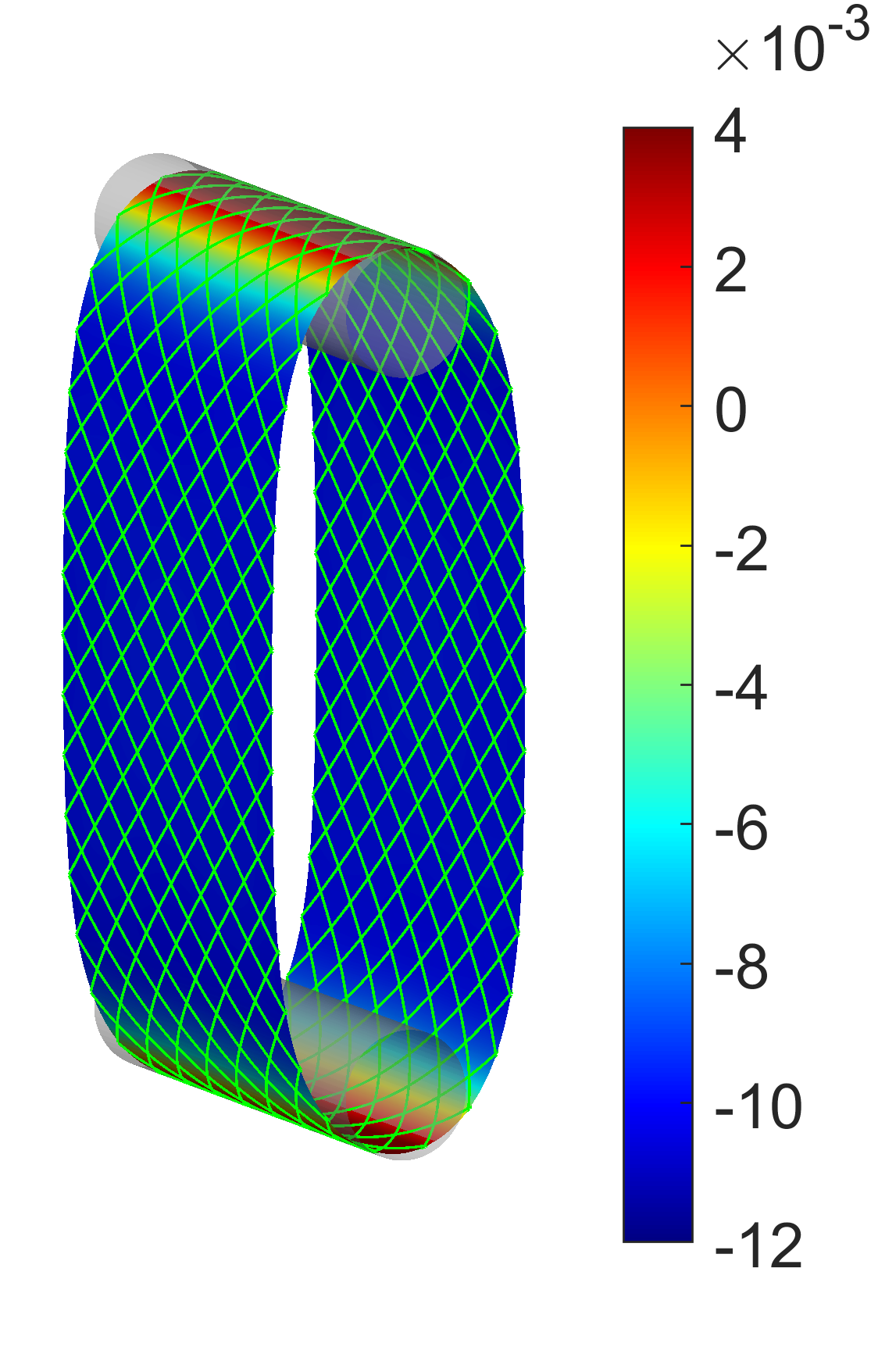}}
\put(-3.1,15.0){\includegraphics[height=0.44\textwidth]{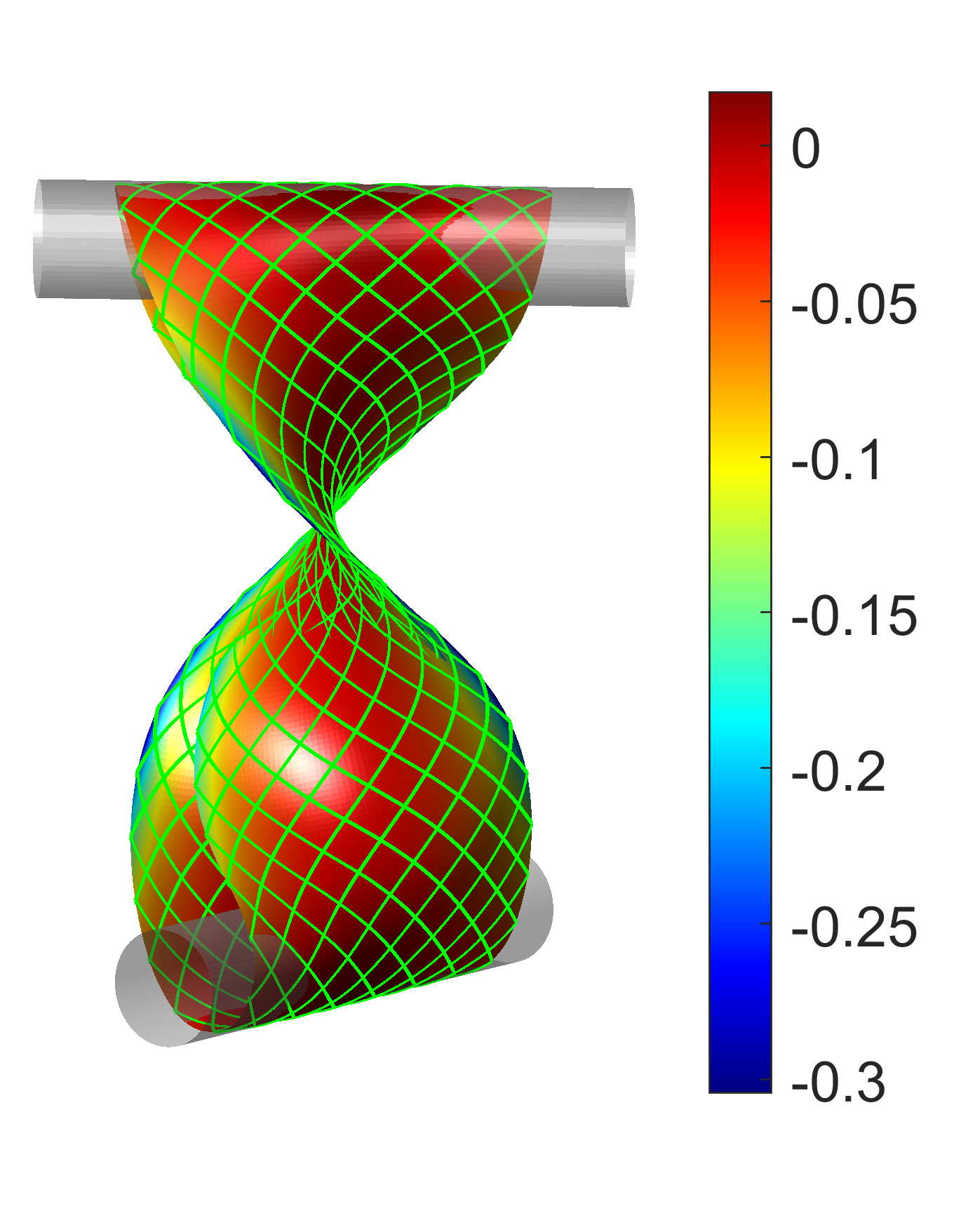}}
\put(2.5,15.0){\includegraphics[height=0.44\textwidth]{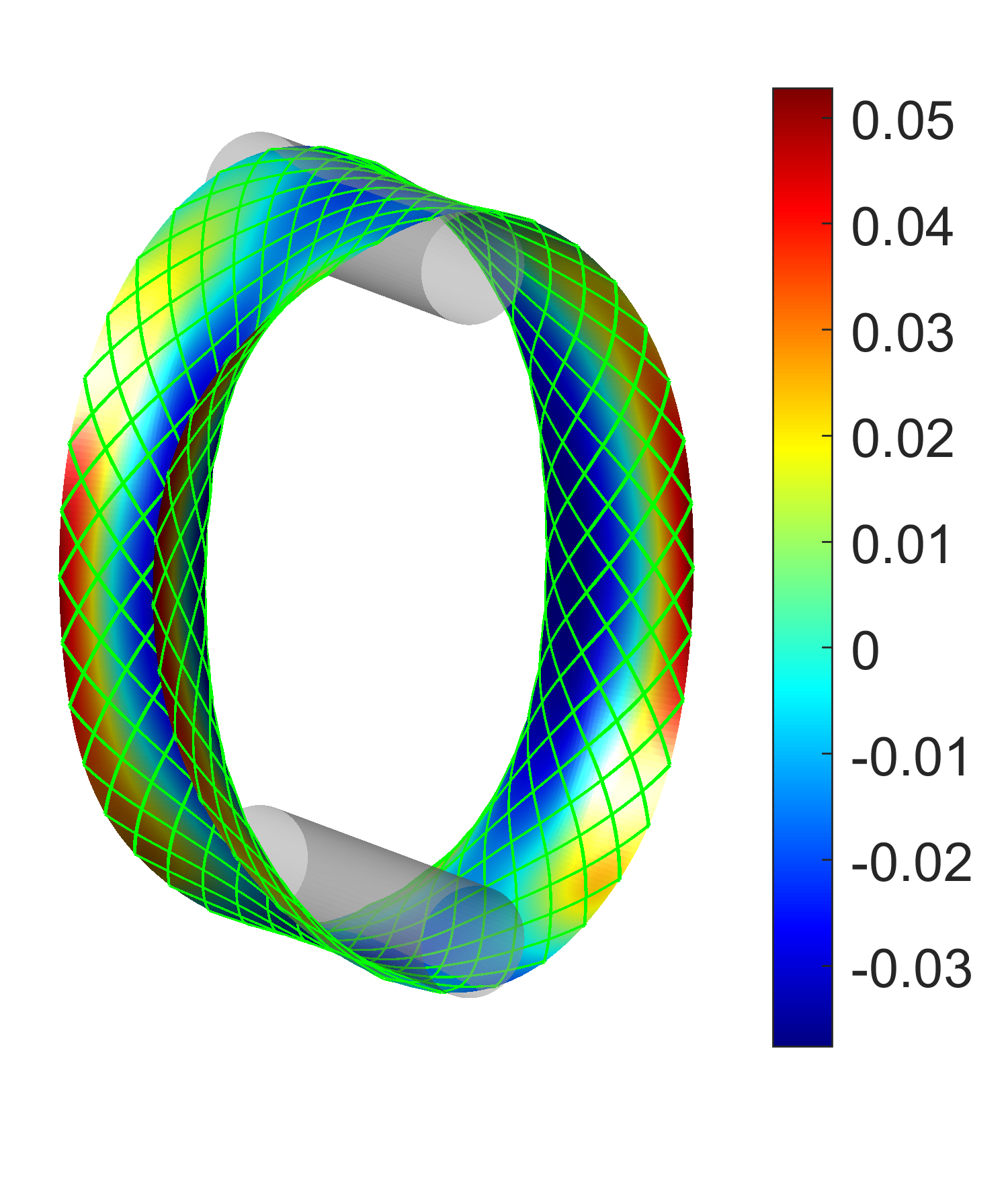}}
\put(-8.1,7.5){\includegraphics[height=0.44\textwidth]{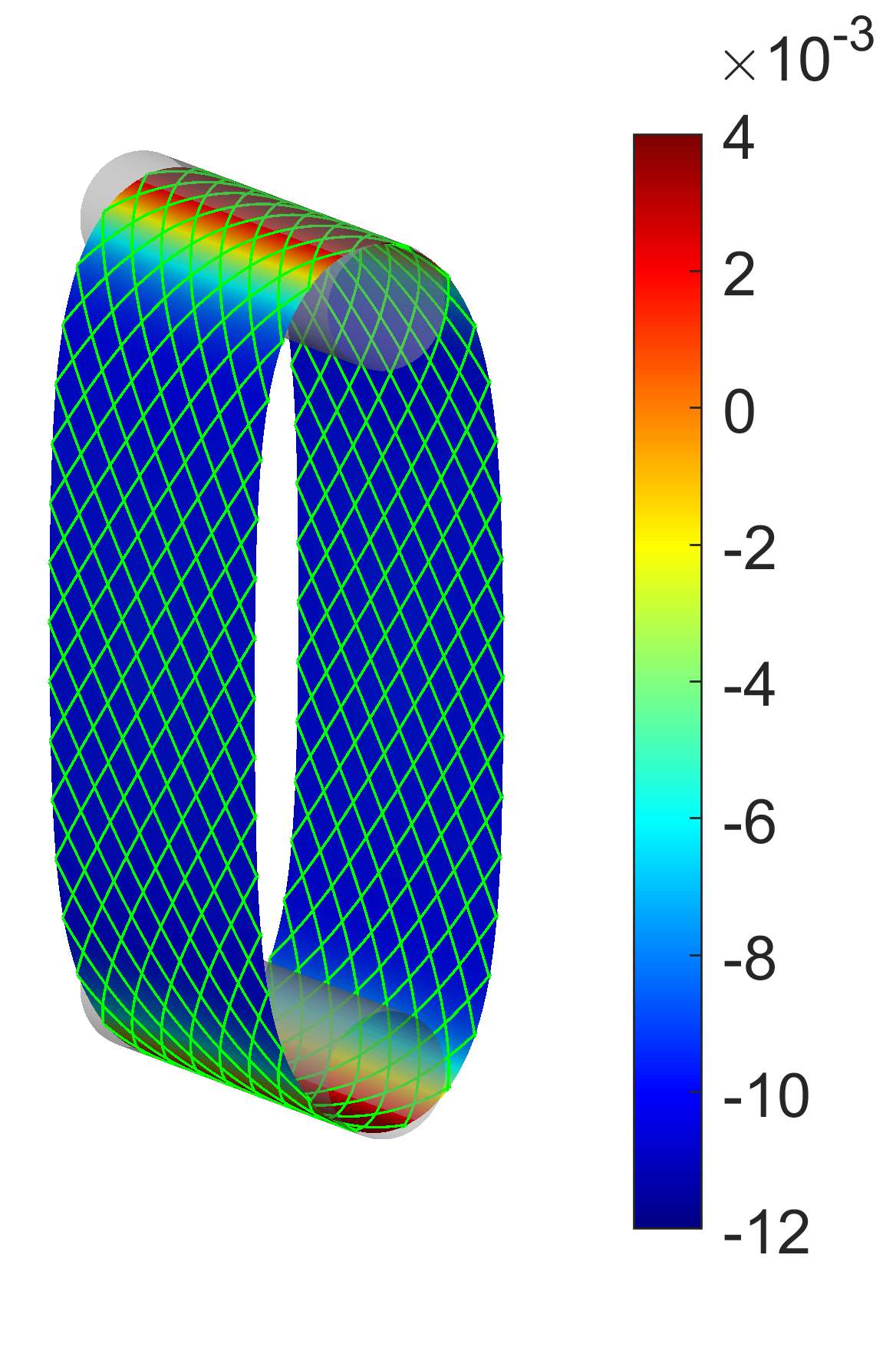}}
\put(-3.1,7.5){\includegraphics[height=0.44\textwidth]{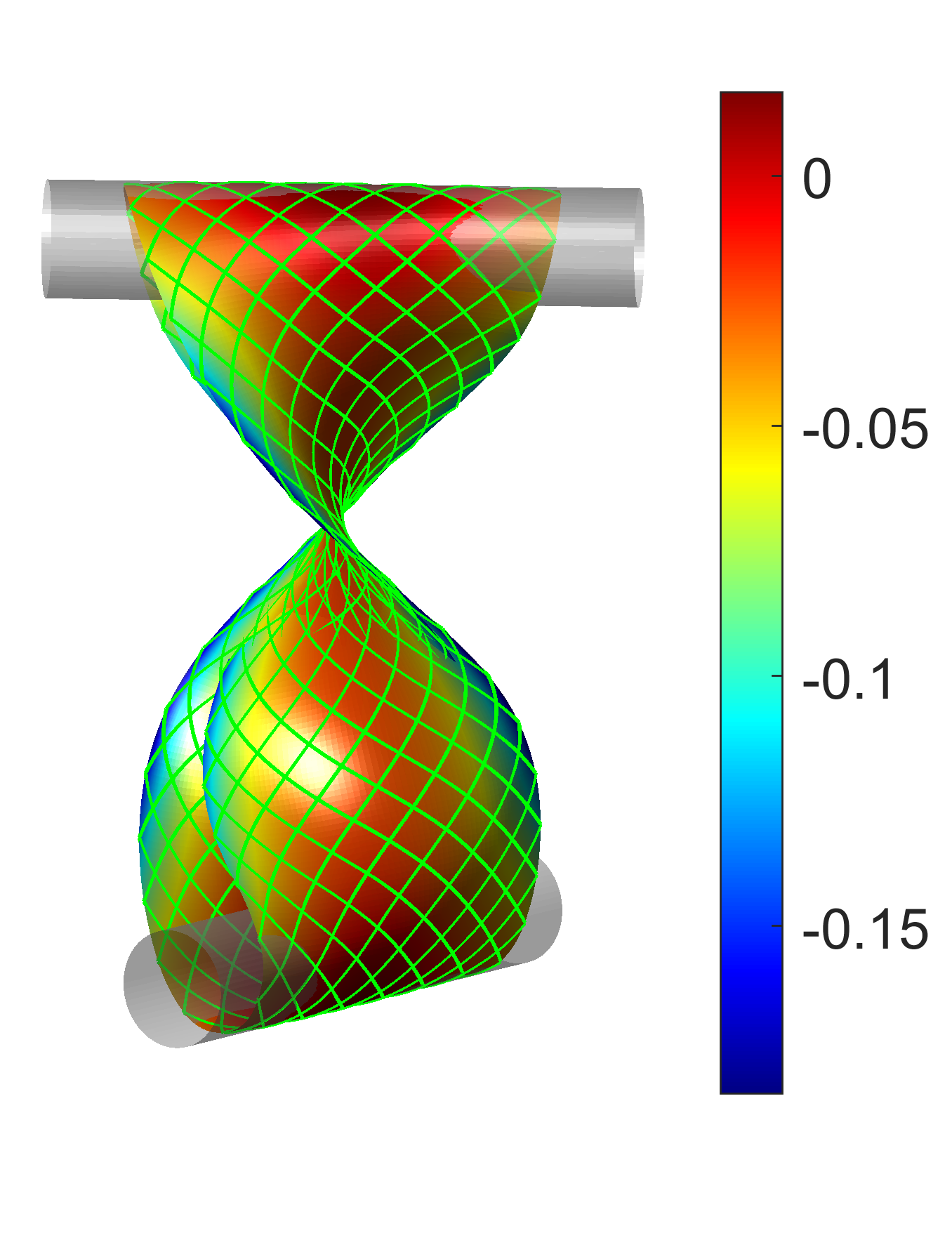}}
\put(2.5,7.5){\includegraphics[height=0.44\textwidth]{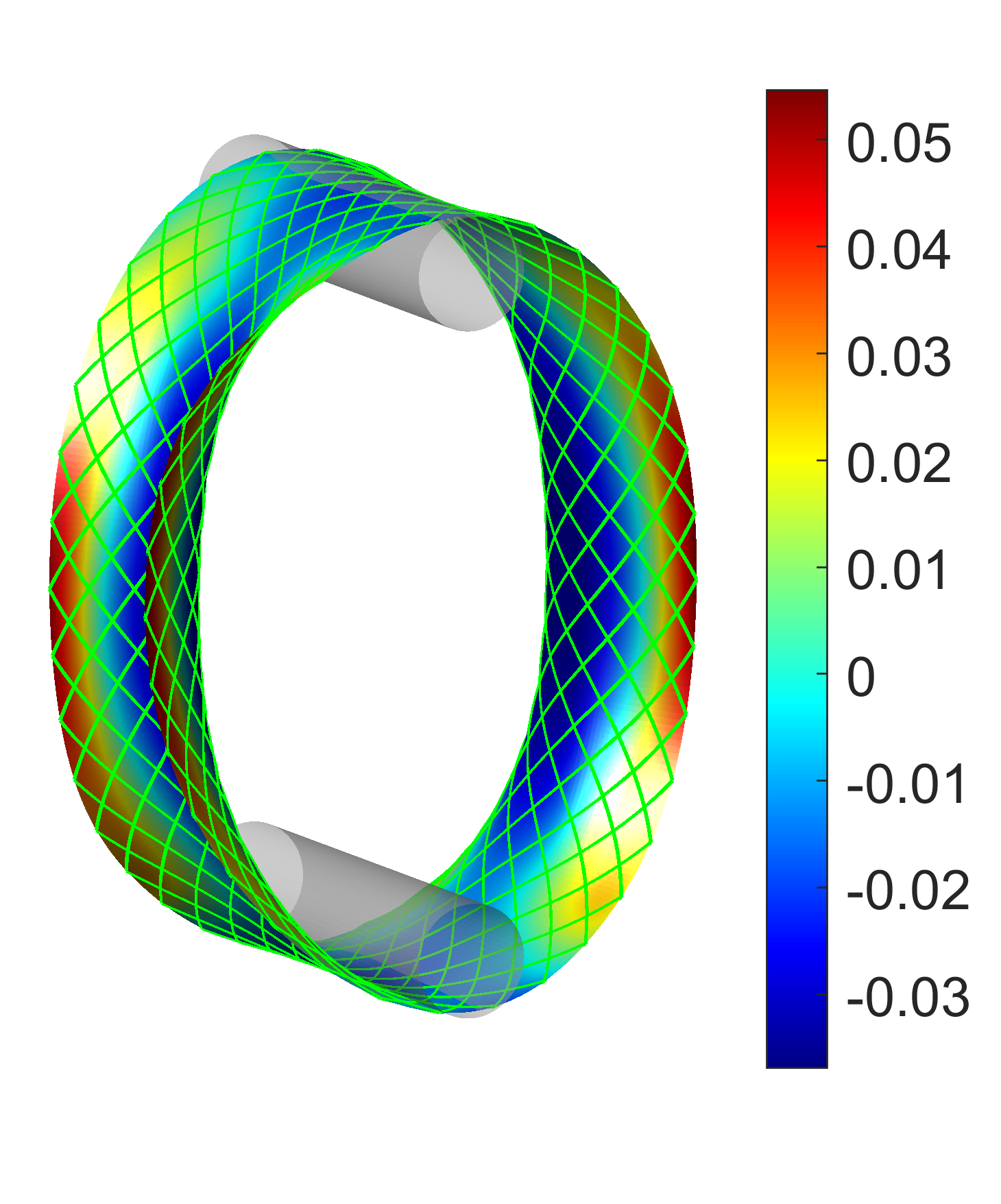}}
\put(-8.1,0.0){\includegraphics[height=0.44\textwidth]{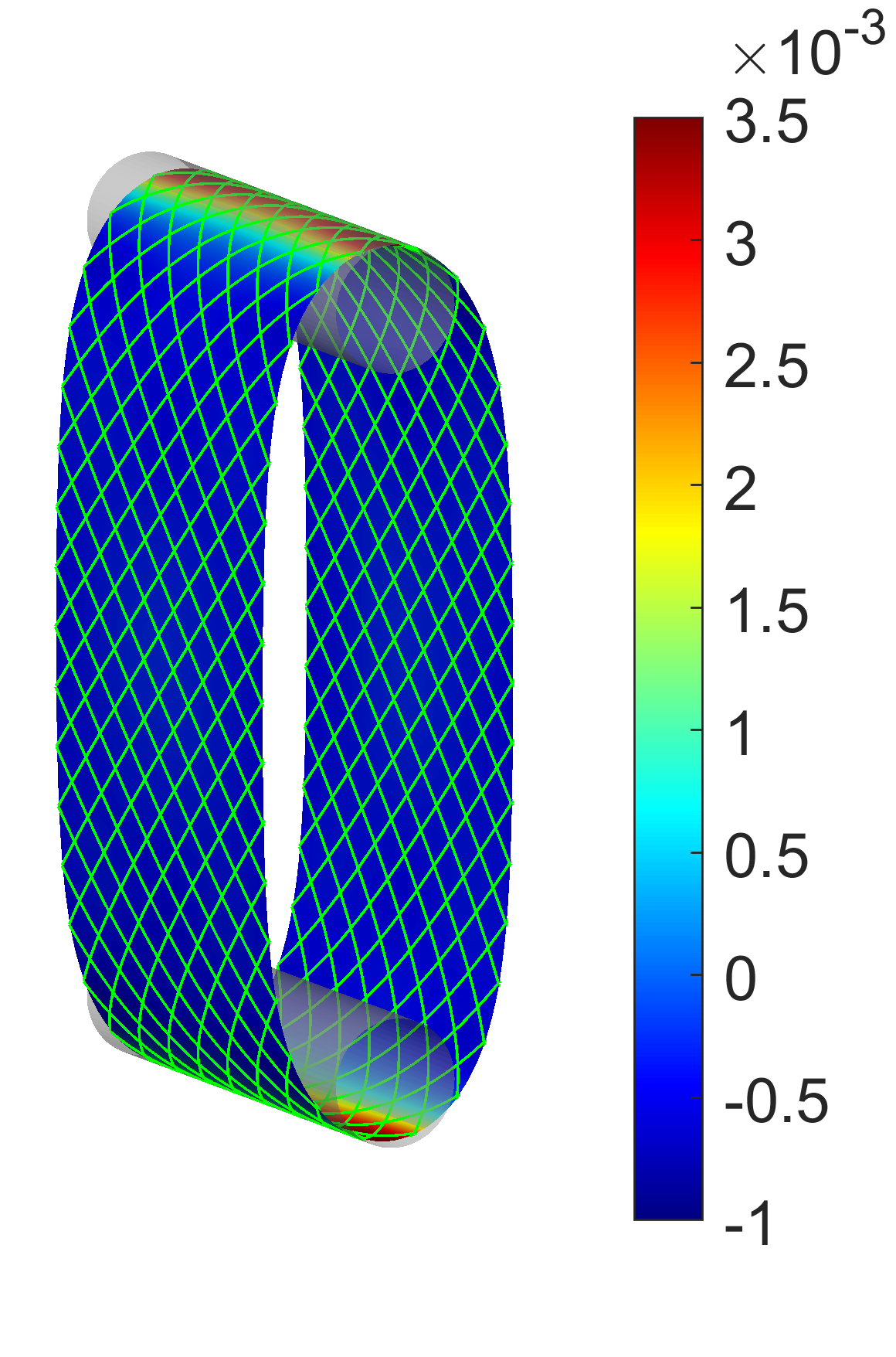}}
\put(-3.1,0.0){\includegraphics[height=0.44\textwidth]{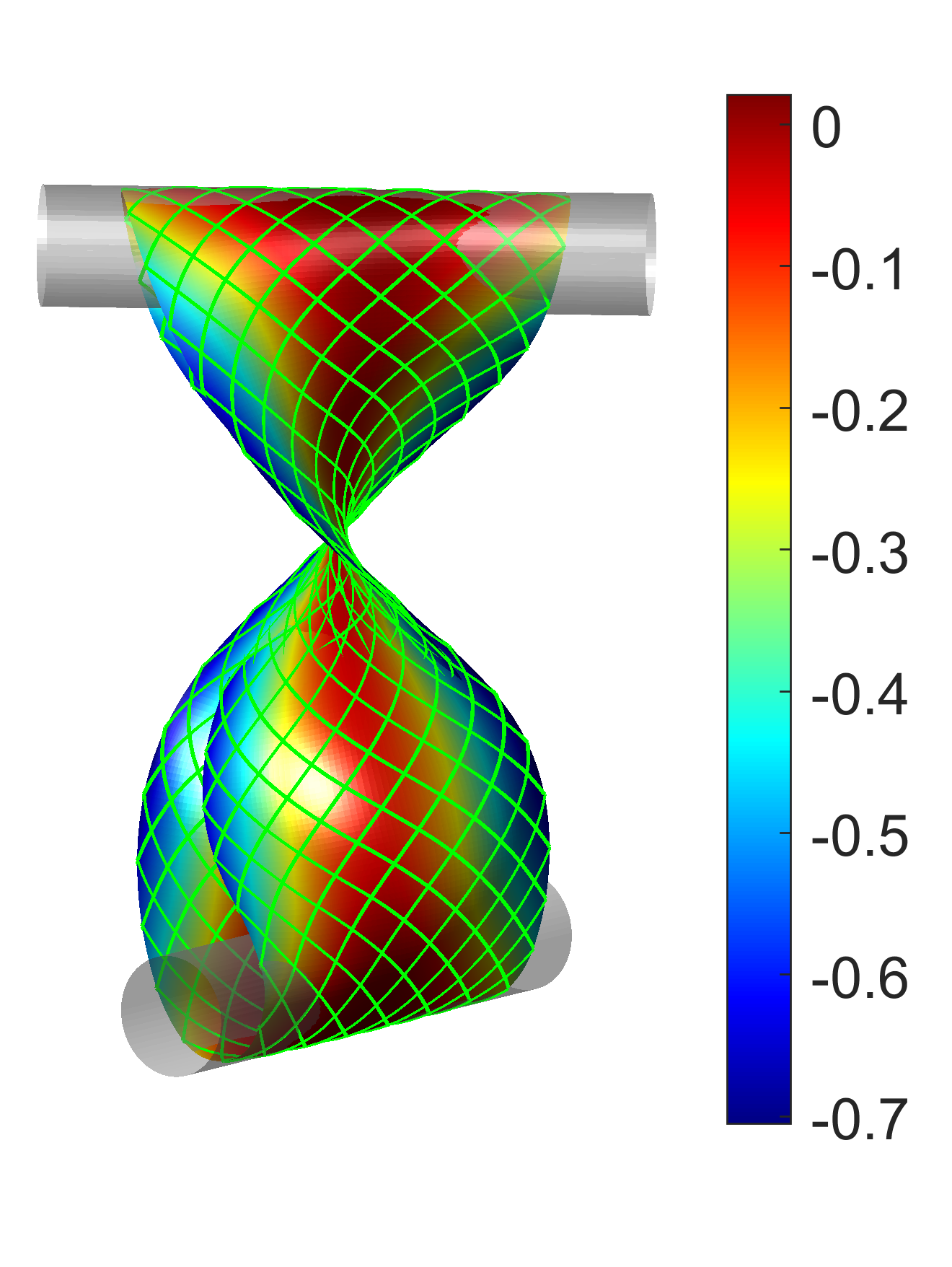}}
\put(2.5,0.0){\includegraphics[height=0.44\textwidth]{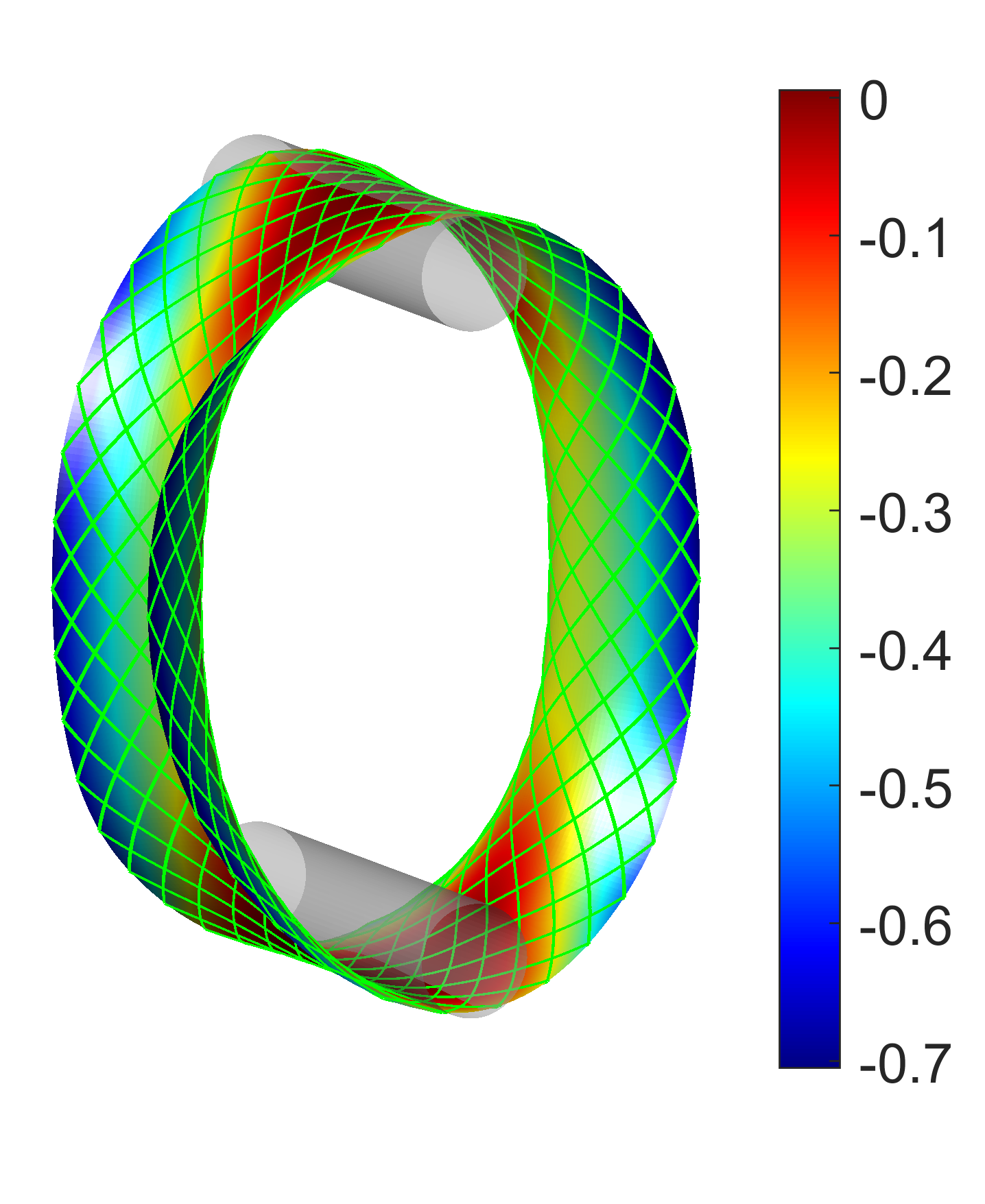}}
%
\put(-7.7,15){(a) shear stress $\sigma_{12} ~ [\mu_0]$}
\put(-7.7,7.5){(b) elastic angle change $\phi_\mre$ [rad]}
\put(-7.7,0){(c)  plastic angle change $\phi_\mrp$ [rad]}
\end{picture}
\caption{Twisting of a textile band: (a) shear stress, (b) elastic angle change and (c) plastic angle change before twisting, at twisting angle $240^\circ$ and after untwisting (left to right). The green lines illustrate the two considered  fiber families. }
\label{f:rubber_xx}
\end{center}
\end{figure}

Before twisting, the textile band is first stretched by moving the cylindrical tubes apart by $\bar{u}^{\max}_z = 4.0\,r_{\!\mrc}$ within $40$ load steps. The stretched configuration has minor angle plasticity and is shown in  Fig.~\ref{f:rubber_xx} (first column).  The textile band is then twisted by applying a rotation  to both the upper and the lower tubes in opposite directions 
 from starting angle $\phi \equiv0^\circ$  up to $\phi \equiv \phi_{\mathrm{max}}/2$ with $\phi_{\mathrm{max}}= 240^\circ$.   Finally, the textile band is untwisted back to starting angle $\phi \equiv 0^\circ$. The twisting and untwisting uses $2$ load steps per degree. 
The contact force resulting from the above loading sequence is shown in Fig.~\ref{f:rubber_load_disp}a.

For an efficient contact simulation, we employ an adaptive penalty parameter for contact between the textile band and the cylindrical tubes as shown in  Fig.~\ref{f:rubber_load_disp}b. Further, in order to restrict the source of dissipation  to angle plasticity, sticking is assumed during textile self-contact (by using a very large friction coefficient). Thus no  energy is dissipated due to textile self-contact, while at the same time this helps to stabilize  the textile band at large deformations (due to an increase of the effective stiffness for in-plane shear and out-of-plane bending at the contact zone).


In order to stabilize the textile band for fiber compression, we employ here the stabilization technique proposed by \cite{shelltextileIGA} with $\epsilon_\mathrm{stab}^\mre = 10\,\mu_0$, and  $\epsilon_\mathrm{stab}^\mrv = 100\,\mu_0$. Further, to deal with the instability due to shell buckling, we use mass-proportional (Rayleigh) damping with proportionality coefficient   $\nu = 10^{-8} \mu_0\,\Delta t$, where $\Delta t$ denotes the (pseudo) time increment, taken as unity in the example.

Fig.~\ref{f:rubber_xx}(a-c) plot the shear stress, elastic angle $\phi_\mre$, and plastic angle $\phi_\mrp$, respectively.  As the figures show residual deformations and stresses remain after unloading. The results demonstrate that the proposed plasticity model can be also used for large 3D deformations.

\section{Conclusion}{\label{s:conclude}}
We have presented a nonlinear shear  elastoplasticity model  for dry \res{woven} fabrics within the theoretical framework of \cite{shelltextile} that is based on general anisotropic Kirchhoff-Love shells with  embedded fibers. 
Therefore the change of the fiber angle -- the angle between two fiber families -- is split into elastic and plastic parts. 
It is shown that these angle changes can be induced by  strain tensors of the displacement gradient. With this, a yield function 
 with isotropic hardening is proposed based on the observed experimental data of \cite{Cao2008a}. The yield function uses seven parameters to  capture the three phases \resc{occurring} in the rotational inter-ply friction between fiber families: initial sticking,  rotational sliding friction with low plastic resistance, and increased hardening due to yarn-yarn locking.  The elastoplasticity formulation  is solved by a predictor-corrector algorithm, which is then implemented within the isogeometric Kirchhoff-Love shell formulation of \cite{shelltextileIGA}. In order to verify the finite element implementation,  the analytical solution of the picture frame test is presented for the proposed  elastoplasticity model. The  analytical solution captures multiple loading and unloading cycles by accounting for the expanding yield surface.


The proposed plasticity model is calibrated from the picture \res{frame} test and validated by the bias extension test for six different samples sizes.  It is observed that the proposed model  consistently predicts the \resc{experimental} data, especially during the sticking and rotational sliding phase. In the increased hardening phase, where the \resc{experimental} data becomes more scattered across different sources, our model prediction curves lie between the mean values of the \resc{experimental} data.

\res{Although the presented work focuses on simulations of woven fabrics, it is highly relevant for the efficient and accurate study of textiles in general, since the formulation employs several recent advances in both theoretical and computational modeling: The underlying shell theory facilitates isogeometric discretization that offers  smooth and accurate, yet efficient surface descriptions in comparison to classical finite element methods. The smoothness enables rotation-free formulations with only three degrees of freedom per control point, yet is capable of capturing the sophisticated kinematics of in-plane and out-of-plane bending. With this, our computational shell model can be used to study the complex deformations appearing in challenging examples such as wrinkling, self-contact and mold draping of textiles. }

\res{Furthermore, this work has demonstrated the ability of the \textit{surface invariant-based} approach to capture finite shear elastoplasticity in anisotropic Kirchhoff-Love shells.  It serves as a basis for developing similar plasticity models for bending, twisting, and stretching of anisotropic shells, gradient continua, and textiles.}

 \res{It is noted that our calibration and validation is based  on the experimental data that is available in the literature, which only captures loading but not unloading. Further calibration and validation therefore calls for the experimental investigation of unloading from various load levels for the same material in consideration. Another important extension of this work is exploring kinematic hardening in shear elastoplasticity of woven fabrics, as this may be required for the accurate description of  cyclic loading. Apart from the shear response, a validated elastoplasticity model for bending  (including out-of-plane bending, in-plane bending, and torsion) is still required and should be investigated in future work. These aspects also call for further experimental investigations: Ideally a wide range of various loading, unloading and cycling tests should be conducted for a given textile material in order to fully calibrate and validate all modes of deformation.}

\appendix

\section{Angle measures as invariants induced by strain tensors}{\label{s:prove_invariant}}

This section shows that the angle strain measures~\eqref{e:definePhi1}-\eqref{e:definePhi3}  can be  induced by strain tensors. To this end,  similarly to Eq.~\eqref{e:DFdef_splitp_defined}, we  extract the part of deformation gradient \eqref{e:defgrad} that excludes the change in fiber lengths at current configuration  $\sS$. I.e.
\eqb{lll}
\bar\bF \dis \ds \sum_{\res{I}=1}^2\bF\,(\bL_\res{I}\otimes\bL^\res{I})\,\lambda_\res{I}^{-1}=\bell_\res{I}\otimes\bL^\res{I} ~.\label{e:DFdef_split0_defined}
\eqe
The corresponding  right Cauchy-Green surface tensors following from  Eq.~\eqref{e:DFdef_split0_defined} and \eqref{e:DFdef_splitp_defined} are
 \begin{empheq}[ ]{align}
\bar\bC &:= \bar\bF^\mrT\,\bar\bF= \bar{a}_{\alpha\beta}\,\bA^\alpha\otimes\bA^\beta =  \theta_\res{IJ}  \, \bL^\res{I}\otimes\bL^\res{J}~,\label{e:Cpdef_split0_defined}\\[2mm]
\bC_{\!\mrp} &:= \bF_{\!\mrp}^\mrT\,\bF_{\!\mrp} = \hat{a}_{\alpha\beta}\,\bA^\alpha\otimes\bA^\beta =  \thetahat_\res{IJ} \, \bL^\res{I}\otimes\bL^\res{J}~,
\label{e:Cpdef_splitp_defined}
\end{empheq}
where   $\theta_\res{IJ} $  and  $\thetahat_\res{IJ}$ are defined by Eq.~\eqref{e:defineGamma_0} and \eqref{e:defineGamma_hat}, respectively, and where $\bar{a}_{\alpha\beta}$ and $\hat{a}_{\alpha\beta} $ are the components of tensors $\bar\bC$ and $\bC_{\!\mrp}$, respectively, with respect to  base $\bA^\alpha\otimes\bA^\beta$ in the reference configuration. From Eq.~\eqref{e:Cpdef_split0_defined} and \eqref{e:Cpdef_splitp_defined}, they can be expressed in terms of the fiber angles as
\eqb{lll}
\bar{a}_{\alpha\beta} = \bA_\alpha\,\bar\bC \,\bA_\beta  =  L^\res{IJ}_{\alpha\beta}\,\theta_\res{IJ} ,\\[3mm]
\hat{a}_{\alpha\beta} = \bA_\alpha\,\bC_{\!\mrp} \,\bA_\beta = L^\res{IJ}_{\alpha\beta}\,\thetahat_\res{IJ}~,
\label{e:F_relate_theta}
\eqe
where $L^\res{IJ}_{\alpha\beta}:=(\bL^\res{I}\otimes\bL^\res{J}): (\bA_\alpha\otimes\bA_\beta)$  denote the curvilinear components of   structural fiber tensors $\bL^\res{I}\otimes\bL^\res{J}$ in the reference configuration. Note that fiber family index \res{$I$ in a fiber pair} and curvilinear coordinate index $\alpha$ can be raised or lowered by \resc{their} corresponding metrics, e.g.~$L_\res{IJ}^{\alpha\beta}= \Theta_\res{IK}\,\Theta_{ \res{LJ}}\, L_{\gamma\delta}^{\res{KL}}\,A^{\alpha\gamma}\,A^{\delta\beta} $.

Also following from Eq.~\eqref{e:Cpdef_split0_defined} and \eqref{e:Cpdef_splitp_defined}, we find that angle measures~\eqref{e:definePhi1}-\eqref{e:definePhi3} can be expressed as
 \begin{empheq}[ ]{align}
\phi  &= \bL_1\,(\bar \bC- \bI)\,\bL_2 = L^{\alpha\beta}_{12}\,(\bar{a}_{\alpha\beta} - A_{\alpha\beta})~,\label{e:definePhi1b}  \\[2mm]
\phi_{\mre} &=  \bL_1\,(\bar \bC-\bC_{\!\mrp})\,\bL_2 = L^{\alpha\beta}_{12}\,(\bar{a}_{\alpha\beta} - \hat{a}_{\alpha\beta} ) ~,\label{e:definePhi2b}\\[2mm]
\phi_{\mrp} &=\bL_1\,(\bC_{\!\mrp}- \bI)\,\bL_2 = L^{\alpha\beta}_{12}\,(\hat{a}_{\alpha\beta} - A_{\alpha\beta})~,
\label{e:definePhi3b}
\end{empheq}
i.e. they are invariants of \res{strain tensors}  $\bar \bC-\bI$, $\bar \bC-\bC_{\!\mrp}$ and $\bC_{\!\mrp}-\bI$, respectively. 

\remark{It can be verified that deformations characterized by \eqref{e:DFdef_splitp_defined} and \eqref{e:DFdef_split0_defined}, preserve the length of fiber \res{$I$} in configurations $\hat\sS$ and $\sS$, respectively. Indeed
 \begin{empheq}[ ]{align}
(\bF_{\!\mrp}\,\bL_\res{I})\cdot(\bF_{\!\mrp}\,\bL_\res{I}) &= \bL_\res{I}\,\bC_{\!\mrp} \,\bL_\res{I} =  \thetahat_\res{II} = \bellhat_\res{I}\cdot\bellhat_\res{I} =   1~,\\[2mm]
(\bar\bF\,\bL_\res{I})\cdot(\bar\bF\,\bL_\res{I}) &= \bL_\res{I}\,\bar\bC \,\bL_\res{I} =  \theta_\res{II} = \bell_\res{I}\cdot\bell_\res{I} =   1~,
\end{empheq}
where Eqs.~\eqref{e:defineGamma_hat} and \eqref{e:defineGamma_current} have been used, and no summation is applied on index \res{$I$}.}

\remark{As the consequence of this length preservation in configuration $\hat\sS$, the surface area changes during plastic deformation (i.e.~$J_{\mrp}  \neq 1$).}

\remark{A material model  based on invariants \eqref{e:definePhi1b}-\eqref{e:definePhi3b} induces only the effective membrane stress~(\ref{e:consticom}.1) that contributes to the first term of Eq.~\eqref{e:Giie_2}.   Indeed, in view of Eqs.~\eqref{e:Cpdef_splitp_defined}, \eqref{e:Cpdef_split0_defined}, \eqref{e:DFdef_split0_defined}, \eqref{e:defined_bahat} and \eqref{e:DFdef_splitp}, we can conclude that  tensors $\bar \bC$ and $\bC_{\!\mrp}$  are  only associated with  the first displacement gradient, which does not affect curvature tensors \eqref{e:Kten} and \eqref{e:Ktenb}. Invariants of this types usually facilitate the decomposition of stretching and bending, which is beneficial in the construction of shell material models. 
}


\section{\res{Finite element formulation}}\label{s:fe}
This section \res{recalls the finite element force vectors for  Kirchhoff-Love shells with embedded fibers \resc{from} \cite{shelltextileIGA}.  It} should be noted that the form of these equations is unchanged for plastic deformations. Only the  constitutive model changes \res{as discussed in Sec.~\ref{s:constitute} and \ref{s:modelelastoplasticity}.}
\subsection{FE discretization}
As seen from  weak form \eqref{e:Giie_2}, the last two terms  in the internal virtual work contain the virtual curvatures, which require at least $C^1$-continuity across the surface. Here, we use an isogeometric discretization to fulfill this requirement. Since this discretization can provide a smooth and accurate description of the geometry, the curvature measures can be computed directly from the geometry. As a result, only  three displacement dofs per control point are needed to properly describe the shell.


The  undeformed element domain $\Omega^e_0$ and the  deformed element domain $\Omega^e$ are interpolated as
\eqb{lll}
\bX = \mN_e\,\mX_e~, \quad $and$ \quad
\bx = \mN_e\,\mx_e~, 
\label{e:interpx}
\eqe
where $\mX_e$ and $\mx_e$ are  the initial and current positions of control points,  respectively. The array 
\eqb{lll}
\mN_e(\xi^\alpha):= [N_1\bone,\, N_2\bone,\, ...,\, N_{n_\mre}\bone]
\label{e:mNedefine}
 \eqe
contains the isogeometric shape functions \citep{borden11} for each control point, and $n_\mre$ is the number of control points defining the element. \res{In Eq.~\eqref{e:mNedefine}, $\bone:= \bI + \bN\otimes\bN = \ba_\alpha\otimes\ba^\alpha + \bn\otimes\bn$ denotes the 3D identity tensor.}

From  Eq.~\eqref{e:interpx} follows the interpolation of other kinematical quantities, such as
\eqb{lll}
\begin{aligned}
\delta\bx ~~\is \mN_e\,\delta\mx_e \,~~~, \\[1mm]
\ba_\alpha ~~\is \mN_{e,\alpha}\,\mx_e~\, ~,\\[1mm]
\delta\ba_\alpha ~~\is  \mN_{e,\alpha}\,\delta\mx_e ~,
\end{aligned}
\quad\quad\quad \quad
\begin{aligned}
\ba_{\alpha,\beta} ~~ \is \mN_{e,\alpha\beta}\,\mx_e ~,\\[1mm]
\ba_{\alpha;\beta}~~  \is \mN_{e;\alpha\beta}\,\mx_e ~, \\[1mm]
\delta\bar\bc_{,\alpha} ~~\is \mC_{e,\alpha}\,\delta\mx_e~.
\end{aligned}
\label{e:dxe}
\eqe
Here, we have defined the elemental arrays
\eqb{lllllll}
\mN_{e,\alpha} \dis  [N_{1,\alpha}\bone,\, N_{2,\alpha}\bone,\, ...,\, N_{{n_\mre},\alpha}\bone]~,\\[3mm]
\mN_{e,\alpha\beta} \dis [N_{1,\alpha\beta}\bone,\, N_{2,\alpha\beta}\bone,\, ...,\, N_{n_\mre,\alpha\beta}\bone]~,\\[3mm]
\mN_{e;\alpha\beta} \dis \mN_{,\alpha\beta} - \Gamma^\gamma_{\alpha\beta}\,\mN_{,\gamma}~,\\[3mm]

\mC_{e,\alpha} \dis 
   \big[ \sL^\gamma_\alpha \,(\otc{\bone}-2\,\bell\otimes\bell) - \sC^\gamma_\alpha\,\bell\otimes\bc - \sN^\gamma_\alpha\,\bell\otimes\bn \big]\,\mN_{,\gamma}  - \ell^{\gamma}\,(\bell\otimes\bc) \,\mN_{,\gamma\alpha} ~,
\label{e:shapeC}
\eqe
where the  scalars $\sL^\gamma_\alpha$, $\sC^\gamma_\alpha$ and $\sN^\gamma_\alpha$ are defined as
\eqb{lll}
\sL^\gamma_{\alpha}\dis -  \ell^\gamma\, \big(\lambda^{-1}\,c_{\beta}  \,{L}^\beta_{,\alpha} + \ell^{\beta}\, \Gamma_{\!\beta\alpha}^{\mrc}  \big )~,  \\[3mm]
\sC^\gamma_{\alpha} \dis  \lambda^{-1}\,{L}^\gamma_{,\alpha} - \ell^{\gamma}\, \big (\lambda^{-1}\,\ell_{\beta}  \,{L}^\beta_{,\alpha} + \ell^{\beta}\, \Gamma_{\!\beta\alpha}^{\uell} \big ) ~, \\[3mm]
\sN^\gamma_{\alpha} \dis c^\gamma\, \ell^{\beta}\,b_{\beta\alpha}~.
\label{e:defineAuxScalars}
 \eqe
Inserting \eqref{e:dxe} into \eqref{e:variations} gives
\eqb{rrl}
\delta \auab 
\is \delta\mx_e^\mrT\, \big(\mN_{e,\alpha}^\mrT\,\mN_{e,\beta} + \mN_{e,\beta}^\mrT\,\mN_{e,\alpha}\big)\,\mx_e~, \\[3mm]
\delta \buab \is \delta\mx_e^\mrT\,\mN_{e;\alpha\beta}^\mrT\,\bn~,\\[3mm]
%
\bar{M}_0^{\alpha\beta}\,\delta\bar{b}_{\alpha\beta} \is  \delta\mx_e^\mrT\, {\mC}_{e;\alpha\beta}^{\gamma\mrT}\, \ba_\gamma\, \bar{M}_0^{\alpha\beta} ~, 
\label{e:variations_app}
\eqe
where\footnote{Eq.~(\ref{e:variations_app}.3) here is equivalent to expression (53.3) in \cite{shelltextileIGA} but offers conciseness thanks to the definition of ${\mC}_{e;\alpha\beta}^{\gamma}$.}
\eqb{rrlllll}
  {\mC}_{e;\alpha\beta}^{\gamma} \dis - c^\gamma_{;\beta}\,\mN_{e,\alpha}  -\delta^\gamma_\alpha \mC_{e,\beta}~.
\label{e:shapeM}
\eqe

\subsection{FE force vectors}\label{s:fvectors}
The internal FE force vectors are obtained by inserting Eqs.~\eqref{e:dxe} and \eqref{e:variations_app} into Eq.~\eqref{e:Giie_2}. This gives
\eqb{l}
G^e_\mathrm{int} = \delta\mx_e^\mrT\,\big(\mf^e_\mathrm{int\tau} + \mf^e_{\mathrm{int}M} + \mf^e_{\mathrm{int}\bar{M}}\big)~,
\label{e:Pinth}
\eqe
where
\eqb{llll}
\mf^e_\mathrm{int\tau} 
\dis  \ds\int_{\Omega^e_0} \tau^{\alpha\beta} \,\mN_{e,\alpha}^\mrT\,\vaub\,\dif A~,\\[5mm]
\mf^e_{\mathrm{int}M} 
\dis  \ds\int_{\Omega^e_0} \, M_0^{\alpha\beta}\,\mN_{e;\alpha\beta}^\mrT\,\bn\, \dif A~,\\[5mm]
\mf^e_{\mathrm{int}\bar{M}} 
\dis  \ds\int_{\Omega^e_0} \, \bar{M}_0^{\alpha\beta}\,\mC_{e;\alpha\beta}^{\gamma\mrT}\,\ba_\gamma  \, \dif A~
\label{e:FEforces}
\eqe
denote the internal FE force vectors associated with membrane stretching, out-of-plane bending, and in-plane bending, respectively. \resc{Similarly}, discretization of the external virtual work in  Eq.~\eqref{e:Giie_3} gives 
\eqb{l}
G^e_\mathrm{ext} = \delta\mx_e^\mrT\,\big(\mf^e_{\mathrm{ext0}} + \mf^e_{\mathrm{ext}p}+\mf^e_{\mathrm{ext}t}+\mf^e_{\mathrm{ext}m} + \mf^e_{\mathrm{ext}\bar{m}}  \big) +  \delta\mx_A\cdot\mf^A_{\mathrm{ext}m_\nu}~,
\label{e:Pexth}
\eqe
with the external FE forces 
\eqb{lllll}
\mf^e_{\mathrm{ext0}}  \dis \ds\int_{\Omega^e_0}\mN_e^\mrT\,\bff_{\!0}\,\dif A~, \\[5mm]
\mf^e_{\mathrm{ext}p}   \dis \ds\int_{\Omega^e}\mN_e^\mrT\,p\,\bn\,\dif a~, \\[5mm]
\mf^e_{\mathrm{ext}t}   \dis \ds\int_{\partial_t\Omega^e}\mN_e^\mrT\,\bt\,\dif s~, \\[5mm]
\mf^e_{\mathrm{ext}m_\tau}   \dis -\ds\int_{\partial_{m\tau}\Omega^e}\mN_{e,\alpha}^\mrT\,\nu^\alpha\,m_\tau\,\bn\,\dif s ~,\\[5mm]
\mf^e_{\mathrm{ext}\bar{m}}   \dis  \ds\int_{\partial_{\bar{m}} \Omega^e} \mN_{e,\alpha}^\mrT\,\ell^{\alpha}\, \bar{m}\,\bc\,\,\dif s ~, ~~\\[5mm]
\mf^A_{\mathrm{ext}m_\nu}   \dis m_\nu\,\bn_A~.~~~~~~~~~~~~~~~~~~~~~
\label{e:fext}
\eqe
Here, the external body force $\bff$ is assumed to consist of a  dead load  $\bff_{\!0}$ and an external surface pressure  $p$, as  $\bff =  \bff_{\!0}/J + p\,\bn$~. 

\remark{The consistent tangent matrices resulting from the linearization of the virtual work can be found  in the literature. For example, the tangents of the forces in \eqref{e:FEforces} and  \eqref{e:fext} can be found in \cite{shelltextileIGA} -- see c.f.~Eq.~(60-64) and cf.~Eq.~(87) therein, respectively. Pressure loads, such as hydrostatic pressure, are discussed in  \cite{membrane}, while \cite{responsedavid2024} discuss external corner forces. }

\section*{Acknowledgments} 
 The funding of the German Research Foundation (DFG) through project SA1822/11-1 is gratefully acknowledged. Thang X. Duong thanks Prof.~Josef Kiendl for his support. 

\bigskip
\bibliographystyle{apalike}
\bibliography{sauerduong,bibliography}

\end{document}